\documentclass[12pt]{article}
\usepackage{epsfig,amssymb,amsmath,psfrag,subfigure,rotate,color}

\allowdisplaybreaks

\def\XXint#1#2#3{{\setbox0=\hbox{$#1{#2#3}{\int}$ }
\vcenter{\hbox{$#2#3$ }}\kern-.6\wd0}}

\def \be  {\begin{equation}}
\def \ee  {\end{equation}}
\def \ba  {\begin{eqnarray}}
\def \ea  {\end{eqnarray}}
\def \baa {\begin{eqnarray*}}
\def \eaa {\end{eqnarray*}}
\def \lab #1 {\label{#1}}

\newcommand\re[1]{(\ref{#1})}
\def\d{\hbox{{d}\kern-.20em\hbox{l}}}

\def \matrix #1 {\left(\begin{array}{cc} #1 \end{array}\right)}

\def \res{\mathop{\rm res}\nolimits}

\newcommand \ket [1] {|{#1}\rangle}
\newcommand \bra [1] {\langle {#1}|}
\newcommand{\bit}[1]{\mbox{\boldmath$#1$}}

\newcommand{\ft}[2]{{\textstyle\frac{#1}{#2}}}
\numberwithin{equation}{section}
\begin{document}

\begin{titlepage}

\thispagestyle{empty}

\vspace*{3cm}

\centerline{\large \bf Towards NMHV amplitudes at strong coupling}
\vspace*{1cm}

\centerline{\sc A.V.~Belitsky}

\vspace{10mm}

\centerline{\it Department of Physics, Arizona State University}
\centerline{\it Tempe, AZ 85287-1504, USA}

\vspace{2cm}

\centerline{\bf Abstract}

\vspace{5mm}

Pentagon Operator Product Expansion provides a non-perturbative framework for analysis of scattering amplitudes in planar maximally supersymmetric gauge theory
building up on their duality to null polygonal superWilson loop and integrability. In this paper, we construct a systematic expansion for the main ingredients of
the formalism, i.e., pentagons, at large 't Hooft coupling as a power series in its inverse value. The calculations are tested against relations provided by the 
so-called Descent Equation which mixes transitions at different perturbative orders. We use leading order results to have a first glimpse into the structure of 
scattering amplitude at NMHV level at strong coupling.

\end{titlepage}

\setcounter{footnote} 0

\newpage

\pagestyle{plain}
\setcounter{page} 1

{
\footnotesize 
\tableofcontents}

\newpage

\section{Introduction}

The formulation of the Pentagon \cite{Basso:2013vsa} Operator Product Expansion \cite{Alday:2010ku} for superWilson loop $\mathcal{W}_N$ on a null polygonal 
contour paved a way for unravelling analytical structure of scattering superamplitude $\mathcal{A}_N$ of $N$ particles, they are dual to
\cite{Alday:2007hr,Drummond:2007cf,Brandhuber:2007yx,CaronHuot:2010ek,Mason:2010yk,Belitsky:2011zm}, at any value of 't Hooft coupling in planar maximally 
supersymmetric Yang-Mills theory. This convergent series is developed in terms of elementary excitations $\psi$ of the color flux-tube propagating on the world-sheet 
stretched on the loop \cite{Basso:2013vsa},
\begin{align}
\mathcal{W}_N = \sum_{\psi_1, \dots, \psi_{3N - 15}}
\bra{0} \hat{\mathcal P} \ket{\psi_1} {\rm e}^{- \tau_1 E_{\psi_1} + \sigma_1 p_{\psi_1} + i \phi_1 m_1}
\bra{\psi_1} \hat{\mathcal P} \ket{\psi_2} 
 {\rm e}^{- \tau_2 E_{\psi_2} + \sigma_2 p_{\psi_2} + i \phi_2 m_2}
\dots
\bra{\psi_{3N-15}} \hat{\mathcal P} \ket{0} 
\, ,
\end{align}
with $(N-5)$ sets of three conformal cross ratios $\tau, \sigma, \phi$ which encode the shape of the boundary. In addition to the dispersion relations $E_{\psi} = E_{\psi} (u)$,  $p_{\psi} = p_{\psi} (u)$ 
(parametrized by the rapidity $u$) for $\psi$, which were known to all loops for quite some time \cite{Basso:2010in}, their form factor couplings to the contour $\bra{0} \hat{\mathcal P} \ket{\psi_i}$ 
and transitions amplitudes $\bra{\psi_i} \hat{\mathcal P} \ket{\psi_j}$ between adjacent squares in a geometric tessellation of the polygon were uncovered in a series of recent papers 
\cite{Basso:2013aha,Belitsky:2014rba,Basso:2014koa,Belitsky:2014sla,Basso:2014nra,Belitsky:2014lta,Belitsky:2015efa,Basso:2014hfa,Basso:2015rta}.

For a few notable exceptions \cite{Basso:2013vsa,Basso:2014koa,Basso:2014jfa,Fioravanti:2015dma}, recent literature was predominantly focused on perturbative analyses of scattering amplitudes at weak 
coupling where a plethora of data is available from different formalisms such as hexagon \cite{Dixon:2011nj,Dixon:2014iba} and heptagon 
\cite{Golden:2014xqa,Golden:2014xqf,Golden:2014pua,Drummond:2014ffa} bootstraps. The reason for this is that at each order in 't Hooft coupling there is only a very small number of flux-tube excitations
which determine the amplitude in question. At strong coupling on the contrary, summation over their infinite number should be performed to reproduce the minimal area result obtained within the 
Thermodynamic Bethe Ansatz \cite{Alday:2007hr,Alday:2009dv,Alday:2010vh} as well as as to systematically decode all higher order corrections in $1/g$. At leading order in the inverse coupling, this was 
effectively demonstrated recently in Ref.\ \cite{Fioravanti:2015dma} following the route outlined in Ref.\ \cite{Basso:2014koa} for MHV amplitudes. Since the dominant contribution at infinite coupling is 
essentially insensitive to the the helicity of external particles involved in scattering, one anticipates to find a factorized overall minimal area prefactor in non-MHV amplitudes as well \cite{Folklore}. 

In this work, we initiate a systematic study of the flux-tube pentagons in the perturbative regime at strong coupling. Presently, we will not attempt however to unravel the structure of nonperturbative 
${\rm e}^{- \pi g}$ corrections, though these can be systematically accounted for upon a more thorough consideration. They will become important in the analysis of the transition region from strong to 
finite and then weak coupling. Compared to other excitations,---fermions, gluons and bound states thereof,--- scalars, also known as holes, possess exponentially vanishing masses at strong coupling, 
potentially producing leading order contribution in the multi-collinear kinematics, i.e., $\tau_i \to \infty$. However, their effect in the amplitude is formally suppressed by inverse coupling relative to 
semiclassical string effects and, for this reason, we will ignore holes in the nonperturbative regime, though they were shown to provide an additive geometry-independent constant contribution to the 
minimal area due to their intricate infrared dynamics when resummed to all orders \cite{Basso:2014jfa}. We further comment on their contribution to NMHV amplitudes in the Conclusions.

For the exception of the scalars, which are not presently discussed, the perturbative string regime corresponds to the one where the rapidity of excitations scales with 't Hooft constant, $u = 2 g \hat{u}$
as $g$ is sent to infinity while $\hat{u}$ is kept fixed. For the gauge fields and bound states, the physical region of $\hat{u}$ corresponds to the interval $(-1,1)$, while for fermions, $\hat{u}$ 
resides on the small fermion sheet containing the point of the fermion at rest and thus varies over two semi-infinite segments $\hat{u} \in (-\infty, -1) \cup (1, \infty)$. It is for these values, the energy 
$E$ and momentum $p$ of these flux-tube excitations are of order one in $g$, i.e., $E, p \sim g^0$. Elsewhere, we are entering kinematics where they scale as a power of $g$ (fractional for 
near-flat or integer for semiclassical regimes) yielding exponentially suppressed contribution to the Wilson loop.

Our subsequent consideration is organized as follows. In the next section, we start with small fermions and solve their parity even and odd flux-tube equations in inverse powers of the
coupling. These are used then to construct direct and mirror S-matrices which enter as the main ingredients into the fermion-fermion pentagons. We continue in Sect.\ \ref{GluonSection}  with a 
similar consideration for gluons and their bound states. Due to complicated analytical structure on the physical sheet, we first pass to the half-mirror, or Goldstone, sheets and perform the strong 
coupling analysis there.  In this manner, we find bound-state--bound-states pentagons. In Sect.\ \ref{FermionGluonSection}, we use the results of the previous two sections to find mixed 
fermion--bound-states pentagons verifying consistency of our findings by means of the exchange relations. Another layer of consistency checks arises from consequences of the Descent 
Equation for superamplitudes in Sect.\ \ref{DESection}. Finally, we apply our construction to resum the entire series of gluon bound states and effective fermion-antifermion strong-coupling 
bound pairs for a particular component of NMHV amplitude, observing anticipated factorization of the minimal area from a helicity-dependent prefactor. Finally, we conclude. Several appendices 
contain compendium of integrals needed for calculations involved as well as a list of results which are two cumbersome to be quoted in the main text.

\section{Small fermion transitions}
\label{SmallFermionSection}

Let us start our consideration with fermions. As we advertised in the Introduction, only the fermion living on the small Riemann sheet survives at strong coupling. In the theory of the flux-tube, the  
direct and mirror scattering matrices for small-fermion--small-(anti)fermion  elementary excitations (and consequently their pentagon transitions) \cite{Basso:2014koa,Belitsky:2014sla} 
\begin{align}
\label{DirectffANDfbarfSmatrices}
S_{\rm \bar{f}f} (u,v) 
&
=
S_{\rm ff} (u,v) 
\nonumber\\
&
= \exp \left( - 2 i f_{\rm ff}^{(1)} (u, v) + 2 i f_{\rm ff}^{(2)} (u, v) \right)
\, , \\
\label{MirrorffANDfbarfSmatrices}
S_{\rm \ast \bar{f}f} (u,v) 
&
= 
\frac{u - v + i}{u - v} S_{\rm \ast ff} (u,v) 
\nonumber\\
&
= \exp \left( 2 f_{\rm ff}^{(3)} (u, v) - 2 f_{\rm ff}^{(4)} (u, v) \right)
\, , 
\end{align}
are determined by means of the dynamical phases 
\begin{align}
\label{DirectFermionfs}
f_{\rm ff}^{(1)} (u, v) 
&
= - \frac{1}{2} \int_0^\infty \frac{dt}{t} \cos (v t) \widetilde\gamma^{\rm f}_{+, u} (2 gt)
\, , \quad
f_{\rm ff}^{(2)} (u, v) 
= - \frac{1}{2} \int_0^\infty \frac{dt}{t} \sin (v t) \gamma^{\rm f}_{-, u} (2 gt)
\, , \\
\label{MirrorFermionfs}
f_{\rm ff}^{(3)} (u, v) 
&
= + \frac{1}{2} \int_0^\infty \frac{dt}{t} \sin (v t) \widetilde\gamma^{\rm f}_{-, u} (2 gt)
\, , \quad
f_{\rm ff}^{(4)} (u, v) 
= - \frac{1}{2} \int_0^\infty \frac{dt}{t} \cos (v t) \gamma^{\rm f}_{+, u} (2 gt)
\, . 
\end{align}
which depend on the solutions to the flux-tube equations with sources specific to the type of excitation under consideration. Let us turn to the solution of 
the $u$-parity even and odd functions $ \gamma^{\rm f}_{-, u}$ and $ \widetilde\gamma^{\rm f}_{-, u}$, respectively, at strong coupling.

\subsection{General solution for even $u$-parity}

An infinite set of even $u$-parity flux-tube equations for the small fermion \cite {Basso:2010in,Belitsky:2014sla} can be cast in the form\footnote{This is achieved by means of the Jacobi-Anger 
summation formulas, and differentiation w.r.t.\ the rapidity $v$.}
\begin{align}
\label{FTEqf1}
&
\int _0^\infty dt \, \sin (v t)
\left[
\frac{\gamma^{\rm f}_{+, u} (2 g t)}{1 - {\rm e}^{- t}}
-
\frac{\gamma^{\rm f}_{-, u} (2 g t)}{{\rm e}^{t} - 1}
\right]
=
\frac{1}{2} \int_0^\infty dt \,  \sin (v t) \cos (u t)
\, , \\
\label{FTEqf2}
&
\int _0^\infty dt \, \cos (v t)
\left[
\frac{\gamma^{\rm f}_{-, u} (2 g t)}{1 - {\rm e}^{- t}}
+
\frac{\gamma^{\rm f}_{+, u} (2 g t)}{{\rm e}^{t} - 1}
\right]
=
0
\, .
\end{align}
These are valid for $|v|< 2g$, while the rapidity of the small fermion resides in the domain $|u|>2g$. As was demonstrated in Refs.\ \cite{Basso:2007wd,Basso:2009gh}, the above equations, or 
rather their analogues for the ground state of the flux tube,---the cusp anomalous dimension,---can be significantly simplified by performing a transformation to  
\begin{align}
\label{ComplexGammaf}
\Gamma^{\rm f}_u (\tau) 
\equiv 
\Gamma^{\rm f}_{+,u} (\tau) + i \Gamma^{\rm f}_{-,u} (\tau) 
=
\left(
1 + i \coth \frac{\tau}{4 g}
\right)
\gamma^{\rm f}_{u} (\tau)
\, ,
\end{align}
where we introduced a complex flux-tube function
\begin{align}
\label{Complexgammaf}
\gamma^{\rm f}_{u} (\tau)
\equiv
\gamma^{\rm f}_{+,u} (\tau) + i \gamma^{\rm f}_{-,u} (\tau) 
\, .
\end{align}
In this way the integrands in the left-hand side read
\begin{align}
\label{gammafToGammafpm}
\frac{\gamma^{\rm f}_{+, u} (2 g t)}{1 - {\rm e}^{- t}}
-
\frac{\gamma^{\rm f}_{-, u} (2 g t)}{{\rm e}^{t} - 1}
&=
\ft{1}{2}
\left[
\Gamma^{\rm f}_{+, u} (2 g t)
+
\Gamma^{\rm f}_{-, u} (2 g t)
\right]
\, , \\
\label{gammafToGammafmp}
\frac{\gamma^{\rm f}_{-, u} (2 g t)}{1 - {\rm e}^{- t}}
+
\frac{\gamma^{\rm f}_{+, u} (2 g t)}{{\rm e}^{t} - 1}
&=
\ft{1}{2}
\left[
\Gamma^{\rm f}_{-, u} (2 g t)
-
\Gamma^{\rm f}_{+, u} (2 g t)
\right]
\, .
\end{align}
By rescaling the rapidity $\hat{u} = u/(2 g)$ and the integration variable $\tau = 2 g t$, the equations cease to possess explicit dependence on the 't Hooft coupling.
The dependence on the latter is induced however via analyticity conditions on their solutions as will be done in Sect. \ref{SectQuantCondFermion}.

To solve the above equations it is instructive to first reduce them to a singular integral equation by means of a Fourier transformation \cite{Basso:2009gh}. Namely, we 
introduce the Fourier transforms of the functions involved
\begin{align}
\label{GgammmaFourier}
\varphi^{\rm f}_{u} (p) = \int_{-\infty}^\infty \frac{d \tau}{2 \pi} {\rm e}^{i p \tau} \, \gamma^{\rm f}_{u} (\tau)
\, , \qquad
\Phi^{\rm f}_{u} (p) = \int_{-\infty}^\infty \frac{d \tau}{2 \pi} {\rm e}^{i p \tau} \, \Gamma^{\rm f}_{u} (\tau)
\, .
\end{align}
Notice that since $\gamma^{\rm f}_{u} (\tau)$ is an analytic function in the complex plane, it admits a convergent expansion in terms of Bessel functions, i.e., $\gamma (\tau) \sim \sum_n J_n (\tau)$,
on the real axis. Thus the support of its Fourier transform is restricted to the interval $|p|<1$, i.e., 
\begin{align}
\varphi^{\rm f}_u (p) |_{|p|>1} = 0 \, ,
\end{align}
while $\Phi^{\rm f}_u (p)$ is nonvanishing on the entire real line. The inverse Fourier transform of the latter can be decomposed in terms of $\tau$-even and odd functions  
\begin{align}
\label{GammafFourier}
\Gamma^{\rm f}_{+, u} (\tau) = \int_{- \infty}^\infty d p \, \cos (p \tau) \, \Phi^{\rm f}_{u} (p)
\, , \qquad
\Gamma^{\rm f}_{-, u} (\tau) = - \int_{- \infty}^\infty d p \, \sin (p \tau) \, \Phi^{\rm f}_{u} (p)
\, ,
\end{align}
respectively, which due to the fact that the function $\Phi^{\rm f}_u$ is real,  $[\Phi^{\rm f}_u (\tau)]^\ast = \Phi^{\rm f}_u (\tau)$, correspond to the real and imaginary part of $\Gamma_u^{\rm f}$.

Now we are in a position to derive an integral equation for the Fourier transform $\Phi^{\rm f}_{u}$. To this end, we replace the two linear combinations 
\re{gammafToGammafpm} and \re{gammafToGammafmp} by their right-hand sides in the flux-tube equations \re{FTEqf1} and \re{FTEqf2}, respectively, 
and rescale the integration variable and rapidities as explained after Eq.\ \re{gammafToGammafmp}. Next, we substitute the definitions \re{GammafFourier} 
into the equations obtained in the previous step and evaluate the emerging $\tau$-integrals which result in simple rational functions of rapidities. Adding up 
the two results, we find that $\Phi^{\rm f}_{u}$ obeys the following equation
\begin{align}
\label{FourierFluxTubeEq}
\Phi^{\rm f}_u (p)
&
+
\int_{- 1}^1 \frac{dk}{\pi} \, \Phi^{\rm f}_u (k) \frac{\mathcal P}{k - p} 
\nonumber\\
&= 
-
\int_{- \infty}^\infty dk \, \Phi^{\rm f}_u (k) \frac{\theta (k^2 - 1)}{k - \hat{u}} 
-
\frac{1}{2 \pi}
\left[
\frac{1}{p - \hat{u}} + \frac{1}{p + \hat{u}}
\right]
\equiv J_u^{\rm f} (p)
\, , 
\end{align}
where we changed the variable $\hat{v}$ to $\hat{v} = p$ and the integral is defined by means of the Cauchy principal value $\mathcal{P}$. Due to the original domain of the validity of \re{FTEqf1} 
and \re{FTEqf2},  this equation defines $\Phi^{\rm f}_u (p)$ for $|p| < 1$ only. However, the inhomogeneity on its right-hand side involves $\Phi^{\rm f}_u (p)$ outside of the interval $(-1,1)$. This 
contribution can be found by noticing that due to the support region of $\varphi^{\rm f}$, its Fourier transform behaves as $\gamma^{\rm f} (\tau) \sim {\rm e}^{|\tau|}$ for large complex $\tau$. As a 
consequence, the integral \re{GgammmaFourier} for $\Phi^{\rm f}_u (p)$ can be computed by means of the Cauchy theorem (with residues emerging from the trigonometric prefactor) by closing 
the contour at infinity. This can be done however only provided $|p|>1$, yielding
\begin{align}
\label{OuterPhif}
\Phi^{\rm f}_u (p) |_{|p| > 1}
=
\theta (p - 1)
\sum_{n \geq 1} c^{{\rm f}, +}_{u} (n, g) {\rm e}^{- 4 \pi n g (p - 1)}
+
\theta (- p - 1)
\sum_{n \geq 1} c^{{\rm f}, -}_{u} (n, g) {\rm e}^{- 4 \pi n g (-p - 1)}
\, , 
\end{align}
where
\begin{align}
c^{\pm}_{u} (n, g) = \mp 4 g \gamma^{\rm f}_{u} (\pm i 4 \pi g n) {\rm e}^{- 4 \pi n g}
\, .
\end{align}
Obviously it is nonperturbative in its origin and still involves unknown coefficients in its decomposition. They will be fixed in Sect.\ \ref{SectQuantCondFermion}.

Summarizing, the solution to the flux-tube equation \re{FourierFluxTubeEq} can be rewritten as a sum of solutions to homogeneous and inhomogeneous equations
following standard methods \cite{Mikh61}
\begin{align}
\Phi^{\rm f}_{u} (p)
&
=
\frac{c}{p + 1}
\left( \frac{1 + p}{1 - p} \right)^{1/4} 
+
\frac{1}{2} J_u^{\rm f} (p)
-
\left( \frac{1 + p}{1 - p} \right)^{1/4} 
\int_{-1}^1 \frac{dk}{2 \pi} \left( \frac{1 - k}{1 + k} \right)^{1/4} \frac{\mathcal{P}}{k - p} J_u^{\rm f} (k)
\, .
\end{align}
One can easily verify a posteriori that this indeed solves \re{FourierFluxTubeEq} making use of integrals \re{Integralq>1} and \re{Integralp<1}.
Substituting the source $J_u^{\rm f}$, and partitioning the denominator, we can evaluate the resulting integrals making use of Eqs.\ \re{Integralq>1} and \re{Integralp<1}, such that
\begin{align}
\label{Phifp<1}
\Phi^{\rm f}_{u} (p) |_{|p| < 1}
=
\phi^{\rm f}_{u} (p)
-
\sqrt{2}  \left( \frac{1 + p}{1 - p} \right)^{1/4}
\int_{- \infty}^\infty
\frac{d k}{2 \pi} \frac{\theta (k^2 - 1)}{k - p}  \left( \frac{k - 1}{k + 1} \right)^{1/4} \Phi^{\rm f}_{u} (k)
\, ,
\end{align}
with
\begin{align}
\phi^{\rm f}_{u} (p)
=
-
\frac{1}{2 \sqrt{2} \pi} \left( \frac{1 + p}{1 - p} \right)^{1/4}
\left[
\frac{1}{p - \hat{u}} \left( \frac{\hat{u} - 1}{\hat{u} + 1} \right)^{1/4}
+
\frac{1}{p + \hat{u}} \left( \frac{\hat{u} + 1}{\hat{u} - 1} \right)^{1/4}
\right]
\, ,
\end{align}
and where in the last term of \re{Phifp<1} one has to substitute the expansion \re{OuterPhif} and the constant $c$ was set to zero to comply with properties of scattering phases. Fourier 
transforming back, we find the all-order expression for $\Gamma_u^{\rm f} (\tau)$,
\begin{align}
\label{GeneralSolGamma}
\Gamma_u^{\rm f} (\tau)
=
\chi_u^{\rm f} (\tau)
&
+
\sum_{n \geq 1} \frac{c_u^{{\rm f}, -} (n, g)}{4 \pi g n - i \tau}
\left[
- i \tau V_0 (- i \tau) U_1^- (4 \pi g n) + 4 \pi g n V_1 (- i \tau) U_0^- (4 \pi g n)
\right]
\\
&
+
\sum_{n \geq 1} \frac{c_u^{{\rm f}, +} (n, g)}{4 \pi g n + i \tau}
\left[
- i \tau V_0 (- i \tau) U_1^+ (4 \pi g n) + 4 \pi g n V_1 (- i \tau) U_0^+ (4 \pi g n)
\right]
\, , \nonumber
\end{align}
where
\begin{align}
\chi_u^{\rm f} (\tau)
=
- 
\frac{1}{4} \left(
\frac{\hat{u} - 1}{\hat{u} + 1}
\right)^{1/4}
W (- i \tau, \hat{u})
- 
\frac{1}{4} \left(
\frac{\hat{u} + 1}{\hat{u} - 1}
\right)^{1/4}
W (- i \tau, - \hat{u})
\, .
\end{align}
The integral representations of the special functions involved are given in Appendix \ref{IntegralsAppendix}. We will turn to defining the expansion coefficients after we address the $u$-parity odd 
case first in the next section.

\subsection{General solution for odd $u$-parity}

Up to minor modification, the odd $u$-parity case is analyzed in a similar manner. Starting with the flux-tube equations  \cite {Basso:2010in,Belitsky:2014sla}
\begin{align}
&
\int _0^\infty dt \, \sin (v t)
\left[
\frac{\widetilde\gamma^{\rm f}_{+, u} (2 g t)}{1 - {\rm e}^{- t}}
+
\frac{\widetilde\gamma^{\rm f}_{-, u} (2 g t)}{{\rm e}^{t} - 1}
\right]
=
0
\, , \\
&
\int _0^\infty dt \, \cos (v t)
\left[
\frac{\widetilde\gamma^{\rm f}_{-, u} (2 g t)}{1 - {\rm e}^{- t}}
-
\frac{\widetilde\gamma^{\rm f}_{+, u} (2 g t)}{{\rm e}^{t} - 1}
\right]
=
\frac{1}{2} \int_0^\infty dt \,  \cos (v t) \sin (u t)
\, ,
\end{align}
we introduce a complex function $\widetilde\gamma$ via the equation analogous to \re{Complexgammaf} that differs by a relative minus sign
\begin{align}
\label{Complexgammatildef}
\widetilde\gamma^{\rm f}_{u} (\tau)
\equiv
\widetilde\gamma^{\rm f}_{+,u} (\tau) - i \widetilde\gamma^{\rm f}_{-,u} (\tau) 
\, ,
\end{align}
and pass to a new function
\begin{align}
\label{ComplexGammatildef}
\widetilde\Gamma^{\rm f}_u (\tau) 
\equiv 
\widetilde\Gamma^{\rm f}_{+,u} (\tau) - i \widetilde\Gamma^{\rm f}_{-,u} (\tau) 
=
\left(
1 + i \coth \frac{\tau}{4 g}
\right)
\widetilde\gamma^{\rm f}_{u} (\tau)
\, ,
\end{align}
such that
\begin{align}
\frac{\widetilde\gamma^{\rm f}_{+, u} (2 g t)}{1 - {\rm e}^{- t}}
+
\frac{\widetilde\gamma^{\rm f}_{-, u} (2 g t)}{{\rm e}^{t} - 1}
&=
\ft{1}{2}
\left[
\widetilde\Gamma^{\rm f}_{+, u} (2 g t)
-
\widetilde\Gamma^{\rm f}_{-, u} (2 g t)
\right]
\, , \\
\frac{\widetilde\gamma^{\rm f}_{-, u} (2 g t)}{1 - {\rm e}^{- t}}
-
\frac{\widetilde\gamma^{\rm f}_{+, u} (2 g t)}{{\rm e}^{t} - 1}
&=
\ft{1}{2}
\left[
\widetilde\Gamma^{\rm f}_{+, u} (2 g t)
+
\widetilde\Gamma^{\rm f}_{-, u} (2 g t)
\right]
\, .
\end{align}
Introducing the Fourier transforms identical to Eqs.\ \re{GgammmaFourier}, with however a sign difference for the $\tau$-odd part,
\begin{align}
\label{GammatildeFourier}
\widetilde\Gamma^{\rm f}_{+, u} (\tau) = \int_{- \infty}^\infty d k \, \cos (k \tau) \, \widetilde\Phi^{\rm f}_{u} (k)
\, , \qquad
\widetilde\Gamma^{\rm f}_{-, u} (\tau) = \int_{- \infty}^\infty d k \, \sin (k \tau) \, \widetilde\Phi^{\rm f}_{u} (k)
\, ,
\end{align}
we obtain the singular integral equation that $\widetilde\Phi^{\rm f}_{u}$ obeys
\begin{align}
\label{FourierFluxTubeTildeEq}
\widetilde\Phi^{\rm f}_u (p)
&
+
\int_{- 1}^1 \frac{dk}{\pi} \, \widetilde\Phi^{\rm f}_u (k) \frac{\mathcal P}{k - p} 
\nonumber\\
&= 
-
\int_{- \infty}^\infty dk \, \widetilde\Phi^{\rm f}_v (k) \frac{\theta (k^2 - 1)}{k - \hat{u}} 
-
\frac{1}{2 \pi}
\left[
\frac{1}{p - \hat{u}} - \frac{1}{p + \hat{u}}
\right]
\, .
\end{align}

As before, we split the solution $\Phi^{\rm f}_v (p)$ into two regions, the interior of the interval $(-1,1)$ and its outside. The latter admits an infinite series representation
\begin{align}
\label{OuterPhiftilde}
\widetilde\Phi^{\rm f}_u (p) |_{|p| > 1}
=
\theta (p - 1)
\sum_{n \geq 1} \tilde{c}^{{\rm f}, +}_{u} (n, g) {\rm e}^{- 4 \pi n g (p - 1)}
+
\theta (- p - 1)
\sum_{n \geq 1} \tilde{c}^{{\rm f}, -}_{u} (n, g) {\rm e}^{- 4 \pi n g (-p - 1)}
\, , 
\end{align}
with the expansion coefficients
\begin{align}
\tilde{c}^{{\rm f}, \pm}_{u} (n, g) = \mp 4 g \widetilde\gamma^{\rm f}_{u} (\pm i 4 \pi g n) {\rm e}^{- 4 \pi n g}
\, ,
\end{align}
which will be fixed in the next section. Making use of the explicit sources, the solution to Eq.\ \re{FourierFluxTubeTildeEq} yields the function inside the interval $(-1,1)$,
\begin{align}
\widetilde\Phi^{\rm f}_{u} (p) |_{|p| < 1}
=
\widetilde\phi^{\rm f}_{u} (p)
-
\sqrt{2}  \left( \frac{1 + p}{1 - p} \right)^{1/4}
\int_{- \infty}^\infty
\frac{d k}{2 \pi} \frac{\theta (k^2 - 1)}{k - p}  \left( \frac{k - 1}{k + 1} \right)^{1/4} \widetilde\Phi^{\rm f}_{u} (k)
\, ,
\end{align}
with
\begin{align}
\widetilde\phi^{\rm f}_{u} (p)
=
-
\frac{1}{2 \sqrt{2} \pi} \left( \frac{1 + p}{1 - p} \right)^{1/4}
\left[
\frac{1}{p - \hat{u}} \left( \frac{\hat{u} - 1}{\hat{u} + 1} \right)^{1/4}
-
\frac{1}{p + \hat{u}} \left( \frac{\hat{u} + 1}{\hat{u} - 1} \right)^{1/4}
\right]
\, .
\end{align}
Fourier transforming back, it immediately produces the all-order expression for $\widetilde\Gamma_u^{\rm f} (\tau)$,
\begin{align}
\label{GeneralSolTildeGamma}
\widetilde\Gamma_u^{\rm f} (\tau)
=
\tilde\chi_u^{\rm f} (\tau)
&
+
\sum_{n \geq 1} \frac{\tilde{c}_u^{{\rm f}, -} (n, g)}{4 \pi g n - i \tau}
\left[
- i \tau V_0 (- i \tau) U_1^- (4 \pi g n) + 4 \pi g n V_1 (- i \tau) U_0^- (4 \pi g n)
\right]
\\
&
+
\sum_{n \geq 1} \frac{\tilde{c}_u^{{\rm f}, +} (n, g)}{4 \pi g n + i \tau}
\left[
- i \tau V_0 (- i \tau) U_1^+ (4 \pi g n) + 4 \pi g n V_1 (- i \tau) U_0^+ (4 \pi g n)
\right]
\, , \nonumber
\end{align}
where
\begin{align}
\label{chiftilde}
\tilde\chi_u^{\rm f} (\tau)
=
- 
\frac{1}{4} \left(
\frac{\hat{u} - 1}{\hat{u} + 1}
\right)^{1/4}
W (- i \tau, \hat{u})
+
\frac{1}{4} \left(
\frac{\hat{u} + 1}{\hat{u} - 1}
\right)^{1/4}
W (- i \tau, - \hat{u})
\, .
\end{align}
Now we are in a position to construct a quantization condition for the unknown coefficients $\tilde{c}_u^{{\rm f}, \pm}$ as well as 
${c}_u^{{\rm f}, \pm}$ from the previous section.

\subsection{Quantization conditions and their solutions}
\label{SectQuantCondFermion}

According to their definitions \re{ComplexGammaf} and \re{ComplexGammatildef}, $\Gamma_u^{\rm f} (\tau)$  and $\widetilde\Gamma_u^{\rm f} (\tau)$, respectively, possess an infinite number of 
fixed zeroes on the imaginary axis at $\tau = 4 \pi i g x_m$ due to the trigonometric multiplier present in both, i.e., 
\begin{align}
\Gamma_u^{\rm f} \left( 4 \pi i g x_m \right) = 0
\, , \qquad
\widetilde\Gamma_u^{\rm f} \left( 4 \pi i g x_m \right) = 0
\, ,
\end{align}
with $x_m = (m - \ft14)$ where $m \in \mathbb{Z}$. They define quantization conditions for the expansion coefficients $c^{\pm}_{u}$. These can be cast in the explicit form
\begin{align}
\label{QuantCondition}
\frac{\chi_u^{\rm f} (4 \pi i g x_m)}{V_0 (4 \pi g x_m)}
&
=
\sum_{n \geq 1} c^{{\rm f}, -}_{u} (n, g) \frac{x_m U_1^- (4 \pi g n) + n \, r (4 \pi g x_m) U_0^- (4 \pi g n)}{n + m}
\nonumber\\
&
+
\sum_{n \geq 1} c^{{\rm f}, +}_{u} (n, g) \frac{x_m U_1^+ (4 \pi g n) + n \, r (4 \pi g x_m) U_0^+ (4 \pi g n)}{n - m}
\, , 
\end{align}
where we divided both sides by $V_0$ and introduced the ratio
\begin{align}
r (z) = \frac{V_1 (z)}{V_0 (z)}
\, .
\end{align} 
Similar relation holds for $\tilde{c}^{\pm}_{u}$, where one has to dress everything with tildes. An equation analogous to \re{QuantCondition}, but for the vacuum state describing the cusp anomalous 
dimension, was proposed and solved in Ref.\ \cite{Basso:2009gh}. Here we will adopt the strategy advocated there and expand Eq.\ \re{QuantCondition} systematically in the inverse powers of the 
't Hooft coupling.

Making use of the asymptotic expansion of the special functions for their large argument as given in Appendix \ref{IntegralsAppendix}, the above quantization conditions split into two depending on the
sign of $x_m$ since the functions involved enjoy different asymptotic behavior subject to the condition $x_m \lessgtr 0$. Then the parity-even expansion coefficients admit the form
\begin{align}
c^{{\rm f}, \pm}_{u} (n, g) = (8 \pi g n)^{\pm 1/4} \left[ a^{{\rm f}, \pm}_{u} (n) + \frac{b^{{\rm f}, \pm}_{u} (n)}{4 \pi g}  + O (1/g^2) \right]
\, ,
\end{align}
with explicit $a$ and $b$ being 
\begin{align}
a^{{\rm f}, +}_{u} (n) 
&
= - \frac{2 \, \Gamma (n + \ft14)}{\Gamma (n + 1) \Gamma^{2} (\ft14)}  \chi_0^{{\rm f}, +} (u) 
\, , \qquad
a^{{\rm f}, -}_{u} (n) 
= 
- \frac{\Gamma (n + \ft34)}{2 \Gamma (n + 1) \Gamma^{2} (\ft34)}  \chi_0^{{\rm f}, -} (u) 
\, , \\
b^{{\rm f}, +}_{u} (n)
&
=
\frac{2 \Gamma (n + \ft14)}{\Gamma (n + 1) \Gamma^{2} (\ft14)}
\bigg\{
\left[
\frac{\pi}{16} + \frac{3}{8} \ln 2
\right]
\chi_0^{{\rm f}, -} (u)
\\
&
-
\left[
\frac{\pi}{16} - \frac{3}{8} \ln 2
\right]
(\chi_0^{{\rm f}, +} (u) - 8 \chi_{10}^{{\rm f}, +} (u))
+
\frac{1}{32 n} 
\left(
3 \chi_0^{{\rm f}, +} (u) 
-
32 \chi_{10}^{{\rm f}, +} (u)
\right)
\bigg\} 
\, , \nonumber\\
b^{{\rm f}, -}_{u} (n)
&
=
-
\frac{\Gamma (n + \ft34)}{2 \Gamma (n + 1) \Gamma^{2} (\ft34)}
\bigg\{
\left[
- \frac{\pi}{16} + \frac{3}{8} \ln 2
\right]
\chi_0^{{\rm f}, +}(u)
\\
&
+
\left[
\frac{\pi}{16} + \frac{3}{8} \ln 2
\right]
(\chi_0^{{\rm f}, -} (u) - 8 \chi_{10}^{{\rm f}, -} (u))
+
\frac{1}{32 n} 
\left(
5 \chi_0^{{\rm f}, -} (u) 
-
32 \chi_{10}^{{\rm f}, -} (u)
\right)
\bigg\} 
\, , \nonumber
\end{align}
respectively. Here, we introduced inhomogeneities arising from the large coupling expansion of the left-hand side of the quantization condition
\begin{align}
\frac{\chi_u^{\rm f} (\pm 4 \pi i g |x_m|)}{V_0 (\pm 4 \pi g |x_m|)}
=
\chi^{{\rm f}, \pm}_0 (u)
+
\frac{1}{4 \pi g}
\frac{
\chi^{{\rm f}, \pm}_{1} (u)}{x_m}
+
O (1/g^2)
\, ,
\end{align}
with explicit order-by-order contributions being
\begin{align}
\chi^{{\rm f}, \pm}_0 (u)
&=
-
\frac{1}{4}
\left[
\frac{1}{\hat{u} \pm 1}
\left(
\frac{\hat{u} + 1}{\hat{u} - 1}
\right)^{1/4}
-
\frac{1}{\hat{u} \mp 1}
\left(
\frac{\hat{u} - 1}{\hat{u} + 1}
\right)^{1/4}
\right]
\, , \\
\chi^{{\rm f}, +}_1 (u)
&=
-
\frac{3}{16}
\left[
\frac{1}{(\hat{u} + 1)^2}
\left(
\frac{\hat{u} + 1}{\hat{u} - 1}
\right)^{1/4}
+
\frac{1}{(\hat{u} - 1)^2}
\left(
\frac{\hat{u} - 1}{\hat{u} + 1}
\right)^{1/4}
\right]
\, , \\
\chi^{{\rm f}, -}_1 (u)
&=
\frac{5}{16}
\left[
\frac{1}{(\hat{u} - 1)^2}
\left(
\frac{\hat{u} + 1}{\hat{u} - 1}
\right)^{1/4}
+
\frac{1}{(\hat{u} + 1)^2}
\left(
\frac{\hat{u} - 1}{\hat{u} + 1}
\right)^{1/4}
\right]
\, .
\end{align}

In complete analogy, the solutions to the parity-odd equation read
\begin{align}
\tilde{c}^{{\rm f}, \pm}_{v} (n, g) = (8 \pi g n)^{\pm 1/4} \left[ \tilde{a}^{{\rm f}, \pm}_{v} (n) + \frac{\tilde{b}^{{\rm f}, \pm}_{v} (n)}{4 \pi g}  + O (1/g^2) \right]
\, ,
\end{align}
where the $\tilde{a}$ and $\tilde{b}$ coefficients are 
\begin{align}
\tilde{a}^{{\rm f}, +}_{u} (n) 
&
= - \frac{2 \ell \, \Gamma (n + \ft14)}{\Gamma (n + 1) \Gamma^{2} (\ft14)} \tilde\chi_0^{{\rm f}, +} (u) 
\, , \qquad
\tilde{a}^{{\rm f}, -}_{u} (n) 
= 
- \frac{\ell \, \Gamma (n + \ft34)}{2 \Gamma (n + 1) \Gamma^{2} (\ft34)} \tilde\chi_0^{{\rm f}, -} (u) 
\, , \\
\tilde{b}^{{\rm f}, +}_{u} (n)
&
=
\frac{2 \ell \, \Gamma (n + \ft14)}{\Gamma (n + 1) \Gamma^{2} (\ft14)}
\bigg\{
\left[
\frac{\pi}{16} + \frac{3}{8} \ln 2
\right]
\tilde\chi_0^{{\rm f}, -} (u)
\\
&
-
\left[
\frac{\pi}{16} - \frac{3}{8} \ln 2
\right]
(\tilde\chi_0^{{\rm f}, +} (u) - 8 \tilde\chi_{10}^{{\rm f}, +} (u))
+
\frac{1}{32 n} 
\left(
3 \tilde\chi_0^{{\rm f}, +} (u) 
-
32 \tilde\chi_{10}^{{\rm f}, +} (u)
\right)
\bigg\} 
\, , \nonumber\\
\tilde{b}^{{\rm f}, -}_{u} (n)
&
=
-
\frac{\ell \, \Gamma (n + \ft34)}{2 \Gamma (n + 1) \Gamma^{2} (\ft34)}
\bigg\{
\left[
- \frac{\pi}{16} + \frac{3}{8} \ln 2
\right]
\tilde\chi_0^{{\rm f}, +}(u)
\\
&
+
\left[
\frac{\pi}{16} + \frac{3}{8} \ln 2
\right]
(\tilde\chi_0^{{\rm f}, -} (u) - 8 \tilde\chi_{10}^{{\rm f}, -} (u))
+
\frac{1}{32 n} 
\left(
5 \tilde\chi_0^{{\rm f}, -} (u) 
-
32 \tilde\chi_{10}^{{\rm f}, -} (u)
\right)
\bigg\} 
\, , \nonumber
\end{align}
respectively, determined by another set of inhomogeneities arising in the left-hand side of the quantization condition
\begin{align}
\frac{\tilde\chi_u^{\rm f} (\pm 4 \pi i g |x_m|)}{V_0 (\pm 4 \pi g |x_m|)}
=
\tilde\chi^{{\rm f}, \pm}_{0} (u)
+
\frac{1}{4 \pi g}
\frac{\tilde\chi^{{\rm f}, \pm}_{1} (u)}{x_m}
+
O (1/g^2)
\, ,
\end{align}
with
\begin{align}
\tilde{\chi}^{{\rm f}, \pm}_0 (u)
&=
\frac{1}{4}
\left[
\frac{1}{\hat{u} \pm 1}
\left(
\frac{\hat{u} + 1}{\hat{u} - 1}
\right)^{1/4}
+
\frac{1}{\hat{u} \mp 1}
\left(
\frac{\hat{u} - 1}{\hat{u} + 1}
\right)^{1/4}
\right]
\, , \\
\tilde{\chi}^{{\rm f}, +}_1 (u)
&=
\frac{3}{16}
\left[
\frac{1}{(\hat{u} + 1)^2}
\left(
\frac{\hat{u} + 1}{\hat{u} - 1}
\right)^{1/4}
-
\frac{1}{(\hat{u} - 1)^2}
\left(
\frac{\hat{u} - 1}{\hat{u} + 1}
\right)^{1/4}
\right]
\, , \\
\tilde{\chi}^{{\rm f}, -}_1 (u)
&=
- \frac{5}{16}
\left[
\frac{1}{(\hat{u} - 1)^2}
\left(
\frac{\hat{u} + 1}{\hat{u} - 1}
\right)^{1/4}
-
\frac{1}{(\hat{u} + 1)^2}
\left(
\frac{\hat{u} - 1}{\hat{u} + 1}
\right)^{1/4}
\right]
\, .
\end{align}
The strong coupling expansion can be performed in a straightforward fashion to any required order. To save space we will not present subleading terms explicitly here.

\subsection{Strong coupling expansion}

Having determined the last unknown ingredients of the solutions, we can sum-up the infinite series in Eqs.\ \re{GeneralSolTildeGamma}, \re{GeneralSolTildeGamma}
and determine the inverse coupling expansion of the flux-tube functions $\Gamma$ and $\widetilde\Gamma$. For further use, let us decompose the latter in terms of even 
and odd components with respect to $\tau$. They are
\begin{align}
\label{Gammaf}
\Gamma_{\pm, u}^{\rm f} (\tau)
&
=
\mp
\frac{1}{4} \left(
\frac{\hat{u} - 1}{\hat{u} + 1}
\right)^{1/4}
W^\pm (\tau, \hat{u})
\mp
\frac{1}{4} \left(
\frac{\hat{u} + 1}{\hat{u} - 1}
\right)^{1/4}
W^\pm (\tau, - \hat{u})
\\
&
\mp
\frac{\chi_0^{{\rm f}, -} (u)}{4 \pi g} \left( \frac{\pi}{8} + \frac{3}{4} \ln 2 \right) V_1^\pm (\tau)
\pm
\frac{\chi_0^{{\rm f}, +} (u)}{4 \pi g} \left( \frac{\pi}{8} - \frac{3}{4} \ln 2 \right) \left[ V_1^\pm (\tau) \mp 4 \tau V_0^\mp (\tau) \right]
+
O (1/g^2)
\, , \nonumber
\end{align}
and
\begin{align}
\label{Gammaftilde}
\widetilde\Gamma_{\pm, u}^{\rm f} (\tau)
&
=
-
\frac{1}{4} \left(
\frac{\hat{u} - 1}{\hat{u} + 1}
\right)^{1/4}
W^\pm (\tau, \hat{u})
+
\frac{1}{4} \left(
\frac{\hat{u} + 1}{\hat{u} - 1}
\right)^{1/4}
W^\pm (\tau, - \hat{u})
\\
&
-
\frac{\tilde\chi_0^{{\rm f}, -} (u)}{4 \pi g} \left( \frac{\pi}{8} + \frac{3}{4} \ln 2 \right) V_1^\pm (\tau)
+
\frac{\tilde\chi_0^{{\rm f}, +} (u)}{4 \pi g} \left( \frac{\pi}{8} - \frac{3}{4} \ln 2 \right) \left[ V_1^\pm (\tau) \mp 4 \tau V_0^\mp (\tau) \right]
+
O (1/g^2)
\, , \nonumber
\end{align}
where we introduced $\tau$-even and -odd functions by decomposing $W (- i \tau, \hat{u})$ as $W (- i \tau, \hat{u}) = W^+ (\tau, \hat{u}) - i W^- (\tau, \hat{u})$ and similarly for $V_n$,
see Eqs.\ \re{Wplusminus} and \re{Vnplusminus}.

The $1/g$ expansion of the dynamical phases for the direct and mirror scattering matrices is now preformed in a straightforward fashion by trading $\gamma$'s for the linear combination 
of $\Gamma$'s according to the equations
\begin{align}
\label{GammaTogamma}
\gamma^{\rm f}_{\pm, u} (\tau)
=
\frac{\Gamma^{\rm f}_{\pm,u} (\tau) \pm \coth \frac{\tau}{4 g} \, \Gamma^{\rm f}_{\mp,u} (\tau) }{1 + \coth^2 \frac{\tau}{4 g}}
\, , \qquad
\widetilde\gamma^{\rm f}_{\pm, u} (\tau)
=
\frac{\widetilde\Gamma^{\rm f}_{\pm,u} (\tau) \mp \coth \frac{\tau}{4 g} \, \widetilde\Gamma^{\rm f}_{\mp,u} (\tau) }{1 + \coth^2 \frac{\tau}{4 g}}
\, ,
\end{align}
and expanding the integrands of \re{DirectFermionfs} and \re{MirrorFermionfs} for fixed $\tau$. Substituting the above solutions into the scattering phases, we find
\begin{align}
\label{ffDynamicalPhasesStrong}
f_{\rm ff}^{(\alpha)} ({u}_1, {u}_2) 
=
\frac{1}{32 g}
\Bigg\{ 
A^{(\alpha)}_{\rm ff} (\hat{u}_1, \hat{u}_2) 
+
\frac{1}{4 g}
\left[
B^{(\alpha)}_{\rm ff} (\hat{u}_1, \hat{u}_2) 
+
\frac{3 \ln2}{2 \pi}  C^{(\alpha)}_{\rm ff} (\hat{u}_1, \hat{u}_2) 
\right]
+
O (1/g^2)
\Bigg\}
\, ,
\end{align}
($\alpha = 1,2,3,4$) with explicit functions deferred to Appendix \ref{FFAppendix} due to their length. We verified their correctness by means of the exchange relations that imply 
that $f_{\rm ff}^{(2)} ({u}_1, {u}_2) = f_{\rm ff}^{(1)} ({u}_2, {u}_1)$ as well as symmetry of the mirror phases $f_{\rm ff}^{(3)} ({u}_1, {u}_2) = f_{\rm ff}^{(3)} ({u}_2, {u}_1)$ 
and $f_{\rm ff}^{(4)} ({u}_1, {u}_2) = f_{\rm ff}^{(4)} ({u}_2, {u}_1)$. Further checks will be performed below.

At leading order, the fermion-fermion S-matrix and its mirror read
\begin{align}
\ln S_{\rm ff} (u_1, u_2) 
&
=
- \frac{i}{8 g (\hat{u}_1 - \hat{u}_2)} 
\left[
\left( \frac{\hat{u}_1 - 1}{\hat{u}_1 + 1} \right)^{1/4} \left( \frac{\hat{u}_2 + 1}{\hat{u}_2 - 1} \right)^{1/4}
\!\!
+
\left( \frac{\hat{u}_1 + 1}{\hat{u}_1 - 1} \right)^{1/4} \left( \frac{\hat{u}_2 - 1}{\hat{u}_2 + 1} \right)^{1/4}
- 2
\right]
, \\
\ln S_{\rm \ast ff} (u_1, u_2) 
&
=
\frac{1}{8 g (\hat{u}_1 - \hat{u}_2)} 
\left[
\left( \frac{\hat{u}_1 - 1}{\hat{u}_1 + 1} \right)^{1/4} \left( \frac{\hat{u}_2 + 1}{\hat{u}_2 - 1} \right)^{1/4}
\!\!
-
\left( \frac{\hat{u}_1 + 1}{\hat{u}_1 - 1} \right)^{1/4} \left( \frac{\hat{u}_2 - 1}{\hat{u}_2 + 1} \right)^{1/4}
+
2 i
\right]
,
\end{align}
with fermion-antifermion related to them via Eqs.\ \re{DirectffANDfbarfSmatrices} and \re{MirrorffANDfbarfSmatrices}. They agree with earlier results of \cite{Fioravanti:2013eia}. 
While the subleading terms are new.

The small-fermion--small-(anti)fermion pentagons 
\begin{align}
P_{\rm f|f} ({u}_1|{u}_2) 
&
= 
\frac{i (1 - \hat{x}_{\rm f} [\hat{u}_1] \hat{x}_{\rm f} [\hat{u}_2])}{2g (\hat{u}_1 - \hat{u}_2)} 
P_{\rm \bar{f}|f} ({u}_1|{u}_2) 
\\
&
= 
\frac{i \sqrt{1 - \hat{x}_{\rm f} [\hat{u}_1] \hat{x}_{\rm f} [\hat{u}_2]}}{2g (\hat{u}_1 - \hat{u}_2)} 
\exp
\left(
- i f_{\rm ff}^{(1)} (u_1, u_2)
+ i f_{\rm ff}^{(2)} (u_1, u_2)
- f_{\rm ff}^{(3)} (u_1, u_2)
+ f_{\rm ff}^{(4)} (u_1, u_2)
\right)
\, , \nonumber
\end{align}
and the measure read
\begin{align}
\mu_{\rm f} (u)
&=
- \frac{1}{\sqrt{1 - \hat{x}_{\rm f}^2[\hat{u}]}} \exp \left( f^{(3)}_{\rm ff} (u, u) - f^{(4)}_{\rm ff} (u, u) \right)
\, ,
\end{align}
in terms of the found phases \re{ffDynamicalPhasesStrong}. Here we introduced a more natural from the point of view of fermions small fermion Zhukowski variable 
$x_{\rm f} [u] = \ft12 ( u - \sqrt{u^2 - (2 g)^2} )$ rescaled with the 't Hooft coupling $x_{\rm f} [u] = g \hat{x}_{\rm f} [\hat{u}]$
\begin{align}
\label{fermionZhuk}
\hat{x}_{\rm f} = \hat{u} - \sqrt{\hat{u}^2 - 1}
\, .
\end{align}
To avoid repetitious formulas, we will not display the $1/g$ expansion of pentagons explicitly which merely reduces to the substitution of Eq.\ \re{ffDynamicalPhasesStrong} with \re{Aff1} -- \re{Cff4} 
into the above formulas, however, we write down the measure to the $O (1/g^2)$ order, which requires taking a limit,
\begin{align}
\mu_{\rm f} (u)
&=
- \frac{1}{\sqrt{1 - \hat{x}_{\rm f}^2[\hat{u}] }} \exp 
\left(\frac{1}{16 g} \frac{1}{\hat{u}^2 - 1} 
\left[
1 - \frac{\pi + 12 \ln 2 (\hat{u}^2 + 1)}{16 \pi g (\hat{u}^2 - 1)}
\right]
+
O (1/g^3)
\right)
\, .
\end{align}

\section{Gluon transitions}
\label{GluonSection}

Now we are turning to the gauge fields and their bound states. The direct and mirror S-matrices for opposite and like helicity gluon stacks can be constructed by a fusion procedure,
as was previously reported in Ref.\ \cite{Basso:2014nra}. To avoid complications in algebra due to presence of an infinite number of cuts on the physical sheet, it was instructive
to pass to the Goldstone sheet \cite{Basso:2011rc}, which is half-way between the real and mirror kinematics. The result of the analysis is summarized in the equations
\begin{align}
\label{Slldirect}
S_{\ell_1 \bar{\ell}_2} (u_1, u_2)
&=
s^{-1}_{\ell_1 \ell_2} (u_1, u_2)  S_{\ell_1 \ell_2} (u_1, u_2)
\\
&=
\exp \left(
2 i \sigma_{\ell_1 \ell_2} (u_1, u_2)
-
2 i f^{(1)}_{\ell_1 \ell_2} (u_1, u_2)
+
2 i f^{(2)}_{\ell_1 \ell_2} (u_1, u_2)
\right)
\, , \nonumber\\
\label{Sllmirror}
S_{\ast\ell_1 \bar{\ell}_2} (u_1, u_2)
&=
S_{\ast\ell_2 \ell_1} (u_2, u_1)
\\
&=
s_{\ast\ell_1 \bar{\ell}_2} (u_1, u_2)
\exp \left(
2 \widehat\sigma_{\ell_1 \ell_2} (u_1, u_2)
+
2 f^{(3)}_{\ell_1 \ell_2} (u_1, u_2)
-
2 f^{(4)}_{\ell_1 \ell_2} (u_1, u_2)
\right)
\, , \nonumber
\end{align}
respectively. Here the rational prefactor for the same-helicity S-matrix is
\begin{align}
\label{XXXsmatrix}
s_{\ell_1 \ell_2}  (u_1, u_2)
&
=
\frac{\Gamma \left( 1 + \frac{\ell_1 + \ell_2}{2} - i u_1 + i u_2 \right)
\Gamma \left(\frac{\ell_1 + \ell_2}{2} - i u_1 + i u_2 \right)
}{\Gamma \left( 1 + \frac{\ell_1 + \ell_2}{2} + i u_1 - i u_2 \right)
\Gamma \left(\frac{\ell_1 + \ell_2}{2} + i u_1 - i u_2 \right)}
\\
&
\times
\frac{\Gamma \left( 1 + \frac{\ell_1 - \ell_2}{2} + i u_1 - i u_2 \right)
\Gamma \left(\frac{\ell_1 + \ell_2}{2} + i u_1 - i u_2 \right)
}{\Gamma \left( 1 + \frac{\ell_1 - \ell_2}{2} - i u_1 + i u_2 \right)
\Gamma \left(\frac{\ell_1 + \ell_2}{2} - i u_1 + i u_2 \right)
} \nonumber
\end{align}
and corresponds to the scattering phase of spin-$\ell$ magnons for compact XXX spin chain. While in the mirror matrix, it takes the form 
\begin{align}
s_{\ast\ell_1 \bar{\ell}_2} (u_1, u_2)
=
(-1)^{\ell_2} 
\frac{\Gamma \left( 1 + \frac{\ell_1 - \ell_2}{2} - i u_1 + i u_2 \right)
\Gamma \left(\frac{\ell_1 + \ell_2}{2} + i u_1 - i u_2 \right)
}{\Gamma \left( 1 + \frac{\ell_1 + \ell_2}{2} - i u_1 + i u_2 \right)
\Gamma \left(\frac{\ell_1 - \ell_2}{2} + i u_1 - i u_2 \right)}
\, .
\end{align}
The dynamical phases in the above equations, in a form slightly different compared to Ref.\ \cite{Basso:2014nra}, read for direct
\begin{align}
\sigma_{\ell_1 \ell_2} (u_1, u_2)
&
=
\int_0^\infty \! \frac{dt}{t ({\rm e}^t - 1)}
\Big[
{\rm e}^{- \ell_1 t/2} \sin (u_1 t) J_0 (2 gt)
-
{\rm e}^{- \ell_2 t/2} \sin (u_2 t) J_0 (2 gt)
\\
&
\qquad\qquad\qquad\qquad\qquad\qquad\qquad\qquad\quad
-
{\rm e}^{- (\ell_1 + \ell_2) t/2} \sin \left((u_1 - u_2) t\right) 
\Big]
, \nonumber\\
\label{f1ll}
f^{(1)}_{\ell_1 \ell_2}  (u_1, u_2)
&
=
\int_0^\infty \frac{dt}{t} {\rm e}^{- \ell_1 t/2} \sin (u_1 t)
\left[
\frac{\gamma^{\rm g}_{-, u_2} (2 g t)}{1 - {\rm e}^{-t}}
+
\frac{\gamma^{\rm g}_{+, u_2} (2 g t)}{{\rm e}^t - 1}
\right]
\, , \\
\label{f2ll}
f^{(2)}_{\ell_1 \ell_2}  (u_1, u_2)
&
=
\int_0^\infty \frac{dt}{t} 
\left(
{\rm e}^{- \ell_1 t/2} \cos (u_1 t)
-
J_0 (2 g t)
\right)
\left[
\frac{\widetilde\gamma^{\rm g}_{+, u_2} (2 g t)}{1 - {\rm e}^{-t}}
+
\frac{\widetilde\gamma^{\rm g}_{-, u_2} (2 g t)}{{\rm e}^t - 1}
\right]
\, .
\end{align}
and mirror cases
\begin{align}
\widehat\sigma_{\ell_1 \ell_2} (u_1, u_2)
&
=
\int_0^\infty \frac{dt}{t (1 - {\rm e}^{- t})}
\Big[
{\rm e}^{- \ell_1 t/2} \cos (u_1 t) J_0 (2 gt)
+
{\rm e}^{- \ell_2 t/2} \cos (u_2 t) J_0 (2 gt)
\\
&
\qquad\qquad\qquad\qquad\qquad\qquad\qquad\qquad\quad \,
-
{\rm e}^{- (\ell_1 + \ell_2) t/2} \cos \left((u_1 - u_2) t\right) 
-
J_0^2 (2 g t)
\Big]
\, , \nonumber\\
\label{f3ll}
f^{(3)}_{\ell_1 \ell_2} (u_1, u_2)
&
=
-
\int_0^\infty \frac{dt}{t} {\rm e}^{- \ell_1 t/2} \sin (u_1 t)
\left[
\frac{\widetilde\gamma^{\rm g}_{-, u_2} (2 g t)}{1 - {\rm e}^{-t}}
-
\frac{\widetilde\gamma^{\rm g}_{+, u_2} (2 g t)}{{\rm e}^t - 1}
\right]
\, , \\
\label{f4ll}
f^{(4)}_{\ell_1 \ell_2} (u_1, u_2)
&
=
+
\int_0^\infty \frac{dt}{t} 
\left(
{\rm e}^{- \ell_1 t/2} \cos (u_1 t)
-
J_0 (2 g t)
\right)
\left[
\frac{\gamma^{\rm g}_{+, u_2} (2 g t)}{1 - {\rm e}^{-t}}
-
\frac{\gamma^{\rm g}_{-, u_2} (2 g t)}{{\rm e}^t - 1}
\right]
\, .
\end{align}
Though it is not obvious from the above representation, the exchange relations \cite{Basso:2013pxa,Belitsky:2014sla} imply certain symmetry properties of the dynamical phases. Namely, under 
the permutation of arguments (and spin labels $\ell$), they change as
\begin{align}
f^{(1)}_{\ell_1 \ell_2}  (u_1, u_2) = f^{(2)}_{\ell_2 \ell_1}  (u_2, u_1)
\, , \qquad
f^{(3)}_{\ell_1 \ell_2}  (u_1, u_2) = f^{(3)}_{\ell_2 \ell_1}  (u_2, u_1)
\, , \qquad
f^{(4)}_{\ell_1 \ell_2}  (u_1, u_2) = f^{(4)}_{\ell_2 \ell_1}  (u_2, u_1)
\, .
\end{align}
These will be used below as a verification of results obtained at strong coupling. The above expression are well suited to the current strong-coupling analysis, however, we have to transform
them first.

\subsection{Passing to Goldstone sheet}

As we just mentioned above, the physical sheet in the complex $u$ plane possesses an infinite number of cuts $[-2g, 2g]$ stacked up with the interval $i$ along the imaginary 
axis. For $\ell$-gluon bound state, they start from $|\Im{\rm m} [u]| = \ell/2$ and go up/downwards. In the strong-coupling limit, one immediately finds oneself in a predicament, 
since all of the cuts collapse into one on the real axis pinching the physical region of rapidities $- 2g < u < 2g$. To overcome this complication one has to stay in the latter
region but keep away from all of the cuts. This is possible provided one passes to the Goldstone sheet by moving upwards through the first Zhukowski cut in the upper half-plane
of $u$. A distinguished feature of this sheet is that it has only a finite number of cuts ranging from $- \ell/2$ to $\ell/2$. So one can safely navigate away from $[-2g + i \ell/2, 2g + i \ell/2]$
to $\Im{\rm m} [u] > \ell/2$ still staying in the strip. When on the Goldstone sheet, one takes the strong coupling limit, and then one can always undo the analytic continuation 
afterwards and go back to the physical sheet. 

According to this discussion, we perform the analytic continuation $u \stackrel{\rm G}{\to} u + i \ell/2 + i 0_+ \to u^{\rm G} = u$ for $|u|< 2g$ and immediately find for 
the flux-tube equations of even
\begin{align}
\label{FTEqG1}
&
\int _0^\infty dt \, \sin (v t)
\left[
\frac{\gamma^{\rm G}_{+, u} (2 g t)}{1 - {\rm e}^{- t}}
-
\frac{\gamma^{\rm G}_{-, u} (2 g t)}{{\rm e}^{t} - 1}
\right]
=
\frac{1}{2} \int_0^\infty dt \sin (v t) \frac{\sinh \frac{\ell t}{2}}{\sinh \frac{t}{2}} {\rm e}^{i u t + t/2} 
\, , \\
\label{FTEqG2}
&
\int _0^\infty dt \, \cos (v t)
\left[
\frac{\gamma^{\rm G}_{-, u} (2 g t)}{1 - {\rm e}^{- t}}
+
\frac{\gamma^{\rm G}_{+, u} (2 g t)}{{\rm e}^{t} - 1}
\right]
=
\frac{1}{2} \int_0^\infty dt \cos (v t) \frac{\sinh \frac{\ell t}{2}}{\sinh \frac{t}{2}}  {\rm e}^{i u t - t/2} 
\, ,
\end{align}
and odd parity
\begin{align}
&
\int _0^\infty dt \, \sin (v t)
\left[
\frac{\widetilde\gamma^{\rm G}_{+, u} (2 g t)}{1 - {\rm e}^{- t}}
+
\frac{\widetilde\gamma^{\rm G}_{-, u} (2 g t)}{{\rm e}^{t} - 1}
\right]
=
\frac{1}{2 i} \int_0^\infty dt \sin (v t) \frac{\sinh \frac{\ell t}{2}}{\sinh \frac{t}{2}} {\rm e}^{i u t - t/2} 
\, , \\
&
\int _0^\infty dt \, \cos (v t)
\left[
\frac{\widetilde\gamma^{\rm G}_{-, u} (2 g t)}{1 - {\rm e}^{- t}}
-
\frac{\widetilde\gamma^{\rm G}_{+, u} (2 g t)}{{\rm e}^{t} - 1}
\right]
=
\frac{1}{2 i} \int_0^\infty dt \cos (v t) \frac{\sinh \frac{\ell t}{2}}{\sinh \frac{t}{2}} {\rm e}^{i u t + t/2} 
\, ,
\end{align}
respectively, in agreement with Ref.\ \cite{Basso:2014nra}. Again we repeat that these are valid for $|v| < 2g$ and $\Im{\rm m} [u] > \ft{\ell}{2}$.
Notice that the sources are now complex. This will lead to minor differences in the analysis that follows.

The stack-(anti)stack S-matrix with both rapidities on the Goldstone sheet then reads
\begin{align}
\label{SGGdirect}
S_{\rm GG} (u_1, u_2)
&
=
s_{\ell_1 \ell_2} (u_1, u_2) S_{\rm G \bar{G}} (u_1, u_2)
\\
&
=
s_{\ell_1 \ell_2} (u_1, u_2) \exp \left( - 2 i f^{(1)}_{\rm GG} (u_1, u_2) + 2 i f^{(2)}_{\rm GG} (u_1, u_2) \right)
\, , \nonumber\\
\label{SGGmirror}
S_{\rm \ast GG} (u_1, u_2)
&
=
S_{\rm \ast G \bar{G}} (u_2, u_1) 
\\
&
= 
s_{\ast\ell_1 \bar{\ell}_2} (u_1, u_2)
\exp \left( 2 f^{(3)}_{\rm GG} (u_1, u_2) - 2 f^{(4)}_{\rm GG} (u_1, u_2) \right)
\, . \nonumber
\end{align}
The mirror symmetry of the flux tube allows one to establish the above relation \re{SGGmirror} between the mirror matrices with opposite and like helicities, which can be easily 
verified from the diagrammatic representation of the latter. Though it is not transparent from the notations in the relation \re{SGGmirror}, we implied one has to interchange  $\ell_1$ 
and $\ell_2$ as well. Here the scattering phases are
\begin{align}
f^{(1)}_{\rm GG} (u_1, u_2) 
&
=
i
\int_0^\infty \frac{dt}{t} {\rm e}^{i u_1 t} \sinh \frac{\ell_1 t}{2}
\left[
\frac{\gamma^{\rm G}_{-, u_2} (2 g t)}{1 - {\rm e}^{-t}}
+
\frac{\gamma^{\rm G}_{+, u_2} (2 g t)}{{\rm e}^{t} - 1}
\right]
\, , \\
f^{(2)}_{\rm GG} (u_1, u_2) 
&
=
-
\int_0^\infty \frac{dt}{t} {\rm e}^{i u_1 t} \sinh \frac{\ell_1 t}{2}
\left[
\frac{\widetilde\gamma^{\rm G}_{+, u_2} (2 g t)}{1 - {\rm e}^{-t}}
+
\frac{\widetilde\gamma^{\rm G}_{+, u_2} (2 g t)}{{\rm e}^{t} - 1}
\right]
\, , \\
f^{(3)}_{\rm GG} (u_1, u_2) 
&
=
- i
\int_0^\infty \frac{dt}{t} {\rm e}^{i u_1 t} \sinh \frac{\ell_1 t}{2}
\left[
\frac{\widetilde\gamma^{\rm G}_{-, u_2} (2 g t)}{1 - {\rm e}^{-t}}
-
\frac{\widetilde\gamma^{\rm G}_{+, u_2} (2 g t)}{{\rm e}^{t} - 1}
\right]
\, , \\
f^{(4)}_{\rm GG} (u_1, u_2) 
&
= -
\int_0^\infty \frac{dt}{t} {\rm e}^{i u_1 t} \sinh \frac{\ell_1 t}{2}
\left[
\frac{\gamma^{\rm G}_{+, u_2} (2 g t)}{1 - {\rm e}^{-t}}
-
\frac{\gamma^{\rm G}_{-, u_2} (2 g t)}{{\rm e}^{t} - 1}
\right]
\, ,
\end{align}
and possess only a finite number of cuts as expected.

\subsection{General solution for even $u$-parity}

In complete analogy with the fermionic case discussed in the preceding sections, we change the basis of functions as in Eq.\ \re{ComplexGammaf} and then Fourier transform their 
linear combination as
\begin{align}
\Gamma_u^{\rm G} (\tau) = \Gamma_{u, +}^{\rm G} (\tau) + i \Gamma_{u, -}^{\rm G} (\tau) 
=
\int_{-\infty}^\infty dk \, {\rm e}^{- i k \tau} \Phi_u^{\rm G} (k)
\, .
\end{align}
Here the function $\Phi_u^{\rm G} (k)$ is complex contrary to the analogous one for the fermion by virtue of a similar property of the sources on the Goldstone sheet. The flux-tube equation
for the former is then rewritten in the form
\begin{align}
\Phi^{\rm G}_u (p)
&
+
\int_{-1}^1 \frac{dk}{\pi} \frac{\mathcal{P}}{k - p} \Phi^{\rm G}_u (k)
=
-
\int_{-\infty}^\infty \frac{dk}{\pi} \frac{\theta (k^2 - 1)}{k - p} \Phi^{\rm G}_u (k)
\\
&
-
\frac{1}{2 \pi}
\sum_{n = 0}^{\ell - 1}
\left[
\frac{1}{p + \hat{u}^{[\ell - 2n - 2]} + i 0} + \frac{1}{p - \hat{u}^{[\ell - 2n - 2]} - i 0}
+
\frac{i}{p + \hat{u}^{[\ell - 2n]} + i 0} - \frac{i}{p - \hat{u}^{[\ell - 2n]} - i 0}
\right]
\, , \nonumber
\end{align}
with the traditional convention used for the shifted (and rescaled) rapidity variable
\begin{align}
\hat{u}^{[\pm \ell]} \equiv \hat{u} \pm \frac{i}{4 g} \ell
\, .
\end{align}
The solution for the interior region reads
\begin{align}
\Phi^{\rm G}_u (p) |_{|p| < 1}
&
=
\phi^{\rm G}_u (p)
-
\frac{1}{\sqrt{2}} \left( \frac{1 + p}{1 - p} \right)^{1/4}
\int_{-\infty}^\infty \frac{dk}{\pi}
\left( \frac{k - 1}{k + 1} \right)^{1/4} \frac{\theta (k^2 - 1)}{k - p} \Phi^{\rm G}_u (k)
\, ,
\end{align}
where%
\footnote{A formula for the partition of the product of principal value poles becomes handy here, 
$$
\frac{\mathcal P}{x - a} \frac{\mathcal P}{x - b} = \frac{\mathcal P}{a - b} \left[ \frac{\mathcal P}{x - a} - \frac{\mathcal P}{x - b} \right] + \pi^2 \delta (a - b) \delta (x - a)
\, .
$$
}
\begin{align}
&
\phi^{\rm G}_u (p)
=
\frac{1}{2 \sqrt{2}} \sum_{n = 0}^{\ell - 1}
\Bigg\{
{\rm e}^{- i \pi/4} \delta \left( p - \hat{u}^{[\ell - 2n - 2]}\right)
+ 
{\rm e}^{i \pi/4} \delta \left( p + \hat{u}^{[\ell - 2n - 2]}\right)
\\
&\qquad\qquad\qquad\qquad\qquad\qquad\qquad\qquad\quad\ \
-
{\rm e}^{i \pi/4} \delta \left( p - \hat{u}^{\ell - 2n}\right) 
- 
{\rm e}^{- i \pi/4} \delta \left( p + \hat{u}^{[\ell - 2n]}\right)
\nonumber\\
&\ 
- \frac{1}{\pi}
\left( \frac{1 + p}{1 - p} \right)^{1/4} 
\left[
{\rm e}^{- i \pi/4}
\frac{\mathcal{P}}{p + \hat{u}^{[\ell - 2n - 2]}}
\left( \frac{1 + \hat{u}^{[\ell - 2n - 2]}}{1 - \hat{u}^{[\ell - 2n - 2]}} \right)^{1/4} 
\!\!
+
{\rm e}^{i \pi/4}
\frac{\mathcal{P}}{p - \hat{u}^{[\ell - 2n - 2]}}
\left( \frac{1 - \hat{u}^{[\ell - 2n - 2]}}{1 + \hat{u}^{[\ell - 2n - 2]}} \right)^{1/4} 
\right]
\nonumber\\
&\ 
- \frac{1}{\pi}
\left( \frac{1 + p}{1 - p} \right)^{1/4} 
\left[
{\rm e}^{- i \pi/4}
\frac{\mathcal{P}}{p + \hat{u}^{[\ell - 2n]}}
\left( \frac{1 + \hat{u}^{[\ell - 2n]}}{1 - \hat{u}^{[\ell - 2n]}} \right)^{1/4} 
\!\!
+
{\rm e}^{i \pi/4}
\frac{\mathcal{P}}{p - \hat{u}_{n, \ell}}
\left( \frac{1 - \hat{u}^{[\ell - 2n]}}{1 + \hat{u}^{[\ell - 2n]}} \right)^{1/4} 
\right]
\Bigg\}
\, . \nonumber
\end{align}
For the exterior domain, we have as before the series representation
\begin{align}
\label{OuterPhiG}
\Phi^{\rm G}_v (p)|_{|p| > 1} =
\theta (p - 1)
\sum_{n \geq 1} c^{\rm G, +}_{v} (n, g) {\rm e}^{- 4 \pi n g (p - 1)}
+
\theta (- p - 1)
\sum_{n \geq 1} c^{\rm G, -}_{v} (n, g) {\rm e}^{- 4 \pi n g (- p - 1)}
\, .
\end{align}
Fourier transforming back to $\Gamma$, we deduce
\begin{align}
\Gamma^{\rm G}_u (\tau) = \chi^{\rm G}_u (\tau)
&
+
\sum_{n \geq 1} \frac{{c}_u^{\rm G, -} (n, g)}{4 \pi g n - i \tau}
\left[
- i \tau V_0 (- i \tau) U_1^- (4 \pi g n) + 4 \pi g n V_1 (- i \tau) U_0^- (4 \pi g n)
\right]
\nonumber\\
&
+
\sum_{n \geq 1} \frac{{c}_u^{\rm G, +} (n, g)}{4 \pi g n + i \tau}
\left[
- i \tau V_0 (- i \tau) U_1^+ (4 \pi g n) + 4 \pi g n V_1 (- i \tau) U_0^+ (4 \pi g n)
\right]
\, ,
\end{align}
with
\begin{align}
&
\chi^{\rm G}_u (\tau)
= \int_{-1}^1 dk \, {\rm e}^{-i \tau k} \phi^{\rm G}_u (k)
=
\frac{1}{4}
\sum_{n = 0}^{\ell - 1}
\Bigg\{
2 \sqrt{2} \cos\left( \tau \hat{u}^{[\ell - 2n - 2]} + \ft{\pi}{4}\right)
- 
2 \sqrt{2} \cos\left( \tau \hat{u}^{[\ell - 2n]} - \ft{\pi}{4}\right)
\nonumber\\
&
\
-
{\rm e}^{-i \pi/4}
\left( \frac{1 + \hat{u}^{[\ell - 2n - 2]}}{1 - \hat{u}^{[\ell - 2n- 2]}} \right)^{1/4} 
W \left(- i \tau, - \hat{u}^{[\ell - 2n - 2]}\right)
-
{\rm e}^{i \pi/4}
\left( \frac{1 - \hat{u}^{[\ell - 2n - 2]}}{1 + \hat{u}^{[\ell - 2n - 2]}} \right)^{1/4} 
W \left(- i \tau, \hat{u}^{[\ell - 2n - 2]}\right)
\nonumber\\
&\qquad
-
{\rm e}^{i \pi/4}
\left( \frac{1 + \hat{u}^{[\ell - 2n]}}{1 - \hat{u}^{[\ell - 2n]}} \right)^{1/4} 
W \left(- i \tau, - \hat{u}^{[\ell - 2n]}\right)
-
{\rm e}^{- i \pi/4}
\left( \frac{1 - \hat{u}^{[\ell - 2n]}}{1 + \hat{u}^{[\ell - 2n]}} \right)^{1/4} 
W \left(- i \tau, \hat{u}^{[\ell - 2n]}\right)
\Bigg\}
. \nonumber
\end{align}

\subsection{General solution for odd $u$-parity}

The flux-tube equations for the Fourier transform of $\widetilde\Gamma_u^{\rm G} (\tau)$,
\begin{align}
\label{tildeGammaG}
\widetilde\Gamma_u^{\rm G} (\tau) = \widetilde\Gamma_{u, +}^{\rm G} (\tau) + i \widetilde\Gamma_{u, -}^{\rm G} (\tau) 
=
\int_{-\infty}^\infty dk \, {\rm e}^{- i k \tau} \widetilde\Phi_u^{\rm G} (k)
\, .
\end{align}
is again put in the form of a singular integral equation
\begin{align}
\widetilde\Phi^{\rm G}_u (p)
&
+
\int_{-1}^1 \frac{dk}{\pi} \frac{\mathcal{P}}{k - p} \widetilde\Phi^{\rm G}_u (k)
=
-
\int_{-\infty}^\infty \frac{dk}{\pi} \frac{\theta (k^2 - 1)}{k - p} \widetilde\Phi^{\rm G}_u (k)
\\
&
+
\frac{1}{2 \pi}
\sum_{n = 0}^{\ell - 1}
\left[
\frac{1}{p + \hat{u}^{[\ell - 2n - 2]} + i 0} - \frac{1}{p - \hat{u}^{[\ell - 2n - 2]} - i 0}
+
\frac{i}{p + \hat{u}^{\ell - 2n} + i 0} + \frac{i}{p - \hat{u}^{\ell - 2n} - i 0}
\right]
\, , \nonumber
\end{align}
whose solution is
\begin{align}
\widetilde\Phi^{\rm G}_u (p) |_{|p| < 1}
&
=
\widetilde\phi^{\rm G}_u (p)
-
\frac{1}{\sqrt{2}} \left( \frac{1 + p}{1 - p} \right)^{1/4}
\int_{-\infty}^\infty \frac{dk}{\pi}
\left( \frac{k - 1}{k + 1} \right)^{1/4} \frac{\theta (k^2 - 1)}{k - p} \widetilde\Phi^{\rm G}_u (k)
\, ,
\end{align}
where
\begin{align}
&
\widetilde\phi^{\rm G}_u (p)
=
\frac{1}{2 \sqrt{2}} \sum_{n = 0}^{\ell - 1}
\Bigg\{
{\rm e}^{- i \pi/4} \delta \left( p - \hat{u}^{[\ell - 2n - 2]} \right)
-  
{\rm e}^{i \pi/4} \delta \left( p + \hat{u}^{[\ell - 2n - 2]} \right)
\\
&\qquad\qquad\qquad\qquad\qquad\qquad\qquad\qquad\quad\ \
- {\rm e}^{i \pi/4} \delta \left( p - \hat{u}^{[\ell - 2n]} \right) 
+ {\rm e}^{- i \pi/4} \delta \left( p + \hat{u}^{[\ell - 2n]} \right)
\nonumber\\
&\ 
+ \frac{1}{\pi}
\left( \frac{1 + p}{1 - p} \right)^{1/4} 
\left[
{\rm e}^{- i \pi/4}
\frac{\mathcal{P}}{p + \hat{u}^{[\ell - 2n - 2]}}
\left( \frac{1 + \hat{u}^{[\ell - 2n - 2]}}{1 - \hat{u}^{[\ell - 2n - 2]}} \right)^{1/4} 
\!\!
-
{\rm e}^{i \pi/4}
\frac{\mathcal{P}}{p - \hat{u}^{[\ell - 2n - 2]}}
\left( \frac{1 - \hat{u}^{[\ell - 2n - 2]}}{1 + \hat{u}^{[\ell - 2n - 2]}} \right)^{1/4} 
\right]
\nonumber\\
&\ 
+ \frac{1}{\pi}
\left( \frac{1 + p}{1 - p} \right)^{1/4} 
\left[
{\rm e}^{i \pi/4}
\frac{\mathcal{P}}{p + \hat{u}^{[\ell - 2n]}}
\left( \frac{1 + \hat{u}^{[\ell - 2n]}}{1 - \hat{u}^{[\ell - 2n]}} \right)^{1/4} 
\!\!
-
{\rm e}^{- i \pi/4}
\frac{\mathcal{P}}{p - \hat{u}^{[\ell - 2n]}}
\left( \frac{1 - \hat{u}^{[\ell - 2n]}}{1 + \hat{u}^{[\ell - 2n]}} \right)^{1/4} 
\right]
\Bigg\}
\, , \nonumber
\end{align}
and the outside function is again determined by the series \re{OuterPhiG}, where one obviously dresses all coefficients with tildes. Fourier transforming it back \re{tildeGammaG}, 
we deduce
\begin{align}
\widetilde\Gamma^{\rm G}_u (\tau) = \widetilde\chi^{\rm G}_u (\tau)
&
+
\sum_{n \geq 1} \frac{\tilde{c}_u^{\rm G, -} (n, g)}{4 \pi g n - i \tau}
\left[
- i \tau V_0 (- i \tau) U_1^- (4 \pi g n) + 4 \pi g n V_1 (- i \tau) U_0^- (4 \pi g n)
\right]
\nonumber\\
&
+
\sum_{n \geq 1} \frac{\tilde{c}_u^{\rm G, +} (n, g)}{4 \pi g n + i \tau}
\left[
- i \tau V_0 (- i \tau) U_1^+ (4 \pi g n) + 4 \pi g n V_1 (- i \tau) U_0^+ (4 \pi g n)
\right]
\, ,
\end{align}
with
\begin{align}
&
\tilde\chi^{\rm G}_u (\tau)
= \int_{-1}^1 dk \, {\rm e}^{-i \tau k} \widetilde\phi^{\rm G}_u (k)
=
\frac{1}{4}
\sum_{n = 0}^{\ell - 1}
\Bigg\{
- i 2 \sqrt{2} \sin\left( \tau \hat{u}^{[\ell - 2n - 2]} + \ft{\pi}{4} \right)
+ i 2 \sqrt{2} \sin\left( \tau \hat{u}^{[\ell - 2n]} - \ft{\pi}{4} \right)
\nonumber\\
&\
+
{\rm e}^{-i \pi/4}
\left( \frac{1 + \hat{u}^{[\ell - 2n - 2]}}{1 - \hat{u}^{[\ell - 2n - 2]}} \right)^{1/4} 
W \left( - i \tau, - \hat{u}^{[\ell - 2n - 2]} \right)
-
{\rm e}^{i \pi/4}
\left( \frac{1 - \hat{u}^{[\ell - 2n - 2]}}{1 + \hat{u}^{[\ell - 2n - 2]}} \right)^{1/4} 
W \left( - i \tau, \hat{u}^{[\ell - 2n - 2]} \right)
\nonumber\\
&\qquad
+
{\rm e}^{i \pi/4}
\left( \frac{1 + \hat{u}^{[\ell - 2n]}}{1 - \hat{u}^{[\ell - 2n]}} \right)^{1/4} 
W \left(- i \tau, - \hat{u}^{[\ell - 2n]}\right)
-
{\rm e}^{- i \pi/4}
\left( \frac{1 - \hat{u}^{[\ell - 2n]}}{1 + \hat{u}^{[\ell - 2n]}} \right)^{1/4} 
W \left(- i \tau, \hat{u}^{[\ell - 2n]} \right)
\Bigg\}
\, . \nonumber
\end{align}

\subsection{Quantization conditions and their solutions}

The quantization condition for the even $u$-parity function
\begin{align}
\Gamma_u^{\rm G} (4 \pi i g x_m) = 0
\, ,
\end{align}
can be solved order-by-order in the inverse 't Hooft coupling with the result
\begin{align}
c^{{\rm G}, \pm}_{u} (n, g) = (8 \pi g n)^{\pm 1/4} \left[ a^{{\rm G}, \pm}_{u} (n) + \frac{b^{{\rm G}, \pm}_{u} (n)}{4 \pi g}  + O (1/g^2) \right]
\, ,
\end{align}
where the explicit $a$ and $b$ coefficients are found to be
\begin{align}
a^{{\rm G}, +}_{u} (n) 
&
= - \frac{2 \ell \, \Gamma (n + \ft14)}{\Gamma (n + 1) \Gamma^{2} (\ft14)}  \chi_0^{{\rm G}, +} (u) 
\, , \qquad
a^{{\rm G}, -}_{u} (n) 
= 
- \frac{\ell \, \Gamma (n + \ft34)}{2 \Gamma (n + 1) \Gamma^{2} (\ft34)}  \chi_0^{{\rm G}, -} (u) 
\, , \\
b^{{\rm G}, +}_{u} (n)
&
=
\frac{2 \ell \, \Gamma (n + \ft14)}{\Gamma (n + 1) \Gamma^{2} (\ft14)}
\bigg\{
\left[
\frac{\pi}{16} + \frac{3}{8} \ln 2
\right]
\chi_0^{{\rm G}, -} (u)
\\
&
-
\left[
\frac{\pi}{16} - \frac{3}{8} \ln 2
\right]
(\chi_0^{{\rm G}, +} (u) - 8 \chi_{10}^{{\rm G}, +} (u))
-
\chi_{11}^{{\rm G}, +} (u)
+
\frac{1}{32 n} 
\left(
3 \chi_0^{{\rm G}, +} (u) 
-
32 \chi_{10}^{{\rm G}, +} (u)
\right)
\bigg\} 
\, , \nonumber\\
b^{{\rm G}, -}_{u} (n)
&
=
-
\frac{\ell \, \Gamma (n + \ft34)}{2 \Gamma (n + 1) \Gamma^{2} (\ft34)}
\bigg\{
\left[
- \frac{\pi}{16} + \frac{3}{8} \ln 2
\right]
\chi_0^{{\rm G}, +}(u)
\\
&
+
\left[
\frac{\pi}{16} + \frac{3}{8} \ln 2
\right]
(\chi_0^{{\rm G}, -} (u) - 8 \chi_{10}^{{\rm G}, -} (u))
+
\chi_{11}^{{\rm G}, -} (u)
+
\frac{1}{32 n} 
\left(
5 \chi_0^{{\rm G}, -} (u) 
-
32 \chi_{10}^{{\rm G}, -} (u)
\right)
\bigg\} 
\, , \nonumber
\end{align}
respectively. Here, we introduced inhomogeneities arising in the left-hand side of the quantization condition,
\begin{align}
\frac{\chi_u^{\rm G} (\pm 4 \pi i g |x_m|)}{V_0 (\pm 4 \pi g |x_m|)}
=
\ell \, \chi^{{\rm G}, \pm}_0 (u)
+
\frac{\ell}{4 \pi g}
\left[
\frac{
\chi^{{\rm G}, \pm}_{10} (u)}{x_m}
+
\chi^{{\rm G}, \pm}_{11} (u)
\right]
+
O (1/g^2)
\, ,
\end{align}
with
\begin{align}
\chi^{{\rm G}, \pm}_0 (u)
&=
\mp
\frac{1}{2 \sqrt{2}}
\left[
\frac{1}{1 \pm \hat{u}}
\left(
\frac{1 + \hat{u}}{1 - \hat{u}}
\right)^{1/4}
+
\frac{1}{1 \mp \hat{u}}
\left(
\frac{1 - \hat{u}}{1 + \hat{u}}
\right)^{1/4}
\right]
\, , \\
\chi^{{\rm G}, \pm}_{11} (u)
&=
\pm
\frac{\pi}{2 \sqrt{2}} \partial_{\hat{u}}
\left[
\frac{1}{1 \pm \hat{u}}
\left(
\frac{1 + \hat{u}}{1 - \hat{u}}
\right)^{1/4}
-
\frac{1}{1 \mp \hat{u}}
\left(
\frac{1 - \hat{u}}{1 + \hat{u}}
\right)^{1/4}
\right]
\, , \\
\chi^{{\rm G}, +}_{10} (u)
&=
-
\frac{3}{8 \sqrt{2}}
\left[
\frac{1}{(1 + \hat{u})^2}
\left(
\frac{1 + \hat{u}}{1 - \hat{u}}
\right)^{1/4}
+
\frac{1}{(1 - \hat{u})^2}
\left(
\frac{1 - \hat{u}}{1 + \hat{u}}
\right)^{1/4}
\right]
\, , \\
\chi^{{\rm G}, -}_{10} (u)
&=
+
\frac{5}{8 \sqrt{2}}
\left[
\frac{1}{(1 - \hat{u})^2}
\left(
\frac{1 + \hat{u}}{1 - \hat{u}}
\right)^{1/4}
+
\frac{1}{(1 + \hat{u})^2}
\left(
\frac{1 - \hat{u}}{1 + \hat{u}}
\right)^{1/4}
\right]
\, .
\end{align}

The defining condition for the odd $u$-parity coefficients
\begin{align}
\widetilde\Gamma_u^{\rm G} (4 \pi i g x_m) = 0
\end{align}
provides the result
\begin{align}
\tilde{c}^{{\rm G}, \pm}_{u} (n, g) = (8 \pi g n)^{\pm 1/4} \left[ \tilde{a}^{{\rm G}, \pm}_{u} (n) + \frac{\tilde{b}^{{\rm G}, \pm}_{u} (n)}{4 \pi g}  + O (1/g^2) \right]
\, ,
\end{align}
with $\tilde{a}$ and $\tilde{b}$ being 
\begin{align}
\tilde{a}^{{\rm G}, +}_{u} (n) 
&
= - \frac{2 \ell \, \Gamma (n + \ft14)}{\Gamma (n + 1) \Gamma^{2} (\ft14)} \tilde\chi_0^{{\rm G}, +} (u) 
\, , \qquad
\tilde{a}^{{\rm G}, -}_{u} (n) 
= 
- \frac{\ell \, \Gamma (n + \ft34)}{2 \Gamma (n + 1) \Gamma^{2} (\ft34)} \tilde\chi_0^{{\rm G}, -} (u) 
\, , \\
\tilde{b}^{{\rm G}, +}_{u} (n)
&
=
\frac{2 \ell \, \Gamma (n + \ft14)}{\Gamma (n + 1) \Gamma^{2} (\ft14)}
\bigg\{
\left[
\frac{\pi}{16} + \frac{3}{8} \ln 2
\right]
\tilde\chi_0^{{\rm G}, -} (u)
\\
&
-
\left[
\frac{\pi}{16} - \frac{3}{8} \ln 2
\right]
(\tilde\chi_0^{{\rm G}, +} (u) - 8 \tilde\chi_{10}^{{\rm G}, +} (u))
-
\tilde\chi_{11}^{{\rm G}, +} (u)
+
\frac{1}{32 n} 
\left(
3 \tilde\chi_0^{{\rm G}, +} (u) 
-
32 \tilde\chi_{10}^{{\rm G}, +} (u)
\right)
\bigg\} 
\, , \nonumber\\
\tilde{b}^{{\rm G}, -}_{u} (n)
&
=
-
\frac{\ell \, \Gamma (n + \ft34)}{2 \Gamma (n + 1) \Gamma^{2} (\ft34)}
\bigg\{
\left[
- \frac{\pi}{16} + \frac{3}{8} \ln 2
\right]
\tilde\chi_0^{{\rm G}, +}(u)
\\
&
+
\left[
\frac{\pi}{16} + \frac{3}{8} \ln 2
\right]
(\tilde\chi_0^{{\rm G}, -} (u) - 8 \tilde\chi_{10}^{{\rm G}, -} (u))
+
\tilde\chi_{11}^{{\rm G}, -} (u)
+
\frac{1}{32 n} 
\left(
5 \tilde\chi_0^{{\rm G}, -} (u) 
-
32 \tilde\chi_{10}^{{\rm G}, -} (u)
\right)
\bigg\} 
\, , \nonumber
\end{align}
where the functions $\tilde\chi (u)$ arise from the expansion of the source $\tilde\chi$ 
\begin{align}
\frac{\tilde\chi_u^{\rm G} (\pm 4 \pi i g |x_m|)}{V_0 (\pm 4 \pi g |x_m|)}
=
\ell \, \tilde\chi^{{\rm G}, \pm}_{0} (u)
+
\frac{\ell}{4 \pi g}
\left[
\frac{\tilde\chi^{{\rm G}, \pm}_{10} (u)}{x_m}
+
\tilde\chi^{{\rm G}, \pm}_{11} (u)
\right]
+
O (1/g^2)
\, ,
\end{align}
with
\begin{align}
\tilde\chi^{{\rm G}, \pm}_0 (u)
&=
\pm
\frac{1}{2 \sqrt{2}}
\left[
\frac{1}{1 \pm \hat{u}}
\left(
\frac{1 + \hat{u}}{1 - \hat{u}}
\right)^{1/4}
-
\frac{1}{1 \mp \hat{u}}
\left(
\frac{1 - \hat{u}}{1 + \hat{u}}
\right)^{1/4}
\right]
\, , \\
\tilde\chi^{{\rm G}, \pm}_{11} (u)
&=
\mp
\frac{\pi}{2 \sqrt{2}} \partial_{\hat{u}}
\left[
\frac{1}{1 \pm \hat{u}}
\left(
\frac{1 + \hat{u}}{1 - \hat{u}}
\right)^{1/4}
+
\frac{1}{1 \mp \hat{u}}
\left(
\frac{1 - \hat{u}}{1 + \hat{u}}
\right)^{1/4}
\right]
\, , \\
\tilde\chi^{{\rm G}, +}_{10} (u)
&=
+
\frac{3}{8 \sqrt{2}}
\left[
\frac{1}{(1 + \hat{u})^2}
\left(
\frac{1 + \hat{u}}{1 - \hat{u}}
\right)^{1/4}
-
\frac{1}{(1 - \hat{u})^2}
\left(
\frac{1 - \hat{u}}{1 + \hat{u}}
\right)^{1/4}
\right]
\, , \\
\tilde\chi^{{\rm G}, -}_{10} (u)
&=
-
\frac{5}{8 \sqrt{2}}
\left[
\frac{1}{(1 - \hat{u})^2}
\left(
\frac{1 + \hat{u}}{1 - \hat{u}}
\right)^{1/4}
-
\frac{1}{(1 + \hat{u})^2}
\left(
\frac{1 - \hat{u}}{1 + \hat{u}}
\right)^{1/4}
\right]
\, .
\end{align}

\subsection{Strong coupling expansion}

Using the just determined expansion coefficients, we can deduce the $1/g$ expansion of the flux-tube functions, which are
\begin{align}
\label{GammaG}
\Gamma_{\pm, u}^{\rm G} (\tau)
&
=
\mp
\frac{\ell}{2 \sqrt{2}} \left(
\frac{1 - \hat{u}}{1 + \hat{u}}
\right)^{1/4}
W^\pm (\tau, \hat{u})
\mp
\frac{\ell}{2 \sqrt{2}} \left(
\frac{1 + \hat{u}}{1 - \hat{u}}
\right)^{1/4}
W^\pm (\tau, - \hat{u})
+
i \ell \delta_{\pm, -} \sin (\tau \hat{u})
\\
&
\mp \frac{\ell \pi}{4 \pi g}
\partial_{\hat{u}}
\left[
\frac{1}{2 \sqrt{2}} \left(
\frac{1 - \hat{u}}{1 + \hat{u}}
\right)^{1/4}
W^\pm (\tau, \hat{u})
-
\frac{1}{2 \sqrt{2}} \left(
\frac{1 + \hat{u}}{1 - \hat{u}}
\right)^{1/4}
W^\pm (\tau, - \hat{u})
+
i \delta_{\pm, +} \cos (\tau \hat{u})
\right]
\nonumber\\
&
\mp
\frac{\ell \chi_0^{{\rm G}, -} (u)}{4 \pi g} \left( \frac{\pi}{8} + \frac{3}{4} \ln 2 \right) V_1^\pm (\tau)
\pm
\frac{\ell \chi_0^{{\rm G}, +} (u)}{4 \pi g} \left( \frac{\pi}{8} - \frac{3}{4} \ln 2 \right) \left[ V_1^\pm (\tau) \mp 4 \tau V_0^\mp (\tau) \right]
+
O (1/g^2)
\, , \nonumber
\end{align}
and
\begin{align}
\label{GammaGtilde}
\widetilde\Gamma_{\pm, u}^{\rm G} (\tau)
&
=
-
\frac{\ell}{2 \sqrt{2}} \left(
\frac{1 - \hat{u}}{1 + \hat{u}}
\right)^{1/4}
W^\pm (\tau, \hat{u})
+
\frac{\ell}{2 \sqrt{2}} \left(
\frac{1 + \hat{u}}{1 - \hat{u}}
\right)^{1/4}
W^\pm (\tau, - \hat{u})
-
i \ell \delta_{\pm, +} \cos (\tau \hat{u})
\\
&
- \frac{\ell \pi}{4 \pi g}
\partial_{\hat{u}}
\left[
\frac{1}{2 \sqrt{2}} \left(
\frac{1 - \hat{u}}{1 + \hat{u}}
\right)^{1/4}
W^\pm (\tau, \hat{u})
+
\frac{1}{2 \sqrt{2}} \left(
\frac{1 + \hat{u}}{1 - \hat{u}}
\right)^{1/4}
W^\pm (\tau, - \hat{u})
+
i \delta_{\pm, -} \sin (\tau \hat{u})
\right]
\nonumber\\
&
-
\frac{\ell \tilde\chi_0^{{\rm G}, -} (u)}{4 \pi g} \left( \frac{\pi}{8} + \frac{3}{4} \ln 2 \right) V_1^\pm (\tau)
+
\frac{\ell \tilde\chi_0^{{\rm G}, +} (u)}{4 \pi g} \left( \frac{\pi}{8} - \frac{3}{4} \ln 2 \right) \left[ V_1^\pm (\tau) \mp 4 \tau V_0^\mp (\tau) \right]
+
O (1/g^2)
\, , \nonumber
\end{align}
for the even and odd $u$-parity, respectively. Here $\delta_{++} = \delta_{--} = 1$ and $\delta_{+-}  = \delta_{-+} = 0$. Substituting these solutions into the scattering phases, we find
\begin{align}
\label{GGdynamicalPhases}
f_{\rm GG}^{(\alpha)} (u_1, u_2) 
=
\frac{\ell_1 \ell_2}{16 g}
\Bigg\{ 
A^{(\alpha)}_{\rm GG} (u_1, u_2) 
+
\frac{1}{4 g}
\left[
B^{(\alpha)}_{\rm GG} (u_1, u_2) 
+
\frac{3 \ln2}{2 \pi}  C^{(\alpha)}_{\rm GG} (u_1, u_2) 
\right]
+
O (1/g^2)
\Bigg\}
\, ,
\end{align}
where the linear dependence on $\ell$'s holds only up to the order in $1/g$ displayed and it becomes nonlinear beyond it. The explicit expressions are deferred to 
Appendix \ref{GGAppendix}.

At leading order, i.e., keeping just $A$'s, we find the known expressions \cite{Basso:2013vsa,Fioravanti:2013eia,Bianchi:2015iza}
\begin{align}
\ln S_{\rm GG} (u_1, u_2)
&
=
\frac{i \ell_1 \ell_2}{4 g (\hat{u}_1 -  \hat{u}_2)}
\left[
-
\left( \frac{1 - \hat{u}_1}{1 + \hat{u}_1} \right)^{1/4}
\left( \frac{1 + \hat{u}_2}{1 - \hat{u}_2} \right)^{1/4}
-
\left( \frac{1 + \hat{u}_1}{1 - \hat{u}_1} \right)^{1/4}
\left( \frac{1 - \hat{u}_2}{1 + \hat{u}_2} \right)^{1/4}
+
2
\right]
\, , \\
\ln S_{\ast\rm GG} (u_1, u_2)
&
=
\frac{\ell_1 \ell_2}{4 g (\hat{u}_1 -  \hat{u}_2)}
\left[
\left( \frac{1 - \hat{u}_1}{1 + \hat{u}_1} \right)^{1/4}
\left( \frac{1 + \hat{u}_2}{1 - \hat{u}_2} \right)^{1/4}
-
\left( \frac{1 + \hat{u}_1}{1 - \hat{u}_1} \right)^{1/4}
\left( \frac{1 - \hat{u}_2}{1 + \hat{u}_2} \right)^{1/4}
-
2 i
\right]
\, ,
\end{align}
for the direct and mirror S-matrices, respectively. The subleading corrections were recently verified by a direct calculation in string perturbation theory\footnote{We
would like to thank Lorenzo Bianchi for bringing these results to our attention.} \cite{Bianchi:2015vgw}.

The pentagon transitions at strong coupling are found by substituting the above result \re{GGdynamicalPhases} into the expressions for pentagons derived in 
Appendix \ref{GaugePentagon},
\begin{align}
\label{PGGmirror}
&
P_{\rm G|G} (u_1|u_2) 
=
w_{\rm GG} (u_1, u_2)
P_{\rm G|\bar{G}} (u_1|u_2) 
\\
&\quad
=
\frac{{\rm e}^{- i f^{(1)}_{\rm GG} (u_1, u_2) + i f^{(2)}_{\rm GG} (u_1, u_2) - f^{(3)}_{\rm GG} (u_1, u_2) + f^{(4)}_{\rm GG} (u_1, u_2)}
}{s_{\ast \ell_1 \bar{\ell}_2} (u_1, u_2)}
\left[
\frac{
\left(
1 - \frac{1}{\hat{x}^{[- \ell_1]} [\hat{u}_1] \hat{x}^{[- \ell_2]} [\hat{u}_2]}
\right)
\left(
1 - \frac{1}{\hat{x}^{[\ell_1]} [\hat{u}_1] \hat{x}^{[\ell_2]} [\hat{u}_2]}
\right)
}{
\left(
1 - \frac{1}{\hat{x}^{[\ell_1]} [\hat{u}_1] \hat{x}^{[- \ell_2]} [\hat{u}_2]}
\right)
\left(
1 - \frac{1}{\hat{x}^{[-\ell_1]} [\hat{u}_1] \hat{x}^{[\ell_2]} [\hat{u}_2]}
\right)
}
\right]^{1/2}
\!\!\! , \nonumber
\end{align}
with $w_{\rm GG}$ given by Eq.\ \re{wGGboundstate}. Here we employed the following conventions for the shifted rapidities and rescaled Zhukowski gluon variable 
$x [u] = g \, \hat{x} [\hat{u}]$,
\begin{align}
\label{GluonZhuk}
\hat{x}^{[\pm \ell]} [ \hat{u}] \equiv \hat{x} [\hat{u} \pm i \ft{\ell}{4 g}]
\, .
\end{align}
Finally, let us quote the bound state measure to order $O (1/g^2)$
\begin{align}
\mu_{\rm G} (u) = \frac{1}{\ell^2} \exp \left( - \frac{\ell^2}{8 g (1 - \hat{u}^2)} 
\left[
1+ \frac{3 \pi + 12 \ln 2 \, (1 + \hat{u}^2) }{16 \pi g (1 - \hat{u}^2)}  
\right]
+
O(1/g^3)
\right) 
\, . 
\end{align}
Let us point out, however, that in the derivation of this expression it is important to realize that the $g \to \infty$ and the square limit, used to obtain the measure
from the pentagon, do not commute. Strong coupling comes first. The above $1/\ell^2$ arises solely from the $1/s_{\ast \ell_1 \bar{\ell}_2} (u_1, u_2)$ prefactor in 
Eq.\ \re{PGGmirror}.

\section{Fermion--gauge bound state transitions}
\label{FermionGluonSection}

The gauge bound state-(anti)fermion S-matrices are easily constructed along the same lines as the ones for a single gauge excitation and read
\begin{align}
\label{Slfdirect}
S_{\ell {\rm f}} (u_1, u_2) 
&= \frac{u_1 - u_2 - i \ft{\ell}{2}}{u_1 - u_2 + i \frac{\ell}{2}} S_{\ell {\rm \bar{f}}} (u_1,u_2) 
\\
&
=
\exp \left( - 2 i f^{(1)}_{\ell {\rm f}} (u_1, u_2) + 2 i f^{(2)}_{\ell {\rm f}} (u_1, u_2) \right)
\, , \nonumber\\
\label{Slfmirror}
S_{\ast\ell {\rm f}} (u_1, u_2) 
&= 
\frac{u_1 - u_2 - i \ft{\ell}{2}}{u_1 - u_2 + i \frac{\ell}{2}} S_{\ast\ell {\rm f}} (u_1, u_2) 
\\
&
=
\frac{g^2 (- 1)^\ell}{x_{\rm f} [u_2] (u_1 - u_2 + i \ft{\ell}{2})}
\exp \left( 2 f^{(3)}_{\ell {\rm f}} (u_1, u_2) - 2 f^{(4)}_{\ell {\rm f}} (u_1, u_2) \right)
\, . \nonumber 
\end{align}
Since the prefactor, as a function of the gauge rapidity, is rational, we do not even need to pass to the Goldstone sheet to fuse this rational factor for the gluon-antifermion S-matrix. 
While the dynamical phases can be easily generalized for any $\ell$
\begin{align}
\label{flf1}
f^{(1)}_{\ell {\rm f}} (u_1, u_2) 
&
=
\int_0^\infty \frac{dt}{t} {\rm e}^{- \ell t/2} \sin(u_1 t)
\left[
\frac{\gamma^{\rm f}_{-, u_2} (2gt)}{1 - {\rm e}^{- t}}
+
\frac{\gamma^{\rm f}_{+, u_2} (2gt)}{{\rm e}^{t} - 1}
\right]
\, , \\
f^{(2)}_{\ell {\rm f}} (u_1, u_2)
&
=
\int_0^\infty \frac{dt}{t} \left( {\rm e}^{- \ell t/2} \cos(u_1 t) - J_0 (2gt) \right)
\left[
\frac{\widetilde\gamma^{\rm f}_{+, u_2} (2gt)}{1 - {\rm e}^{- t}}
+
\frac{\widetilde\gamma^{\rm f}_{-, u_2} (2gt)}{{\rm e}^{t} - 1}
\right]
\, , \\
f^{(3)}_{\ell {\rm f}} (u_1, u_2)
&
=
-
\int_0^\infty \frac{dt}{t} {\rm e}^{- \ell t/2} \sin(u_1 t)
\left[
\frac{\widetilde\gamma^{\rm f}_{-, u_2} (2gt)}{1 - {\rm e}^{- t}}
-
\frac{\widetilde\gamma^{\rm f}_{+, u_2} (2gt)}{{\rm e}^{t} - 1}
\right]
\, , \\
\label{flf4}
f^{(4)}_{\ell {\rm f}} (u_1, u_2)
&
=
\int_0^\infty \frac{dt}{t} \left( {\rm e}^{- \ell t/2} \cos(u_1 t) - J_0 (2gt) \right)
\left[
\frac{\gamma^{\rm f}_{+, u_2} (2gt)}{1 - {\rm e}^{- t}}
-
\frac{\gamma^{\rm f}_{-, u_2} (2gt)}{{\rm e}^{t} - 1}
\right]
\, ,
\end{align}
after using exchange relations for the phases given in terms of gauge bound state flux-tube functions, see Eqs.\ (A.33), (A.34), (A.40) and (A.41) of Ref.\ \cite{Belitsky:2014lta}. 

\subsection{Passing to Goldstone sheet}

Let us pass to the Goldstone sheet since this is where we will perform the strong coupling expansion. The scattering matrices read\footnote{To derive the last line in Eq.\ \re{SastGf} the following 
formula is useful
$$
\int_0^\infty \frac{dt}{t} \left[ \sin(u_1^{[- \ell]}t) \sin(u_2t) + \cos(u_1^{[- \ell]}t) \cos(u_2t) - \cos(u_2t) J_0 (2gt) \right] = \ln \frac{g^2}{x_{\rm f} [u_2] (u_1^{[-\ell]} - u_2)}
\, .
$$
}
\begin{align}
S_{\rm Gf} (u_1, u_2) 
&= 
\frac{u_1 - u_2 - i \ft{\ell}{2}}{u_1 - u_2 + i \ft{\ell}{2}}
S_{\rm G\bar{f}} (u_1, u_2) 
\\
&= 
\exp \left( - 2i f^{(1)}_{\rm Gf} (u_1, u_2) + 2 i f^{(2)}_{\rm Gf} (u_1, u_2) \right)
\, , \nonumber\\
\label{SastGf}
S_{\rm \ast Gf} (u_1, u_2) 
&=
\frac{u_1 - u_2 - i \ft{\ell}{2}}{u_1 - u_2 + i \ft{\ell}{2}}
S_{\rm \ast G\bar{f}} (u_1, u_2) 
\\
&= -
\frac{u_1 - u_2 - i \ft{\ell}{2}}{u_1 - u_2 + i \ft{\ell}{2}}
\exp \left( 2 f^{(3)}_{\rm Gf} (u_1, u_2) - 2 f^{(4)}_{\rm Gf} (u_1, u_2) \right)
\, , \nonumber 
\end{align}
with corresponding dynamical phases being
\begin{align}
f^{(1)}_{\rm Gf} (u_1, u_2)
&
=
i
\int_0^\infty \frac{dt}{t} {\rm e}^{i u_1 t}
\sinh \frac{\ell t}{2}
\left[
\frac{\gamma_{-, u_2}^{\rm f} (2gt)}{1 - {\rm e}^{- t}}
+
\frac{\gamma_{+, u_2}^{\rm f} (2gt)}{{\rm e}^t - 1} 
\right]
\nonumber\\
&
=
- \frac{1}{2}
\int_0^\infty \frac{dt}{t} \cos (u_2 t) \,
\widetilde\gamma_{+, u_1}^{\rm G} (2gt)
\, , 
\\
f^{(2)}_{\rm Gf} (u_1, u_2)
&
=
-
\int_0^\infty \frac{dt}{t} {\rm e}^{i u_1 t}
\sinh \frac{\ell t}{2}
\left[
\frac{\widetilde\gamma_{+, u_2}^{\rm f} (2gt)}{1 - {\rm e}^{- t}}
+
\frac{\widetilde\gamma_{-, u_2}^{\rm f} (2gt)}{{\rm e}^t - 1} 
\right]
\nonumber\\
&
=
- \frac{1}{2}
\int_0^\infty \frac{dt}{t} \sin (u_2 t) \,
\gamma_{-, u_1}^{\rm G} (2gt)
\, , 
\\
f^{(3)}_{\rm Gf} (u_1, u_2)
&
=
- i
\int_0^\infty \frac{dt}{t} {\rm e}^{i u_1 t}
\sinh \frac{\ell t}{2}
\left[
\frac{\widetilde\gamma_{-, u_2}^{\rm f} (2gt)}{1 - {\rm e}^{- t}}
-
\frac{\widetilde\gamma_{+, u_2}^{\rm f} (2gt)}{{\rm e}^t - 1} 
\right]
\nonumber\\
&
=
\frac{1}{2}
\int_0^\infty \frac{dt}{t} \sin (u_2 t) \,
\widetilde\gamma_{-, u_1}^{\rm G} (2gt)
\, , 
\\
f^{(4)}_{\rm Gf} (u_1, u_2)
&
=
-
\int_0^\infty \frac{dt}{t} {\rm e}^{i u_1 t}
\sinh \frac{\ell t}{2}
\left[
\frac{\gamma_{+, u_2}^{\rm f} (2gt)}{1 - {\rm e}^{- t}}
-
\frac{\gamma_{-, u_2}^{\rm f} (2gt)}{{\rm e}^t - 1} 
\right]
\nonumber\\
&
=
- \frac{1}{2}
\int_0^\infty \frac{dt}{t} \cos (u_2 t) \,
\gamma_{+, u_1}^{\rm G} (2gt)
\, .
\end{align}
Everywhere above it is implied that $\Im{\rm m}[u_1] > \ell/2$. We also showed results in terms of the gauge bound state flux-tube functions. Both of these expressions will be used in 
the next section along with the strong expansion constructed earlier to verify their consistency.

\subsection{Strong coupling expansion}

Employing the strong-coupling expansion of flux-tube functions worked out earlier, we can calculate the dynamical phases of the gluon-fermion pentagons. They admit the following form
\begin{align}
\label{fGfdynamicalPhases}
f^{(\alpha)}_{\rm Gf} (u_1, u_2)
=
\frac{\ell}{16 \sqrt{2} g}
\Bigg\{ 
A^{(\alpha)}_{\rm Gf} (\hat{u}_1, \hat{u}_2) 
+
\frac{1}{4 g}
\left[
B^{(\alpha)}_{\rm Gf} (\hat{u}_1, \hat{u}_2) 
+
\frac{3 \ln2}{2 \pi}  C^{(\alpha)}_{\rm Gf} (\hat{u}_1, \hat{u}_2) 
\right]
+
O (1/g^2)
\Bigg\}
\, ,
\end{align}
with explicit expressions presented in Appendix \ref{FGAppendix}. The latter are the same obtained either by using gauge bound state, i.e., Eqs.\ \re{GammaG}, \re{GammaGtilde}, or small 
fermion, Eqs.\ \re{Gammaf}, \re{Gammaftilde}, flux-tube functions providing a very nice check on the formalism. 

Using explicit solutions, we find immediately at strong coupling
\begin{align}
\ln S_{\rm Gf} (u_1, u_2)
&
=
\frac{i \ell}{4 \sqrt{2} g  (\hat{u}_1 - \hat{u}_2)}
\left[
\left( \frac{1 - \hat{u}_1}{1 + \hat{u}_2} \right)^{1/4} \left( \frac{\hat{u}_2 + 1}{\hat{u}_2 - 1} \right)^{1/4}
\!\!
+
\left( \frac{1 + \hat{u}_1}{1 - \hat{u}_1} \right)^{1/4} \left( \frac{\hat{u}_2 - 1}{\hat{u}_2 + 1} \right)^{1/4}
- \sqrt{2}
\, 
\right]
\! , \\
\ln S_{\rm \ast Gf} (u_1, u_2)
&
=
\frac{\ell}{4 \sqrt{2} g  (\hat{u}_1 - \hat{u}_2)}
\left[
\left( \frac{1 - \hat{u}_1}{1 + \hat{u}_1} \right)^{1/4} \left( \frac{\hat{u}_2 + 1}{\hat{u}_2 - 1} \right)^{1/4}
\!\!
-
\left( \frac{1 + \hat{u}_1}{1 - \hat{u}_1} \right)^{1/4} \left( \frac{\hat{u}_2 - 1}{\hat{u}_2 + 1} \right)^{1/4}
+ i \sqrt{2}
\,
\right]
\! ,
\end{align}
confirming leading order results of Ref.\ \cite{Fioravanti:2015dma}. With Eq.\ \re{fGfdynamicalPhases}, we can now uncover subleading terms.

The pentagon transitions at strong coupling are computed making use of the formulas derived in Appendix \ref{GFpentagonsAppendix}.
For the gauge bound states transitioning into small fermion (and vice versa), the results are 
\begin{align}
\label{BoundGFpentagon}
P_{\rm G|f} (u_1|u_2)
&
=
(-1)^{\ell}
\frac{\hat{u}_1 - \hat{u}_2 - i \ft{\ell}{4 g}}{\hat{u}_1 - \hat{u}_2 + i \ft{\ell}{4 g} }
\left[ 
\frac{\hat{x} [\hat{u}_2] - \hat{x}^{[-\ell]} [\hat{u}_1]}{\hat{x} [\hat{u}_2] - \hat{x}^{[+\ell]} [\hat{u}_1]}
\right]^{1/2}
\\
&
\times
\exp 
\left(
- i f^{(1)}_{\rm Gf} (u_1, u_2) + i f^{(2)}_{\rm Gf} (u_1, u_2) - f^{(3)}_{\rm Gf} (u_1, u_2) + f^{(4)}_{\rm Gf} (u_1, u_2) 
\right)
\, , \nonumber\\
\label{PfG}
P_{\rm f|G} (u_2|u_1)
&
=
i (-1)^\ell
\left[ 
\frac{\hat{x} [\hat{u}_2] - \hat{x}^{[-\ell]} [\hat{u}_1]}{\hat{x} [\hat{u}_2] - \hat{x}^{[+\ell]} [\hat{u}_1]}
\right]^{1/2}
\\
&
\times
\exp 
\left(
i f^{(1)}_{\rm Gf} (u_1, u_2) - i f^{(2)}_{\rm Gf} (u_1, u_2) - f^{(3)}_{\rm Gf} (u_1, u_2) + f^{(4)}_{\rm Gf} (u_1, u_2) 
\right)
\, , \nonumber
\end{align}
and consist in substituting the phases from Appendix \ref{FGAppendix} along with Taylor expanding prefactors following the conventions introduced in Eqs.\ \re{fermionZhuk} and \re{GluonZhuk},
for rescaled small fermion and gauge Zhukowski variables, respectively.

\section{Constraints from Descent Equation}
\label{DESection}

Before we turn to applications, let us provide an additional layer of constraints on the form of the strong-coupling expansion for pentagons. This is offered by the Descent Equation 
\cite{CaronHuot:2011kk,Bullimore:2011kg}  which was recently studied within the context of the pentagon OPE in Ref.\ \cite{Belitsky:2015kda}. 

For the fermion-fermion pentagon, one can immediately find, making use of the results derived in Sect.~\ref{SmallFermionSection}, that it verifies the condition derived in 
\cite{Belitsky:2015kda} when one passes to the small fermion kinematics which dominates the strong coupling limit,
\begin{align}
\int
\frac{d \hat{x}_{\rm f}}{\hat{x}_{\rm f}} (1 -  \hat{x}_{\rm f}^2) \mu_{\rm f} (u) {\rm e}^{- \tau^\prime [ E_{\rm f} (u) - 1]} \delta \big(p_{\rm f} (u)\big) 
P_{{\rm f} | {\rm f}} (- u | v ) 
=
\frac{4 i}{\Gamma (g)}
\, ,
\end{align}
with the right-hand side defined by the cusp anomalous dimension that admits the following strong-coupling expansion
\cite{Beisert:2006ez,AldAruBenEdeKle07,KotLip07,KosSerVol07,BecAngFor07,Basso:2007wd,GubKlePol03,Kru06,FroTse03,FroTirTse07,CasKri07,Belitsky:2007kf}
\begin{align}
\Gamma (g) = 2 g - \frac{3 \ln 2}{2 \pi} + O (1/g)
\, .
\end{align}
Above, we changed from the rapidity variable $\hat{u}$ to the Zhukowski $\hat{x}_{\rm f}$ via $\hat{u} = ( \hat{x}_{\rm f} + \hat{x}_{\rm f}^{-1} )/2$ in the integration measure and adopted
the small fermion energy and momentum dispersion relation from Ref.\ \cite{Basso:2010in} 
\begin{align}
E_{\rm f} (u) = 1 + O (\hat{x}_{\rm f}^2)
\, , \qquad
p_{\rm f} (u) = \frac{\Gamma (g)}{2 g} \hat{x}_{\rm f} + O (\hat{x}_{\rm f}^3)
\, .
\end{align}

Another check involves the fermion-gluon pentagon, see Eq.\ (39) in Ref.\ \cite{Belitsky:2015kda}. Passing in that relation to the  fermion and Goldstone sheets for fermions and gauge excitations,
respectively, we find 
\begin{align}
\int 
\frac{d \hat{x}_{\rm f}}{\hat{x}_{\rm f}} (1 -  \hat{x}_{\rm f}^2) \mu_{\rm f} (u) 
{\rm e}^{- \tau^\prime [ E_{\rm f} (u) - 1]} \delta \big(p_{\rm f} (u)\big) 
&
\int d \mu_{\rm G} (v) \left[  P_{{\rm f}|{\rm G}} (- u | v) \left[ \frac{\hat{x}^+[\hat{v}]}{\hat{x}^- [\hat{v}]} \right]^{1/2} - i \right]
\\
= - \frac{2 i g^2}{\Gamma (g)}
&
\int d \mu_{\rm G} (v) \left[  \frac{g}{\hat{x}^-[\hat{v}]}  - \frac{g}{\hat{x}^+[\hat{v}]} - \ft{i}{2} \left( E_{\rm G} (v) + i p_{\rm G} (v) \right) \right]
\, , \nonumber
\end{align}
where we introduced a differential of the integration measure for later convenience that includes the propagating ``phase'' factor
\begin{align}
\label{DiffMeasure}
d \mu_{\rm p} (v) = \frac{d v}{2 \pi} \mu_{\rm p} (v) {\rm e}^{- \tau E_{\rm p} (v)  + i \sigma p_{\rm p} (v)}
\, ,
\end{align}
with ${\rm p} = {\rm G}$ for the case at hand. A simple counting of powers of the 't Hooft coupling immediately exhibits the fact that this equation relates contributions at different orders
in $g^2$, i.e., its left-hand side requires effects an order higher in coupling compared to its right-hand side. Using the explicit strong coupling solutions from the previous section 
(for $\ell = 1$), we can expand the left-hand side in the vicinity of $\hat{x}_{\rm f} = 0$,
\begin{align}
P_{\rm f|G} (- u|v) \left[ \frac{\hat{x}^+ [\hat{v}]}{\hat{x}^- [\hat{v}]} \right]^{1/2}
=
i
+
\frac{i \hat{x}_{\rm f}}{2 g}
\left[
\frac{g}{\hat{x}^- [v]} - \frac{g}{\hat{x}^+ [v]}
-
\ft{i}{2} (E_{\rm G} (v) + i p_{\rm G} (v))
\right]
+ O (\hat{x}_{\rm f}^2)
\, ,
\end{align}
reproducing the one on the right. Here the energy and momentum of a single gauge excitation are \cite{Basso:2010in}
\begin{align}
E_{\rm G} (v) 
\simeq \frac{1}{\sqrt{2}}
\left[
\left(
\frac{1 + \hat{v}}{1 - \hat{v}}
\right)^{1/4}
+
\left(
\frac{1 - \hat{v}}{1 + \hat{v}}
\right)^{1/4}
\right]
\, , \quad
p_{\rm G} (v) 
\simeq \frac{1}{\sqrt{2}}
\left[
\left(
\frac{1 + \hat{v}}{1 - \hat{v}}
\right)^{1/4}
-
\left(
\frac{1 - \hat{v}}{1 + \hat{v}}
\right)^{1/4}
\right]
\, ,
\end{align}
at leading order, with subleading terms in coupling which can be extracted from Appendix D.2 of Ref.\ \cite{Basso:2010in}.

\section{Application}

As an immediate application of the just derived strong-coupling results, we consider the $\chi_1 \chi_4^3$ component $\mathcal{W}_6^{(\chi_1 \chi_4^3)}$ of the NMHV hexagon,---a function of three 
conformal cross ratios $\tau, \sigma, \phi$,---in the OPE limit $\tau \to \infty$. Though we systematically constructed the $1/g$ expansion in the previous sections, we will restrict our consideration 
below to leading effects in $g$ only in order to observe the emergence of the classical string area from the summation of the pentagon OPE series. The study of subleading terms is much more 
cumbersome and is postponed to a future study.

We start our analysis with the consideration of the contribution of the fermion, that carries the Grassmann quantum numbers of the $\mathcal{W}_6^{(\chi_1 \chi_4^3)}$ component of the
hexagon, along with the bound state of $\ell$ gluons , i.e., the states $\ket{\ell (u) {\rm f} (v)}$. Thus, we have to resum the series
\begin{align}
\label{POPEGFseries}
\mathcal{W}_6^{(\chi_1 \chi_4^3)}
=
\sum_{\ell = 1}^\infty
{\rm e}^{ i (\ell + 1/2) \phi}
\mathcal{W}_{\ell {\rm f}}
\end{align}
where the individual contributions admit the form
\begin{align}
\mathcal{W}_{\ell {\rm f}}
=
\int_{C_{\rm f}} \int_{C_{\rm G}}  \frac{ d \mu_{\rm G} (u) d \mu_{\rm f} (v) (- i) x_{\rm f} [v]}{|P_{\rm G|f} (u|v)|^2}
\, .
\end{align}
To make notations in the integrand more compact, here and below $|P_{\rm p|p^\prime} (u|v)|^2$ stands for $|P_{\rm p|p^\prime} (u|v)|^2 = P_{\rm p|p^\prime} (u|v) 
P_{\rm p^\prime|p} (v|u)$. Above, the differential measures were introduced in Eq.\ \re{DiffMeasure} and the integration contour for the small fermion is $C_{\rm f} = 
(-\infty, -2g) \cup (2 g, \infty)$. For the gluon it is bound to the interval $C_{\rm G} = (-2g, 2g)$, since outside of it the gauge excitation behaves as a giant hole, i.e., its energy 
and momentum scale as a first power of 't Hooft coupling $g$, and induce only exponentially suppressed contribution to the Wilson loop. By virtue of the complementarity 
of the fermionic and gluonic domains, we cannot hit the pole in \re{BoundGFpentagon} at strong coupling. 

Then, at leading order in strong coupling $|P_{\rm f|G}|^2 \sim 1$ and the integral over rapidities factorizes by virtue of this property. Therefore, the sum over all values of $\ell$ in 
Eq.\ \re{POPEGFseries} can be evaluated in a closed form\footnote{Here we employed the well-known series representation of the dilogarithm ${\rm Li}_2 (z) = \sum_{\ell = 1}^\infty
z^\ell/\ell^2$. Let us point out that comparing the obtained expression with Eq.\ (F.46) of Ref.\ \cite{Alday:2010ku}, one has to realize that the parameter $\mu$ in this reference
is related to the angle $\phi$ via $\mu = - {\rm e}^{i \phi}$ as stated after Eq.\ (F.51). So the argument of the dilogarithm comes with a plus sign.},
\begin{align}
\label{W6Li2gluon}
\mathcal{W}_{6}^{(\chi_1 \chi_4^3)}
=
{\rm e}^{i \phi/2}
\int
d \mu_{\rm f} (v) (- i) x_{\rm f} [v]
\left(
1 -
\int \frac{d u}{2 \pi} \mu_{\rm G} (u)
{\rm Li}_2 \left(
{\rm e}^{- \tau E_{\rm G} (u) + i \sigma p_{\rm G} (u) + i \phi}
\right)
\right)
\, .
\end{align}
The second term in braces is of order $g$ and is the first term in the expansion of the exponential of the minimal area. To restore the latter, one has to resum all
one-fermion--multiple gauge bound states contributions. For $N$ of these bound states accompanying the fermion, we find
\begin{align}
\frac{{\rm e}^{i \phi/2}}{N!}
\sum_{\ell_1, \dots, \ell_N = 1}^\infty {\rm e}^{i \phi(\ell_1 + \dots + \ell_N)}
&
\int d \mu_{\rm f} (v) (- i) x_{\rm f} [v]
\int \frac{d \mu_{{\rm G}} (u_1) \dots d \mu_{{\rm G}} (u_N)
}{\prod_{j=1}^N |P_{{\rm G}| {\rm f}} (u_j | v)|^2 \prod_{k > j =1}^N |P_{{\rm G}|{\rm G}} (u_k | u_j)|^2}
\, .
\end{align}
Again by virtue of the scaling $P_{\rm G|G} \sim 1$, we observe factorization and, after the summation over $N$, we deduce
\begin{align}
\mathcal{W}_{6}^{(\chi_1 \chi_4^3)}
=
{\rm e}^{i \phi/2}
\int
d \mu_{\rm f} (v) (- i) x_{\rm f} [v]
\exp
\left(
-
\,
\int \frac{d u}{2 \pi} \mu_{\rm G} (u)
{\rm Li}_2 \left(
{\rm e}^{- \tau E_{\rm G} (u) + i \sigma p_{\rm G} (u) + i \phi}
\right)
\right)
\, .
\end{align}
Let us clarify that here and in Eq.\ \re{W6Li2gluon}, $ \mu_{\rm G}$ stands for the single-gluon measure. Adding to this the effect of antigluon bound states, we modify the exponent by 
an addendum that differs from the displayed term by a mere sign change in front of $\phi$. In this manner, we recover the gluon portion of the minimal area in the $\tau \to \infty$ limit of 
the NMHV amplitude, which obviously contains an overall factor of integrated fermionic measure that is responsible for quantum numbers of the component of the superWilson loop 
under study.

The contribution to MHV amplitude at strong coupling receives an extra effect from an elusive excitation of mass two \cite{Alday:2007hr,Alday:2009dv,Alday:2010vh}. As was first explained in 
Ref.\ \cite{Basso:2014koa} within the OPE framework, it is not an elementary but rather a virtual composite state of small fermion-antifermion pair that comes on-shell as a bound state at infinite 
coupling. This idea was further pursued in an effective framework of Ref.\ \cite{Fioravanti:2015dma} that assumed the existence of bound states of these mesons to reproduce the result from 
Thermodynamic Bethe Ansatz \cite{Alday:2007hr,Alday:2009dv,Alday:2010vh}.

For the case at hand, we thus continue with the contribution of $| \bar{\rm f} (u_1) {\rm f}(v_1) {\rm f} (v_2) \rangle$ state to the NMHV hexagon
\begin{align}
\mathcal{W}_{\rm (\bar{f}f) f}
&
=
\frac{1}{2! 1!}
\int d u_1 d v_1 d v_2 
\frac{\mu_{\rm f} (u_1) \mu_{\rm f} (v_1) \mu_{\rm f} (v_2)}{| P_{\rm f|\bar{f}} (u_1|v_1) P_{\rm f|\bar{f}} (u_1|v_2) P_{\rm f|f} (v_1|v_2) |^2} 
\frac{x_{\rm f} [v_1] x_{\rm f} [v_2]}{x_{\rm f} [u_1]} \mathcal{R}_1 (u_1, v_1, v_2)
\, ,
\end{align}
where $\mathcal{R}_1$ is a matrix part of the transition. The form of the latter for any internal symmetry group quantum numbers was recently worked out in Ref.\ \cite{Basso:2015uxa}. 
What is important for the current analysis is that it has the following generic form
\begin{align}
\mathcal{R}_N (\bit{u}_N, \bit{v}_{N+1})
=
\frac{P_{N}  (\bit{u}_N, \bit{v}_{N+1})}{\prod_{j > i} [(v_j-v_i)^2 + 1] \prod_{l > k} [(u_l-u_k)^2 + 1] \prod_{m, n} [(u_m - v_n )^2 + 4]}
\, , 
\end{align}
and possesses poles expected for nonsinglet transitions \cite{Basso:2014koa,Belitsky:2014lta}. The polynomial in the numerator is of degree $2^{2 N-1}$ in variables $\bit{u}_N = (u_1, \dots, u_N)$ 
and $\bit{v}_{N+1} = (v_1, \dots, v_{N + 1})$. The lowest nontrivial one is
\begin{align}
P_1 (u_1, v_1, v_2) 
= 
40 + 6 u_1^2 + 4 v_1^2 - 2 v_1 v_2 + 4 v_2^2 - 6 u_1 (v_1 + v_2)
\, .
\end{align}

Rescaling the fermionic rapidities with the coupling constant, $u_j = 2g \hat{u}_j$ etc., one observes that the $\mathcal{W}_{\rm (\bar{f}f) f}$ would be suppressed compared to the 
contribution of gluons analyzed above. However, there is a subtlety here, pointed out in Ref.\ \cite{Basso:2014koa}, that the integration contour gets pinched by the 
aforementioned poles as $g \to \infty$. Thus one has to move the integration contour for $u_1$ to the lower half-plane picking up two poles along the way $u_1 = v_j - 2i$ ($j = 1,2$). 
The latter induce leading order effect in coupling, on the same footing as gauge fields, and read
\begin{align}
\label{Wffbarf}
\mathcal{W}_{\rm (\bar{f}f) f}
=
\!
\int\! d \mu_{\rm f} (v_2) (- i) x_{\rm f} [v_2] 
\int\! d \mu_{\rm f\bar{f}} (v_1)
\frac{x^{[+2]}_{\rm f}[v_1]}{x^{[-2]}_{\rm f}[v_1]}
\frac{- 1}{|P_{\rm f|\bar{f}} (v^{[-2]}_1 | v_2)|^2 |P_{\rm f|f} (v^{[+2]}_1 | v_2)|^2 (v_1 - v_2) (v^{[+2]}_1 - v_2)}
, 
\end{align}
where we dropped subleading contributions from the deformed contour. Here the composite fermion-antifermion measure is \cite{Basso:2014koa}
\begin{align}
\label{muffbar}
\mu_{\rm f\bar{f}} (v) = - \frac{\mu_{\rm f} (v + i) \mu_{\rm f} (v + i)}{|P_{\rm f|\bar{f}} (v + i| v - i)|^2}
\, .
\end{align}
with the energy/momentum of the composite excitation being $E_{\rm f\bar{f}} (v) = E_{\rm f} (v + i)+ E_{\rm f} (v - i)$/$p_{\rm f\bar{f}} (v) = p_{\rm f} (v + i)+ p_{\rm f} (v - i)$.
Making use of the explicit expressions for the pentagons at strong coupling, one finds that the expression accompanying measures in Eq.\ \re{Wffbarf} goes to minus one 
at leading order, yielding a product representation of the single fermion accompanied by the $(\bar{\rm f} {\rm f})$-pair propagating in the OPE channel.

Generally, for $N$ $(\bar{\rm f} {\rm f})$-pairs, we have
\begin{align}
\mathcal{W}_{{\rm (\bar{f} f)}^N {\rm f}}
=
\frac{1}{N! (N-1)!}
\int 
&
\frac{\prod_{i = 1}^{N} d \mu_{\rm f} (u_i) \prod_{j = 1}^{N + 1} d \mu_{\rm f} (v_j)}{| 
\prod_{j > i} P_{\rm f|f} (v_j|v_i)^2 \prod_{l > k} P_{\rm f|f} (u_l|u_k) \prod_{m, n} P_{\rm f|\bar{f}} (u_m|v_n )
|^2} 
\nonumber\\
&\times
\left(
\prod_{i = 1}^N 
\frac{x_{\rm f} [v_i] }{x_{\rm f} [u_i]} 
\right)
(- i) x_{\rm f} [v_{N + 1}] \mathcal{R}_N (\bit{u}_N, \bit{v}_{N+1})
\, .
\end{align}
The polynomial $\mathcal{R}_N$ obeys a very important property: taking the residue of $\mathcal{R}_N$, for instance, at $u_1 = v_1 + 2i$, yields
\begin{align}
\res\limits_{u_1 = v_1 + 2i} \mathcal{R}_N (\bit{u}_N, \bit{v}_{N+1})
&=
\frac{1}{\prod_{j > 1} [v_j - v_1] [v_j-v_1 - i] \prod_{k > 1} [u_k - v_1 - 2 i] [u_k - v_1 - i]}
\\
&\times
\frac{P_{N-1}  (\bit{u}_{N-1}, \bit{v}_N)}{\prod_{j > i>1} [(v_j-v_i)^2 + 1] \prod_{l > k>1} [(u_l-u_k)^2 + 1] \prod_{m, n \neq 1} [(u_m - v_n )^2 + 4]}
\, , \nonumber
\end{align}
with the polynomial of a lower degree. Thus, we do not need the explicit form of $\mathcal{R}_N$ here. Consecutively  taking the residues, we find
\begin{align}
&
\mathcal{W}_{{\rm (\bar{f}f)}^N{\rm f}}
=
\frac{(- 1)^N}{N!}
\int d \mu_{\rm f} (v_{N+1}) (- i) x_{\rm f} [v_{N+1}] 
\int \prod_{j=1}^N  d \mu_{\rm f\bar{f}} (v_j) \frac{x^{[+2]}_{\rm f}[v_j]}{x^{[-2]}_{\rm f}[v_j]}
\frac{1}{\prod\limits_{N+1 > n \neq m} |P_{\rm f|\bar{f}} (v_m + 2 i|v_n )|^2}
\nonumber\\
&\qquad
\times
\frac{1}{\prod\limits_{N+1 > j > i} (v_j - v_i) (v_j - v_i - i) |P_{\rm f|f} (v_i|v_j)|^2}
\frac{1}{\prod\limits_{N > j > i} (v_j - v_i) (v_j - v_i + i) |P_{\rm f|f} (v_i + 2 i|v_j + 2i)|^2}
\nonumber\\
&\qquad\quad
+ \dots
\, . 
\end{align}
Here the displayed expression is responsible for the exponentiation of the $(\bar{\rm f} {\rm f})$-pair exchange. The ellipsis stand for effect of other poles which induce terms
proportional to lesser powers of the composite measure \re{muffbar}. The solution of this combinatorial problem yields contributions corresponding to scattering of
fermion bound states \cite{Fioravanti:2015dma} which together with single pair propagating in the OPE channel results in dilogarithm expected from Thermodynamic 
Bethe Ansatz \cite{Alday:2007hr,Alday:2009dv,Alday:2010vh} at leading order in strong coupling,
\begin{align}
\mathcal{W}_6^{(\chi_1 \chi_4^3)} 
=
{\rm e}^{i \phi/2}
\int
d \mu_{\rm f} (v) (- i) x_{\rm f} [v]
\exp
\left(
\,
\int \frac{d u}{2 \pi} \mu_{\rm f\bar{f}} (u)
\, 
{\rm Li}_2 \left(
- {\rm e}^{- \tau E_{\rm f\bar{f}} (u) + i \sigma p_{\rm f\bar{f}} (u)}
\right)
\right)
\, .
\end{align}
In a similar fashion, one can work out mixed terms with both fermionic pairs and gluon bound states. The outcome of this consideration is that the complete leading order result is given by a
single exponent with the argument determined by the sum of individual contributions discussed above. A detailed consideration is deferred to a future publication.

\section{Conclusions}

In this paper we initiated a systematic study of the strong coupling expansion for pentagon transitions in the OPE approach to the null polygonal superWilson loop. The framework  is a generalization 
of a previous consideration \cite{Basso:2007wd,Basso:2009gh} for the cusp anomalous dimension, i.e., the vacuum energy density of the flux tube. While we addressed $1/g$ perturbative series, 
we did not include exponentially suppressed contributions in our analysis. These can be recovered in a straightforward fashion from explicit all-order representation of the flux-tube functions 
for relevant excitations. Presently, we considered gauge-field bound states and fermions. Their flux-tube functions can be used to find all other pentagon transitions (to complete the 
list of the ones explicitly given in the main text) in the perturbative string regime by means of exchange relations except the one for the hole transitions which require a separate calculation. 
The contribution of  the latter was not addressed here with the focus being rather on the emergence of the minimal area in NMHV amplitudes. It was argued in Ref.\ \cite{Basso:2014jfa} that 
all multi-scalar exchanges have to be resummed and were shown to induce kinematic-independent leading order effects in addition to the area for MHV case. For NMHV case, this question 
was recently addressed in Ref.\ \cite{Belitsky:2015lzw}. We demonstrated there the factorization of contributions of near-massless scalars from the helicity-dependent massive particles 
carrying the quantum numbers of Grassmann components in question of the superWilson loop and provided a concise formula for their resummed short-distance behavior.

\section*{Acknowledgments}

This research was supported by the U.S. National Science Foundation under the grants PHY-1068286 and PHY-1403891.

\appendix

\section{Special functions and integrals}
\label{IntegralsAppendix}

In the body of the paper, we introduced the following special functions. The function $W$ is related to the hypergeometric function of two variables $\Phi_1$ \cite{PBM89} and reads
\begin{align}
W (z, u) = \frac{\sqrt{2}}{\pi} \int_{-1}^1 dk  \left( \frac{1 + k}{1 - k} \right)^{1/4} {\rm e}^{z k} \frac{\mathcal{P}}{k - u}
\, .
\end{align}
While $V$ and $U$ are related to the confluent hypergeometric function of the second kind and admit the following integral representations \cite{Basso:2009gh}
\begin{align}
V_n (z) 
&
= \frac{\sqrt{2}}{\pi} \int_{-1}^1 dk  \left( \frac{1 + k}{1 - k} \right)^{1/4} \frac{{\rm e}^{k z}}{(1 + k)^n}
\, , \\
U^\pm_n (z) 
&
= \frac{1}{2} \int_{1}^\infty dk  \left( \frac{k + 1}{k - 1} \right)^{\mp1/4} \frac{{\rm e}^{- k (z - 1)}}{(k \mp 1)^n}
\, .
\end{align}

Depending on the sign of $z$, these functions develop different asymptotic behavior at $z \to \pm \infty$. Up to exponentially suppressed contributions, the power series in $1/z$ can be
constructed from the following integrals,
\begin{align}
\label{WAsymptot}
W (\pm |z|, u) |_{|z| \to \infty}
&
\simeq
\pm \frac{{\rm e}^{|z|} (2 |z|)^{\pm 1/4 - 1}}{2^{- 3/2} \pi (1 \mp u)}
\int_0^\infty d \beta \, {\rm e}^{- \beta} \beta^{\mp 1/4} \left( 1 - \frac{\beta}{2 |\tau|} \right)^{\pm 1/4} \left( 1 - \frac{\beta}{(1 \mp u) |z|} \right)^{-1}
\, , \\
\label{VnAsymptot}
V_n (\pm |z|) |_{|z| \to \infty}
&
\simeq
\frac{{\rm e}^{|z|} (2 |z|)^{\pm (1/4 - n/2) - 1 + n/2}}{2^{n - 3/2} \pi}
\int_0^\infty d \beta \, {\rm e}^{- \beta} \beta^{\mp (1/4 - n/2) - n/2} \left( 1 - \frac{\beta}{2 |z|} \right)^{\pm (1/4 - n/2) - n/2}
\, ,
\end{align}
obtained from above by a simple transformation of the integration variable. Similarly an equivalent representation for $U_n$ ($n = 0,1$) reads,
\begin{align}
\label{U0Asymptot}
U^\pm_0 (z)
&
=
(2 z)^{- (4 \pm 1)/4}
\int_0^\infty d \beta \, {\rm e}^{- \beta} \beta^{\pm 1/4} \left( 1 + \frac{\beta}{2 z} \right)^{\mp 1/4}
\, , \\
\label{U1Asymptot}
U^\pm_1 (z)
&
=
\frac{1}{2} (2 z)^{- (2 \mp 1)/4}
\int_0^\infty d \beta \, {\rm e}^{- \beta} \beta^{- (2 \pm 1)/4} \left( 1 + \frac{\beta}{2 z} \right)^{- (2 \mp 1)/4}
\, .
\end{align}
Explicitly, one finds
\begin{align}
\left. \frac{W (\pm |z|, u)}{V_0 (\pm |z|)} \right|_{|z| \to \infty} 
&
= - \frac{1}{u \mp 1} \pm \frac{(4 \mp 1)}{4 z (u \mp 1)^2} + O (1/z^2)
\, , \\
\left. \frac{V_1 (\pm |z|)}{V_0 (\pm |z|)} \right|_{|z| \to \infty} 
&
= - 2 (1 \mp 1) |z| \pm \frac{1}{2} + \frac{4 \mp 1}{16 |z|}+ O (1/z^2)
\, ,
\end{align}
and 
\begin{align}
U^\pm_0 (z) |_{z \to \infty} 
&= (2z)^{- (4 \pm 1)/4} \Gamma \left(\frac{4 \pm 1}{4}\right) \left[ 1 \mp \frac{4 \pm 1}{32 z} + O (1/z^2) \right]
\, , \\
U^\pm_1 (z) |_{z \to \infty} 
&= (2z)^{- (2 \mp 1)/4} \Gamma \left(\frac{2 \mp 1}{4}\right) \left[ 1 \pm \frac{4 \pm 5}{32 z} + O (1/z^2) \right]
\, ,
\end{align}
with subleading terms eagerly evaluated from Eqs.\ \re{WAsymptot}, \re{VnAsymptot} and \re{U0Asymptot}, \re{U1Asymptot} by Taylor expanding the integrand and computing the resulting 
integrals using the definition of the Euler Gamma function.

In the main text, we also introduced different parity components of $V$ and $W$ for the imaginary value of their argument $z = - i \tau$. They are $W (- i \tau, u)  = W^+ (\tau, u) - i W^- (\tau, u)$,
\begin{align}
\label{Wplusminus}
W^+ (\tau, u) 
&= 
\frac{\sqrt{2}}{\pi} \int_{-1}^1 d k \left( \frac{1 + k}{1 - k} \right)^{1/4} \cos (\tau k) \frac{\mathcal{P}}{k - u}
\, , \\
W^- (\tau, u) 
&= 
\frac{\sqrt{2}}{\pi} \int_{-1}^1 d k \left( \frac{1 + k}{1 - k} \right)^{1/4} \sin (\tau k) \frac{\mathcal{P}}{k - u}
\, , \nonumber
\end{align}
and $V_n (- i \tau)  = V_n^+ (\tau) - i V_n^- (\tau)$, 
\begin{align}
\label{Vnplusminus}
V_n^+ (\tau) 
&= 
\frac{\sqrt{2}}{\pi} \int_{-1}^1 d k \left( \frac{1 + k}{1 - k} \right)^{1/4} \frac{\cos (\tau k)}{(1 + k)^n}
\, , \\
V_n^- (\tau) 
&= 
\frac{\sqrt{2}}{\pi} \int_{-1}^1 d k \left( \frac{1 + k}{1 - k} \right)^{1/4} \frac{\sin (\tau k)}{(1 + k)^n}
\, . \nonumber
\end{align}

Finally, the only two integrals that are needed to solve the singular integral equations as well to derive the explicit expressions for all dynamical phases quoted below in 
Appendix \ref{PhasesAppendix} are the following
\begin{align}
\label{Integralq>1}
\int_{-1}^1 \frac{dk}{\pi} \left( \frac{1 - k}{1 + k} \right)^{1/4} \frac{1}{k - q}
&
=
\sqrt{2} \left( \frac{q - 1}{q + 1} \right)^{1/4} - \sqrt{2}
\, , \\
\label{Integralp<1} 
\int_{-1}^1 \frac{dk}{\pi} \left( \frac{1 - k}{1 + k} \right)^{1/4} \frac{\mathcal{P}}{k - p}
&
=
\left( \frac{1 - p}{1 + p} \right)^{1/4} - \sqrt{2} 
\, .
\end{align}
These are valid for $|q| > 1$ and $|p| < 1$, respectively, 

\section{Gauge pentagons: to Goldstone sheet and back}

In this appendix we will construct  pentagons for gauge field bound states. The initial point of this consideration is transitions for a single gluon undergoing a transformation 
into the same or another flux-tube excitation. On the physical sheet, the analytical properties of the flux-tube functions are quite complex due to the presence of an infinite 
number of cuts $[-2g, 2g]$ that are equidistantly separated along the imaginary axis starting at $\Im{\rm m}[u] = \pm \ft12$ and going to infinity. So it becomes problematic to construct the 
bound state observables by fusing single-particle once staying in the kinematical region of rapidities $- 2g < u < 2g$. A way out of this complication is to make an analytic 
continuation to the Goldstone (or half-mirror) sheet which has just two cuts for the gauge field at $\Im{\rm m}[u] = \pm \ft12$ \cite{Basso:2011rc}. Therefore, as proposed in 
Ref.\ \cite{Basso:2014nra}, for technical and practical reasons it is instructive to move upwards through the first cut of the gluon excitation to the half-mirror (Goldstone) sheet 
\begin{align}
u \to u^{\rm G} = u + i \ft{\ell}{2} + i 0 \to u
\, ,
\end{align}
(this results in the change $x^-[u] \to g^2/x^-[u]$ while $x^+[u]$ stays intact) and fuse elementary excitations there, keeping the imaginary part of their rapidities above the cut, i.e., 
$\Im{\rm m} [u] > \ft{1}{2}$. Once $\ell$ of these gluons are fused together, one can always move back to the physical sheet, now passing to it through the top cut of
the bound state $[-2g + i \ft{\ell}{2}, 2g + i \ft{\ell}{2}]$. This implies that the Zhukowsky variables obeys the following transformation rules
\begin{align}
x^{[- \ell]} [u] \to \frac{g^2}{x^{[- \ell]} [u]}
\, , \qquad
x^{[+ \ell]} [u] \to x^{[+ \ell]} [u]
\, .
\end{align}

\subsection{Bound-state--bound-state pentagons}
\label{GaugePentagon}

We start with the gluon-gluon and gluon-antigluon pentagons. These are given as usual by the ratio of the direct and mirror S-matrices (see Eqs.\ \re{Slldirect} and \re{Sllmirror} 
of the main text where one sets $\ell = 1$) \cite{Basso:2013vsa,Basso:2013aha} 
\begin{align}
P^2_{\rm g | g} (u_1| u_2) = w^{-1}_{\rm gg} (u_1, u_2) \frac{S_{\rm gg} (u_1, u_2)}{S_{\rm \ast gg} (u_1, u_2)}
\, , \qquad
P^2_{\rm g | \bar{g}} (u_1| u_2) = w_{\rm gg} (u_1, u_2) \frac{S_{\rm g\bar{g}} (u_1, u_2)}{S_{\rm \ast g\bar{g}} (u_1, u_2)}
\, ,
\end{align}
with the prefactor being
\begin{align}
w_{\rm gg} (u_1, u_2)
&
=
\frac{g^2 (u_1 - u_2) (u_1 - u_2 - i)}{x^+ [u_1] x^- [u_1] x^+ [u_2] x^- [v_2]}
\left(
1 - \frac{g^2}{x^+ [u_1] x^- [u_2]}
\right)^{-1}
\\
&
\times
\left(
1 - \frac{g^2}{x^- [u_1] x^+ [u_2]}
\right)^{-1}
\left(
1 - \frac{g^2}{x^+ [u_1] x^+ [u_2]}
\right)^{-1}
\left(
1 - \frac{g^2}{x^- [u_1] x^- [u_2]}
\right)^{-1}
\, . \nonumber
\end{align}
Going to the Goldstone sheet, we find the latter changes to
\begin{align}
w_{\rm GG} (u_1, u_2)
=
\frac{u_1 - u_2}{u_1 - u_2 + i}
\frac{
\left(
1 - \frac{g^2}{x^+ [u_1] x^- [u_2]}
\right)
\left(
1 - \frac{g^2}{x^- [u_1] x^+ [u_2]}
\right)
}{
\left(
1 - \frac{g^2}{x^+ [u_1] x^+ [u_2]}
\right)
\left(
1 - \frac{g^2}{x^- [u_1] x^- [u_2]}
\right)
}
\, ,
\end{align}
and the S-matrices turn into Eqs.\ \re{SGGdirect} and \re{SGGmirror} with $\ell=1$, respectively. Now, as explained in the preamble, the fusion is straightforward on this sheet as one is
away from all the cuts and obtains
\begin{align}
\label{wGGboundstate}
w_{\rm G G} (u_1, u_2)
&
=
\prod_{k_1 = 1}^{\ell_1}
\prod_{k_2 = 1}^{\ell_2}
w_{\rm GG} (u_1^{[2 k_1 - \ell_1 - 1]}, u_2^{2 k_2 - \ell_2 - 1})
\\
&
=
s_{\ast\ell_1 \bar{\ell}_2} (u_1, u_2)
\frac{
\left(
1 - \frac{g^2}{x^{[\ell_1]} [u_1] x^{[- \ell_2]} [u_2]}
\right)
\left(
1 - \frac{g^2}{x^{[-\ell_1]} [u_1] x^{[\ell_2]} [u_2]}
\right)
}{
\left(
1 - \frac{g^2}{x^{[- \ell_1]} [u_1] x^{[- \ell_2]} [u_2]}
\right)
\left(
1 - \frac{g^2}{x^{[\ell_1]} [u_1] x^{[\ell_2]} [u_2]}
\right)
}
\, . \nonumber
\end{align}
Here $\Im{\rm m}[u_k] > \ell_k/2$ ($k = 1,2$). We obviously abused notations in the first line by calling the single gluon and bound state dressing factors by the same symbol. It will 
be always clear from the context what we are dealing with.

Passing to the physical sheet, but now through the top Zhukowski cut of $\ell$-bound state, i.e., $x^{[- \ell_k]} [u_k] \to g^2/x^{[-\ell_k]}[u_k ]$, we find
\begin{align}
w_{\ell_1 \ell_2} (u_1, u_2)
&
=
s_{\ast\ell_1 \bar{\ell}_2} (u_1, u_2)
\frac{g^2 \left( (u_1 - u_2)^2 + \frac{(\ell_1 + \ell_2)^2}{4} \right)}{x^{[\ell_1]}  [u_1] x^{[-\ell_1]}  [u_1] x^{[\ell_2]}  [u_2] x^{[- \ell_2]}  [u_2]}
\left(
1 - \frac{g^2}{x^{[\ell_1]} [u_1] x^{[- \ell_2]} [u_2]}
\right)^{-1}
\\
&
\times
\left(
1 - \frac{g^2}{x^{[-\ell_1]} [u_1] x^{[\ell_2]} [u_2]}
\right)^{-1}
\left(
1 - \frac{g^2}{x^{[- \ell_1]} [u_1] x^{[- \ell_2]} [u_2]}
\right)^{- 1}
\left(
1 - \frac{g^2}{x^{[\ell_1]} [u_1] x^{[\ell_2]} [u_2]}
\right)^{- 1}
\, . \nonumber
\end{align}
which determines the gauge stack-(anti)stack pentagons when it is accompanied by the ratio of bound-state S-matrices \re{Slldirect} and \re{Sllmirror} 
\begin{align}
\label{Pell1ell2}
P^2_{\ell_1 | \ell_2} (u_1| u_2) = w^{-1}_{\ell_1\ell_2} (u_1, u_2) \frac{S_{\ell_1 \ell_2} (u_1, u_2)}{S_{\ast\ell_1 \ell_2} (u_1, u_2)}
\, , \qquad
P^2_{\ell_1 | \bar{\ell}_2} (u_1| u_2) = w_{\ell_1 \ell_2} (u_1, u_2) \frac{S_{\ell_1\bar{\ell}_2} (u_1, u_2)}{S_{\ast\ell_1\bar{\ell}_2} (u_1, u_2)}
\, .
\end{align}
To compare with known results, let us give them in the explicit form. Using the relation
\begin{align}
\exp \Big(
2 i \sigma_{\ell_1 \ell_2} (u_1, u_2) 
&
- 2 \widehat\sigma_{\ell_1 \ell_2} (u_1, u_2)
\Big)
\\
&=
\frac{\Gamma^2 \left( 1 + \frac{\ell_1 + \ell_2}{2} + i u_1 - i u_2 \right)}{\Gamma^2 \left( 1 + \frac{\ell_1}{2} + i u_1 \right) \Gamma^2 \left( 1 + \frac{\ell_2}{2} - i u_2 \right)}
\frac{x^{[\ell_1]}  [u_1] x^{[-\ell_1]}  [u_1] x^{[\ell_2]}  [u_2] x^{[- \ell_2]}  [u_2]}{g^2 \left( (u_1 - u_2)^2 + \frac{(\ell_1 + \ell_2)^2}{4} \right)}
\nonumber\\
&\times
\exp
\left(
2
\int_0^\infty \frac{dt}{t ({\rm e}^t - 1)} \left( J_0 (2gt) - 1 \right)
\left(
J_0 (2 g t) + 1 - {\rm e}^{- i u_1 t - \ell_1 t/2} - {\rm e}^{i u_2 t - \ell_2 t/2}
\right)
\right)
,
\nonumber
\end{align}
we can cast the helicity-violating pentagon in the form
\begin{align}
P_{\ell_1 | \bar{\ell}_2} (u_1| u_2)
&
=
\frac{\Gamma \left( 1 + \frac{\ell_1 + \ell_2}{2} + i u_1 - i u_2 \right)}{\Gamma \left( 1 + \frac{\ell_1}{2} + i u_1 \right) \Gamma \left( 1 + \frac{\ell_2}{2} - i u_2 \right)}
\left(
1 - \frac{g^2}{x_1^{[\ell_1]} x_2^{[- \ell_2]}}
\right)^{-1/2}
\\
&
\times
\left(
1 - \frac{g^2}{x_1^{[-\ell_1]} x_2^{[\ell_2]}}
\right)^{-1/2}
\left(
1 - \frac{g^2}{x_1^{[- \ell_1]} x_2^{[- \ell_2]}}
\right)^{- 1/2}
\left(
1 - \frac{g^2}{x_1^{[\ell_1]} x_2^{[\ell_2]}}
\right)^{- 1/2}
\nonumber\\
&
\times
\exp
\left(
\int_0^\infty \frac{dt}{t ({\rm e}^t - 1)} \left( J_0 (2gt) - 1 \right)
\left(
J_0 (2 g t) + 1 - {\rm e}^{- i u_1 t - \ell_1 t/2} - {\rm e}^{i u_2 t - \ell_2 t/2}
\right)
\right)
\nonumber\\
&\times
\exp\left( - i f^{(1)}_{\ell_1 \ell_2} (u_1, u_2) + i f^{(2)}_{\ell_1 \ell_2} (u_1, u_2) + f^{(3)}_{\ell_1 \ell_2} (u_1, u_2) - f^{(4)}_{\ell_1 \ell_2} (u_1, u_2) \right)
\, . \nonumber
\end{align}
Here the dynamical phases are given in the text in Eqs.\ \re{f1ll}, \re{f2ll}, \re{f3ll} and \re{f4ll}. While making use of the relation 
\begin{align}
s_{\ell_1 \ell_2} (u_1, u_2) = \frac{s_{\ast\ell_2 \bar{\ell}_1} (u_2, u_1)}{s_{\ast\ell_1\bar{\ell}_2} (u_1, u_2)}
\end{align}
we can take the square of the right-hand side of Eq.\ \re{Pell1ell2} to find
\begin{align}
P_{\ell_1 | \ell_2} (u_1| u_2)
&
=
\frac{(- 1)^{\ell_2} \Gamma \left( \frac{\ell_1 + \ell_2}{2} - i u_1 + i u_2 \right) \Gamma \left( \frac{\ell_1 - \ell_2}{2} + i u_1 - i u_2 \right)
}{
\Gamma \left(1 + \frac{\ell_1}{2} + i u_1 \right)
\Gamma \left(1 + \frac{ \ell_2}{2} - i u_2 \right)
\Gamma \left( 1 + \frac{\ell_1 - \ell_2}{2} - i u_1 + i u_2 \right)
}
\\
&\times
\left(
x_1^{[\ell_1]} x_2^{[- \ell_2]} - g^2
\right)^{1/2}
\left(
x_1^{[-\ell_1]} x_2^{[\ell_2]} - g^2
\right)^{1/2}
\left(
x_1^{[- \ell_1]} x_2^{[- \ell_2]} - g^2
\right)^{1/2}
\left(
x_1^{[\ell_1]} x_2^{[\ell_2]} - g^2
\right)^{1/2}
\nonumber\\
&\times
\exp
\left(
\int_0^\infty \frac{dt}{t ({\rm e}^t - 1)} \left( J_0 (2gt) - 1 \right)
\left(
J_0 (2 g t) + 1 - {\rm e}^{- i u_1 t - \ell_1 t/2} - {\rm e}^{i u_2 t - \ell_2 t/2}
\right)
\right)
\nonumber\\
&\times
\exp\left( - i f^{(1)}_{\ell_1 \ell_2} (u_1, u_2) + i f^{(2)}_{\ell_1 \ell_2} (u_1, u_2) + f^{(3)}_{\ell_1 \ell_2} (u_1, u_2) - f^{(4)}_{\ell_1 \ell_2} (u_1, u_2) \right)
\, . \nonumber
\end{align}
Both of these expressions agree with Ref.\ \cite{Basso:2014nra}.

\subsection{Bound-state--fermion pentagons}
\label{GFpentagonsAppendix}

Next we turn to the gauge bound-state--(anti)fermion pentagons. These are constructed from the single gauge field-(anti)fermion transitions which read \cite{Belitsky:2014lta}
\begin{align}
P^2_{\rm g|f} (u_1 | u_2)= w_{\rm gf} (u_1, u_2) \frac{S_{\rm gf} (u_1, u_2)}{S_{\ast \rm gf} (u_1, u_2)}
\, , \qquad
P^2_{\rm g|\bar{f}} (u_1 | u_2)= w^{-1}_{\rm gf} (u_1, u_2) \frac{S_{\rm gf} (u_1, u_2)}{S_{\ast \rm gf} (u_1, u_2)}
\, ,
\end{align}
with
\begin{align}
w_{\rm gf} (u_1, u_2) = (u_1 - u_2 + \ft{i}{2}) \frac{x_{\rm f} [u_2]}{x^{+} [u_1] x^{-} [u_1]}
\left(1 - \frac{x_{\rm f} [u_2]}{x^+ [u_1]} \right)^{-1} \left(1 - \frac{x_{\rm f} [u_2]}{x^{-} [u_1]} \right)^{-1}
\, ,
\end{align}
and scattering matrices quoted in the body of the paper in Eqs.\ \re{Slfdirect} and \re{Slfdirect} for $\ell = 1$. Going to the Goldstone sheet, we find
\begin{align}
w_{\rm Gf} (u_1, u_2) = - \frac{u_1 - u_2 + \ft{i}{2}}{u_1 - u_2 - \ft{i}{2}} \frac{x^- [u_1] x[u_2] - g^2}{x^+ [u_1] x[u_2] - g^2}
\, .
\end{align}
The fusion of the $w$ factor produces
\begin{align}
w_{\rm Gf} (u_1, u_2) = \prod_{k = 1}^\ell w_{\rm Gf} (u_1^{[2 k - \ell - 1]}, u_2)
=
(- 1)^\ell \frac{\left( u_1 - u_2 + i \frac{\ell}{2} \right) \left( x^{[- \ell]} [u_1] x [u_2] - g^2 \right)}{\left( u_1 - u_2 - i \frac{\ell}{2} \right) \left( x^{[+ \ell]} [u_1] x [u_2] - g^2 \right)}
\, .
\end{align}
Again, we abused the notation here by calling the bound state and single gauge field prefactors by the same letter. Going back to the physical sheet, we find
\begin{align}
w_{\ell {\rm f}} (u_1, u_2) = (-1)^{\ell + 1} (u_1 - u_2 + i \ft{\ell}{2}) \frac{x_{\rm f} [u_2]}{x^{[\ell]} [u_1] x^{[-\ell]} [u_1]}
\left(1 - \frac{x_{\rm f} [u_2]}{x^{[\ell]} [u_1]} \right)^{-1} \left(1 - \frac{x_{\rm f} [u_2]}{x^{[- \ell]} [u_1]} \right)^{-1}
\, .
\end{align}
Analogously, for the gauge bound state-antifermion case, we get
\begin{align}
w_{\ell {\rm \bar{f}}} (u_1, u_2) = w^{- 1}_{\ell {\rm f}} (u_1, u_2) \, .
\end{align}
In this manner we derive the stack-(anti)fermion pentagons 
\begin{align}
P^2_{\ell |\rm f} (u_1 | u_2)= w_{\ell \rm f} (u_1, u_2) \frac{S_{\ell \rm f} (u_1, u_2)}{S_{\ast \ell \rm f} (u_1, u_2)}
\, , \qquad
P^2_{\ell |\rm \bar{f}} (u_1 | u_2)= w^{-1}_{\ell \rm f} (u_1, u_2) \frac{S_{\ell \rm f} (u_1, u_2)}{S_{\ast \ell \rm f} (u_1, u_2)}
\, ,
\end{align}
which read, respectively,
\begin{align}
P_{\ell |\rm f} (u_1 | u_2)
&=
\frac{i}{g} \frac{\left( u - v + i \ft{\ell}{2} \right) x_{\rm f} [u_2]}{\big( x^{[+\ell]} [u_1] - x_{\rm f} [u_2] \big)^{1/2}\big( [x^{[-\ell]} [u_1] - x_{\rm f} [u_2] \big)^{1/2}}
\\
&\times
\exp\left( - i f^{(1)}_{\ell \rm f} (u_1, u_2) + i f^{(2)}_{\ell \rm f} (u_1, u_2) + f^{(3)}_{\ell \rm f} (u_1, u_2) - f^{(4)}_{\ell \rm f} (u_1, u_2) \right)
\, , \nonumber\\
P_{\ell |\rm \bar{f}} (u_1 | u_2)
&=
i g \big( x^{[+\ell]} [u_1] - x_{\rm f} [u_2] \big)^{1/2}\big( [x^{[-\ell]} [u_1] - x_{\rm f} [u_2] \big)^{1/2}
\\
&\times
\exp\left( - i f^{(1)}_{\ell \rm f} (u_1, u_2) + i f^{(2)}_{\ell \rm f} (u_1, u_2) + f^{(3)}_{\ell \rm f} (u_1, u_2) - f^{(4)}_{\ell \rm f} (u_1, u_2) \right)
\, , \nonumber
\end{align}
with dynamical phases quoted in Eqs.\ \re{flf1} -- \re{flf4}. These expressions are in agreement with Ref.\ \cite{Basso:2015rta} up to a different choice of normalization conventions.

In the main text, we also use the pentagon with flipped flux-tube excitations, i.e., $P_{{\rm f} | \ell}$, and continued to the Goldstone sheet, $P_{{\rm f} | {\rm G}}$. This transition can be obtained
in two steps, First, one uses the fact that on the physical sheet,
\begin{align}
P_{{\rm f} | \ell} (u_2 | u_1) = P_{\ell | {\rm f}} (- u_1 | - u_2) 
\, .
\end{align}
Then use the following obvious properties of dynamical phases
\begin{align}
f^{(1,2)}_{\rm pp'} (- u_1, - u_2) = - f^{(1,2)}_{\rm pp'} (u_1, u_2) 
\, , \qquad
f^{(3,4)}_{\rm pp'} (- u_1, - u_2) = + f^{(3,4)}_{\rm pp'} (u_1, u_2) 
\, ,
\end{align}
and only after that continuing the gauge bound state to the Goldstone sheet. In this fashion, we find Eq.\ \re{PfG}.

\section{Dynamical phases}
\label{PhasesAppendix}

In this appendix we summarize dynamical phases for fermion-fermion, gluon bound-state--bound state and fermion-gluon bound state transitions to the first nontrivial oder in $1/g$.
In a similar fashion, one can find the rest of transitions by means of the exchange relations, except for the hole-hole case, which requires a separate study.

\subsection{Fermion-fermion case}
\label{FFAppendix}

For the fermion-fermion phases, the leading contributions are
\begin{align}
\label{Aff1}
A^{(1)}_{\rm ff} (\hat{u}_1, \hat{u}_2) 
&
=
- 
\frac{1}{\hat{u}_1 - \hat{u}_2}
\left[
\left( \frac{\hat{u}_1 - 1}{\hat{u}_1 + 1} \right)^{1/4}
\left( \frac{\hat{u}_2 + 1}{\hat{u}_2 - 1} \right)^{1/4}
+
\left( \frac{\hat{u}_1 + 1}{\hat{u}_1 - 1} \right)^{1/4}
\left( \frac{\hat{u}_2 - 1}{\hat{u}_2 + 1} \right)^{1/4}
-
2
\right]
\nonumber\\
&
- 
\frac{1}{\hat{u}_1 + \hat{u}_2}
\left[
\left( \frac{\hat{u}_1 - 1}{\hat{u}_1 + 1} \right)^{1/4}
\left( \frac{\hat{u}_2 - 1}{\hat{u}_2 + 1} \right)^{1/4}
+
\left( \frac{\hat{u}_1 + 1}{\hat{u}_1 - 1} \right)^{1/4}
\left( \frac{\hat{u}_2 + 1}{\hat{u}_2 - 1} \right)^{1/4}
-
2
\right]
\, ,  \\
A^{(3)}_{\rm ff} (\hat{u}_1, \hat{u}_2) 
&
=
\frac{1}{\hat{u}_1 - \hat{u}_2}
\left[
\left( \frac{\hat{u}_1 - 1}{\hat{u}_1 + 1} \right)^{1/4}
\left( \frac{\hat{u}_2 + 1}{\hat{u}_2 - 1} \right)^{1/4}
-
\left( \frac{\hat{u}_1 + 1}{\hat{u}_1 - 1} \right)^{1/4}
\left( \frac{\hat{u}_2 - 1}{\hat{u}_2 + 1} \right)^{1/4}
\right]
\nonumber\\
&
- 
\frac{1}{\hat{u}_1 + \hat{u}_2}
\left[
\left( \frac{\hat{u}_1 - 1}{\hat{u}_1 + 1} \right)^{1/4}
\left( \frac{\hat{u}_2 - 1}{\hat{u}_2 + 1} \right)^{1/4}
-
\left( \frac{\hat{u}_1 + 1}{\hat{u}_1 - 1} \right)^{1/4}
\left( \frac{\hat{u}_2 + 1}{\hat{u}_2 - 1} \right)^{1/4}
\right]
\, ,  \\
A^{(4)}_{\rm ff} (\hat{u}_1, \hat{u}_2) 
&
=
-
\frac{1}{\hat{u}_1 - \hat{u}_2}
\left[
\left( \frac{\hat{u}_1 - 1}{\hat{u}_1 + 1} \right)^{1/4}
\left( \frac{\hat{u}_2 + 1}{\hat{u}_2 - 1} \right)^{1/4}
-
\left( \frac{\hat{u}_1 + 1}{\hat{u}_1 - 1} \right)^{1/4}
\left( \frac{\hat{u}_2 - 1}{\hat{u}_2 + 1} \right)^{1/4}
\right]
\nonumber\\
&
- 
\frac{1}{\hat{u}_1 + \hat{u}_2}
\left[
\left( \frac{\hat{u}_1 - 1}{\hat{u}_1 + 1} \right)^{1/4}
\left( \frac{\hat{u}_2 - 1}{\hat{u}_2 + 1} \right)^{1/4}
-
\left( \frac{\hat{u}_1 + 1}{\hat{u}_1 - 1} \right)^{1/4}
\left( \frac{\hat{u}_2 + 1}{\hat{u}_2 - 1} \right)^{1/4}
\right]
\, ,
\end{align}
while the subleading coefficients read
\begin{align}
&
B^{(1)}_{\rm ff} (\hat{u}_1, \hat{u}_2) 
=
-
\frac{1}{(\hat{u}_1 - \hat{u}_2)^2}
\left[
\left( \frac{\hat{u}_1 - 1}{\hat{u}_1 + 1} \right)^{1/4}
\left( \frac{\hat{u}_2 + 1}{\hat{u}_2 - 1} \right)^{1/4}
-
\left( \frac{\hat{u}_1 + 1}{\hat{u}_1 - 1} \right)^{1/4}
\left( \frac{\hat{u}_2 - 1}{\hat{u}_2 + 1} \right)^{1/4}
\right]
\nonumber\\
&\qquad\qquad\qquad
-
\frac{1}{(\hat{u}_1 + \hat{u}_2)^2}
\left[
\left( \frac{\hat{u}_1 - 1}{\hat{u}_1 + 1} \right)^{1/4}
\left( \frac{\hat{u}_2 - 1}{\hat{u}_2 + 1} \right)^{1/4}
-
\left( \frac{\hat{u}_1 + 1}{\hat{u}_1 - 1} \right)^{1/4}
\left( \frac{\hat{u}_2 + 1}{\hat{u}_2 - 1} \right)^{1/4}
\right]
\nonumber\\
&\qquad
-
\frac{2 - \hat{u}_1^2 - \hat{u}_2^2}{4 (1- \hat{u}_1^2) (1- \hat{u}_2^2)}
\frac{1}{\hat{u}_1 - \hat{u}_2}
\left[
\left( \frac{\hat{u}_1 - 1}{\hat{u}_1 + 1} \right)^{1/4}
\left( \frac{\hat{u}_2 + 1}{\hat{u}_2 - 1} \right)^{1/4}
+
\left( \frac{\hat{u}_1 + 1}{\hat{u}_1 - 1} \right)^{1/4}
\left( \frac{\hat{u}_2 - 1}{\hat{u}_2 + 1} \right)^{1/4}
\right]
\nonumber\\
&\qquad
-
\frac{2 - \hat{u}_1^2 - \hat{u}_2^2}{4 (1- \hat{u}_1^2) (1- \hat{u}_2^2)}
\frac{1}{\hat{u}_1 + \hat{u}_2}
\left[
\left( \frac{\hat{u}_1 - 1}{\hat{u}_1 + 1} \right)^{1/4}
\left( \frac{\hat{u}_2 - 1}{\hat{u}_2 + 1} \right)^{1/4}
+
\left( \frac{\hat{u}_1 + 1}{\hat{u}_1 - 1} \right)^{1/4}
\left( \frac{\hat{u}_2 + 1}{\hat{u}_2 - 1} \right)^{1/4}
\right]
\, , \\
&
B^{(3)}_{\rm ff} (\hat{u}_1, \hat{u}_2) 
=
\frac{1}{(\hat{u}_1 - \hat{u}_2)^2}
\left[
\left( \frac{\hat{u}_1 - 1}{\hat{u}_1 + 1} \right)^{1/4}
\left( \frac{\hat{u}_2 + 1}{\hat{u}_2 - 1} \right)^{1/4}
+
\left( \frac{\hat{u}_1 + 1}{\hat{u}_1 - 1} \right)^{1/4}
\left( \frac{\hat{u}_2 - 1}{\hat{u}_2 + 1} \right)^{1/4}
-
2
\right]
\nonumber\\
&\qquad\qquad\qquad
-
\frac{1}{(\hat{u}_1 + \hat{u}_2)^2}
\left[
\left( \frac{\hat{u}_1 - 1}{\hat{u}_1 + 1} \right)^{1/4}
\left( \frac{\hat{u}_2 - 1}{\hat{u}_2 + 1} \right)^{1/4}
+
\left( \frac{\hat{u}_1 + 1}{\hat{u}_1 - 1} \right)^{1/4}
\left( \frac{\hat{u}_2 + 1}{\hat{u}_2 - 1} \right)^{1/4}
-
2
\right]
\nonumber\\
&\qquad
+
\frac{2 - \hat{u}_1^2 - \hat{u}_2^2}{4 (1- \hat{u}_1^2) (1- \hat{u}_2^2)}
\frac{1}{\hat{u}_1 - \hat{u}_2}
\left[
\left( \frac{\hat{u}_1 - 1}{\hat{u}_1 + 1} \right)^{1/4}
\left( \frac{\hat{u}_2 + 1}{\hat{u}_2 - 1} \right)^{1/4}
-
\left( \frac{\hat{u}_1 + 1}{\hat{u}_1 - 1} \right)^{1/4}
\left( \frac{\hat{u}_2 - 1}{\hat{u}_2 + 1} \right)^{1/4}
\right]
\nonumber\\
&\qquad
-
\frac{2 - \hat{u}_1^2 - \hat{u}_2^2}{4 (1- \hat{u}_1^2) (1- \hat{u}_2^2)}
\frac{1}{\hat{u}_1 + \hat{u}_2}
\left[
\left( \frac{\hat{u}_1 - 1}{\hat{u}_1 + 1} \right)^{1/4}
\left( \frac{\hat{u}_2 - 1}{\hat{u}_2 + 1} \right)^{1/4}
-
\left( \frac{\hat{u}_1 + 1}{\hat{u}_1 - 1} \right)^{1/4}
\left( \frac{\hat{u}_2 + 1}{\hat{u}_2 - 1} \right)^{1/4}
\right]
\, , \\
&
B^{(4)}_{\rm ff} (\hat{u}_1, \hat{u}_2) 
=
- \frac{1}{(\hat{u}_1 - \hat{u}_2)^2}
\left[
\left( \frac{\hat{u}_1 - 1}{\hat{u}_1 + 1} \right)^{1/4}
\left( \frac{\hat{u}_2 + 1}{\hat{u}_2 - 1} \right)^{1/4}
+
\left( \frac{\hat{u}_1 + 1}{\hat{u}_1 - 1} \right)^{1/4}
\left( \frac{\hat{u}_2 - 1}{\hat{u}_2 + 1} \right)^{1/4}
-
2
\right]
\nonumber\\
&\qquad\qquad\qquad
-
\frac{1}{(\hat{u}_1 + \hat{u}_2)^2}
\left[
\left( \frac{\hat{u}_1 - 1}{\hat{u}_1 + 1} \right)^{1/4}
\left( \frac{\hat{u}_2 - 1}{\hat{u}_2 + 1} \right)^{1/4}
+
\left( \frac{\hat{u}_1 + 1}{\hat{u}_1 - 1} \right)^{1/4}
\left( \frac{\hat{u}_2 + 1}{\hat{u}_2 - 1} \right)^{1/4}
-
2
\right]
\nonumber\\
&\qquad
-
\frac{2 - \hat{u}_1^2 - \hat{u}_2^2}{4 (1- \hat{u}_1^2) (1- \hat{u}_2^2)}
\frac{1}{\hat{u}_1 - \hat{u}_2}
\left[
\left( \frac{\hat{u}_1 - 1}{\hat{u}_1 + 1} \right)^{1/4}
\left( \frac{\hat{u}_2 + 1}{\hat{u}_2 - 1} \right)^{1/4}
-
\left( \frac{\hat{u}_1 + 1}{\hat{u}_1 - 1} \right)^{1/4}
\left( \frac{\hat{u}_2 - 1}{\hat{u}_2 + 1} \right)^{1/4}
\right]
\nonumber\\
&\qquad
-
\frac{2 - \hat{u}_1^2 - \hat{u}_2^2}{4 (1- \hat{u}_1^2) (1- \hat{u}_2^2)}
\frac{1}{\hat{u}_1 + \hat{u}_2}
\left[
\left( \frac{\hat{u}_1 - 1}{\hat{u}_1 + 1} \right)^{1/4}
\left( \frac{\hat{u}_2 - 1}{\hat{u}_2 + 1} \right)^{1/4}
-
\left( \frac{\hat{u}_1 + 1}{\hat{u}_1 - 1} \right)^{1/4}
\left( \frac{\hat{u}_2 + 1}{\hat{u}_2 - 1} \right)^{1/4}
\right]
\, ,
\end{align}
and
\begin{align}
C^{(1)}_{\rm ff} (\hat{u}_1, \hat{u}_2) 
&
=
\frac{1 + \hat{u}_1 \hat{u}_2}{(1 - \hat{u}_1^2)(1 - \hat{u}_2^2)}
\left[
\left( \frac{\hat{u}_1 - 1}{\hat{u}_1 + 1} \right)^{1/4}
\left( \frac{\hat{u}_2 + 1}{\hat{u}_2 - 1} \right)^{1/4}
-
\left( \frac{\hat{u}_1 + 1}{\hat{u}_1 - 1} \right)^{1/4}
\left( \frac{\hat{u}_2 - 1}{\hat{u}_2 + 1} \right)^{1/4}
\right]
\nonumber\\
&
+
\frac{1 - \hat{u}_1 \hat{u}_2}{(1 - \hat{u}_1^2)(1 - \hat{u}_2^2)}
\left[
\left( \frac{\hat{u}_1 - 1}{\hat{u}_1 + 1} \right)^{1/4}
\left( \frac{\hat{u}_2 - 	1}{\hat{u}_2 + 1} \right)^{1/4}
-
\left( \frac{\hat{u}_1 + 1}{\hat{u}_1 - 1} \right)^{1/4}
\left( \frac{\hat{u}_2 + 1}{\hat{u}_2 - 1} \right)^{1/4}
\right]
\, ,
\\
C^{(3)}_{\rm ff} (\hat{u}_1, \hat{u}_2) 
&
=
-
\frac{1 + \hat{u}_1 \hat{u}_2}{(1 - \hat{u}_1^2)(1 - \hat{u}_2^2)}
\left[
\left( \frac{\hat{u}_1 - 1}{\hat{u}_1 + 1} \right)^{1/4}
\left( \frac{\hat{u}_2 + 1}{\hat{u}_2 - 1} \right)^{1/4}
+
\left( \frac{\hat{u}_1 + 1}{\hat{u}_1 - 1} \right)^{1/4}
\left( \frac{\hat{u}_2 - 1}{\hat{u}_2 + 1} \right)^{1/4}
\right]
\nonumber\\
&
+
\frac{1 - \hat{u}_1 \hat{u}_2}{(1 - \hat{u}_1^2)(1 - \hat{u}_2^2)}
\left[
\left( \frac{\hat{u}_1 - 1}{\hat{u}_1 + 1} \right)^{1/4}
\left( \frac{\hat{u}_2 - 	1}{\hat{u}_2 + 1} \right)^{1/4}
+
\left( \frac{\hat{u}_1 + 1}{\hat{u}_1 - 1} \right)^{1/4}
\left( \frac{\hat{u}_2 + 1}{\hat{u}_2 - 1} \right)^{1/4}
\right]
\, ,
\\
\label{Cff4}
C^{(4)}_{\rm ff} (\hat{u}_1, \hat{u}_2) 
&
=
\frac{1 + \hat{u}_1 \hat{u}_2}{(1 - \hat{u}_1^2)(1 - \hat{u}_2^2)}
\left[
\left( \frac{\hat{u}_1 - 1}{\hat{u}_1 + 1} \right)^{1/4}
\left( \frac{\hat{u}_2 + 1}{\hat{u}_2 - 1} \right)^{1/4}
+
\left( \frac{\hat{u}_1 + 1}{\hat{u}_1 - 1} \right)^{1/4}
\left( \frac{\hat{u}_2 - 1}{\hat{u}_2 + 1} \right)^{1/4}
\right]
\nonumber\\
&
+
\frac{1 - \hat{u}_1 \hat{u}_2}{(1 - \hat{u}_1^2)(1 - \hat{u}_2^2)}
\left[
\left( \frac{\hat{u}_1 - 1}{\hat{u}_1 + 1} \right)^{1/4}
\left( \frac{\hat{u}_2 - 	1}{\hat{u}_2 + 1} \right)^{1/4}
+
\left( \frac{\hat{u}_1 + 1}{\hat{u}_1 - 1} \right)^{1/4}
\left( \frac{\hat{u}_2 + 1}{\hat{u}_2 - 1} \right)^{1/4}
\right]
\, .
\end{align}

\subsection{Gluon-gluon case}
\label{GGAppendix}

For the gauge-gauge case, the $1/g$ contribution to phases are
\begin{align}
A^{(1)}_{\rm GG} (\hat{u}_1, \hat{u}_2) 
&
=
\frac{2 \mathcal P}{ \hat{u}_1 - \hat{u}_2} + 2 \pi i \delta (\hat{u}_1 + \hat{u}_2)
\nonumber\\
&
- 
\frac{\mathcal P}{\hat{u}_1 - \hat{u}_2}
\left[
\left( \frac{1 - \hat{u}_1}{1 + \hat{u}_1} \right)^{1/4}
\left( \frac{1 + \hat{u}_2}{1 - \hat{u}_2} \right)^{1/4}
+
\left( \frac{1 + \hat{u}_1}{1 - \hat{u}_1} \right)^{1/4}
\left( \frac{1 - \hat{u}_2}{1 + \hat{u}_2} \right)^{1/4}
\right]
\nonumber\\
&
- 
\frac{\mathcal P}{\hat{u}_1 + \hat{u}_2}
\left[
\left( \frac{1 - \hat{u}_1}{1 + \hat{u}_1} \right)^{1/4}
\left( \frac{1 - \hat{u}_2}{1 + \hat{u}_2} \right)^{1/4}
+
\left( \frac{1 + \hat{u}_1}{1 - \hat{u}_1} \right)^{1/4}
\left( \frac{1 + \hat{u}_2}{1 - \hat{u}_2} \right)^{1/4}
\right]
\, ,  \\
A^{(3)}_{\rm GG} (\hat{u}_1, \hat{u}_2) 
&
=
- \frac{2 i \mathcal P}{ \hat{u}_1 + \hat{u}_2} - 2 \pi \delta (\hat{u}_1 + \hat{u}_2)
\nonumber\\
&
+
\frac{\mathcal P}{\hat{u}_1 - \hat{u}_2}
\left[
\left( \frac{1 - \hat{u}_1}{1 + \hat{u}_1} \right)^{1/4}
\left( \frac{1 + \hat{u}_2}{1 - \hat{u}_2} \right)^{1/4}
-
\left( \frac{1 + \hat{u}_1}{1 - \hat{u}_1} \right)^{1/4}
\left( \frac{1 - \hat{u}_2}{1 + \hat{u}_2} \right)^{1/4}
\right]
\nonumber\\
&
- 
\frac{\mathcal P}{\hat{u}_1 + \hat{u}_2}
\left[
\left( \frac{1 - \hat{u}_1}{1 + \hat{u}_1} \right)^{1/4}
\left( \frac{1 - \hat{u}_2}{1 + \hat{u}_2} \right)^{1/4}
-
\left( \frac{1 + \hat{u}_1}{1 - \hat{u}_1} \right)^{1/4}
\left( \frac{1 + \hat{u}_2}{1 - \hat{u}_2} \right)^{1/4}
\right]
\, , \\
A^{(4)}_{\rm GG} (\hat{u}_1, \hat{u}_2) 
&
=
- \frac{2 i \mathcal P}{ \hat{u}_1 + \hat{u}_2} - 2 \pi \delta (\hat{u}_1 + \hat{u}_2)
\nonumber\\
&
-
\frac{\mathcal P}{\hat{u}_1 - \hat{u}_2}
\left[
\left( \frac{1 - \hat{u}_1}{1 + \hat{u}_1} \right)^{1/4}
\left( \frac{1 + \hat{u}_2}{1 - \hat{u}_2} \right)^{1/4}
-
\left( \frac{1 + \hat{u}_1}{1 - \hat{u}_1} \right)^{1/4}
\left( \frac{1 - \hat{u}_2}{1 + \hat{u}_2} \right)^{1/4}
\right]
\nonumber\\
&
- 
\frac{\mathcal P}{\hat{u}_1 + \hat{u}_2}
\left[
\left( \frac{1 - \hat{u}_1}{1 + \hat{u}_1} \right)^{1/4}
\left( \frac{1 - \hat{u}_2}{1 + \hat{u}_2} \right)^{1/4}
-
\left( \frac{1 + \hat{u}_1}{1 - \hat{u}_1} \right)^{1/4}
\left( \frac{1 + \hat{u}_2}{1 - \hat{u}_2} \right)^{1/4}
\right]
\, ,
\end{align}
and $1/g^2$ corrections take the form
\begin{align}
&
B^{(1)}_{\rm GG} (\hat{u}_1, \hat{u}_2) 
=
\frac{\mathcal P}{[(\hat{u}_1 - \hat{u}_2)^2]_+}
\left[
\left( \frac{1 - \hat{u}_1}{1 + \hat{u}_1} \right)^{1/4}
\left( \frac{1 + \hat{u}_2}{1 - \hat{u}_2} \right)^{1/4}
-
\left( \frac{1 + \hat{u}_1}{1 - \hat{u}_1} \right)^{1/4}
\left( \frac{1 - \hat{u}_2}{1 + \hat{u}_2} \right)^{1/4}
\right]
\nonumber\\
&\qquad\qquad\quad \
+
\frac{\mathcal P}{[(\hat{u}_1 + \hat{u}_2)^2]_+}
\left[
\left( \frac{1 - \hat{u}_1}{1 + \hat{u}_1} \right)^{1/4}
\left( \frac{1 - \hat{u}_2}{1 + \hat{u}_2} \right)^{1/4}
-
\left( \frac{1 + \hat{u}_1}{1 - \hat{u}_1} \right)^{1/4}
\left( \frac{1 + \hat{u}_2}{1 - \hat{u}_2} \right)^{1/4}
+ 
2 i
\right]
\nonumber\\
&\qquad
+
\frac{2 - \hat{u}_1^2 - \hat{u}_2^2}{4 (1- \hat{u}_1^2) (1- \hat{u}_2^2)}
\frac{\mathcal{P}}{\hat{u}_1 - \hat{u}_2}
\left[
\left( \frac{1 - \hat{u}_1}{1 + \hat{u}_1} \right)^{1/4}
\left( \frac{1 + \hat{u}_2}{1 - \hat{u}_2} \right)^{1/4}
+
\left( \frac{1 + \hat{u}_1}{1 - \hat{u}_1} \right)^{1/4}
\left( \frac{1 - \hat{u}_2}{1 + \hat{u}_2} \right)^{1/4}
\right]
\nonumber\\
&\qquad
+
\frac{2 - \hat{u}_1^2 - \hat{u}_2^2}{4 (1- \hat{u}_1^2) (1- \hat{u}_2^2)}
\frac{\mathcal{P}}{\hat{u}_1 + \hat{u}_2}
\left[
\left( \frac{1 - \hat{u}_1}{1 + \hat{u}_1} \right)^{1/4}
\left( \frac{1 - \hat{u}_2}{1 + \hat{u}_2} \right)^{1/4}
+
\left( \frac{1 + \hat{u}_1}{1 - \hat{u}_1} \right)^{1/4}
\left( \frac{1 + \hat{u}_2}{1 - \hat{u}_2} \right)^{1/4}
\right]
\nonumber\\
&\qquad\qquad\quad \
- 2 \pi \delta^\prime (\hat{u}_1 + \hat{u}_2)
\, , \\
&
B^{(3)}_{\rm GG} (\hat{u}_1, \hat{u}_2) 
=
-
\frac{\mathcal P}{[(\hat{u}_1 - \hat{u}_2)^2]_+}
\left[
\left( \frac{1 - \hat{u}_1}{1 + \hat{u}_1} \right)^{1/4}
\left( \frac{1 + \hat{u}_2}{1 - \hat{u}_2} \right)^{1/4}
+
\left( \frac{1 + \hat{u}_1}{1 - \hat{u}_1} \right)^{1/4}
\left( \frac{1 - \hat{u}_2}{1 + \hat{u}_2} \right)^{1/4}
-
2
\right]
\nonumber\\
&\qquad\qquad\quad \
+
\frac{\mathcal P}{[(\hat{u}_1 + \hat{u}_2)^2]_+}
\left[
\left( \frac{1 - \hat{u}_1}{1 + \hat{u}_1} \right)^{1/4}
\left( \frac{1 - \hat{u}_2}{1 + \hat{u}_2} \right)^{1/4}
+
\left( \frac{1 + \hat{u}_1}{1 - \hat{u}_1} \right)^{1/4}
\left( \frac{1 + \hat{u}_2}{1 - \hat{u}_2} \right)^{1/4}
\right]
\nonumber\\
&\qquad
-
\frac{2 - \hat{u}_1^2 - \hat{u}_2^2}{4 (1- \hat{u}_1^2) (1- \hat{u}_2^2)}
\frac{\mathcal{P}}{\hat{u}_1 - \hat{u}_2}
\left[
\left( \frac{1 - \hat{u}_1}{1 + \hat{u}_1} \right)^{1/4}
\left( \frac{1 + \hat{u}_2}{1 - \hat{u}_2} \right)^{1/4}
-
\left( \frac{1 + \hat{u}_1}{1 - \hat{u}_1} \right)^{1/4}
\left( \frac{1 - \hat{u}_2}{1 + \hat{u}_2} \right)^{1/4}
\right]
\nonumber\\
&\qquad
+
\frac{2 - \hat{u}_1^2 - \hat{u}_2^2}{4 (1- \hat{u}_1^2) (1- \hat{u}_2^2)}
\frac{\mathcal{P}}{\hat{u}_1 + \hat{u}_2}
\left[
\left( \frac{1 - \hat{u}_1}{1 + \hat{u}_1} \right)^{1/4}
\left( \frac{1 - \hat{u}_2}{1 + \hat{u}_2} \right)^{1/4}
-
\left( \frac{1 + \hat{u}_1}{1 - \hat{u}_1} \right)^{1/4}
\left( \frac{1 + \hat{u}_2}{1 - \hat{u}_2} \right)^{1/4}
\right]
\nonumber\\
&\qquad\qquad\quad \
- 2 \pi i \, \delta^\prime (\hat{u}_1 + \hat{u}_2)
\, , \\
&
B^{(4)}_{\rm GG} (\hat{u}_1, \hat{u}_2) 
=
\frac{\mathcal P}{[(\hat{u}_1 - \hat{u}_2)^2]_+}
\left[
\left( \frac{1 - \hat{u}_1}{1 + \hat{u}_1} \right)^{1/4}
\left( \frac{1 + \hat{u}_2}{1 - \hat{u}_2} \right)^{1/4}
+
\left( \frac{1 + \hat{u}_1}{1 - \hat{u}_1} \right)^{1/4}
\left( \frac{1 - \hat{u}_2}{1 + \hat{u}_2} \right)^{1/4}
-
2
\right]
\nonumber\\
&\qquad\qquad\quad \
+
\frac{\mathcal P}{[(\hat{u}_1 + \hat{u}_2)^2]_+}
\left[
\left( \frac{1 - \hat{u}_1}{1 + \hat{u}_1} \right)^{1/4}
\left( \frac{1 - \hat{u}_2}{1 + \hat{u}_2} \right)^{1/4}
+
\left( \frac{1 + \hat{u}_1}{1 - \hat{u}_1} \right)^{1/4}
\left( \frac{1 + \hat{u}_2}{1 - \hat{u}_2} \right)^{1/4}
\right]
\nonumber\\
&\qquad
+
\frac{2 - \hat{u}_1^2 - \hat{u}_2^2}{4 (1- \hat{u}_1^2) (1- \hat{u}_2^2)}
\frac{\mathcal{P}}{\hat{u}_1 - \hat{u}_2}
\left[
\left( \frac{1 - \hat{u}_1}{1 + \hat{u}_1} \right)^{1/4}
\left( \frac{1 + \hat{u}_2}{1 - \hat{u}_2} \right)^{1/4}
-
\left( \frac{1 + \hat{u}_1}{1 - \hat{u}_1} \right)^{1/4}
\left( \frac{1 - \hat{u}_2}{1 + \hat{u}_2} \right)^{1/4}
\right]
\nonumber\\
&\qquad
+
\frac{2 - \hat{u}_1^2 - \hat{u}_2^2}{4 (1- \hat{u}_1^2) (1- \hat{u}_2^2)}
\frac{\mathcal{P}}{\hat{u}_1 + \hat{u}_2}
\left[
\left( \frac{1 - \hat{u}_1}{1 + \hat{u}_1} \right)^{1/4}
\left( \frac{1 - \hat{u}_2}{1 + \hat{u}_2} \right)^{1/4}
-
\left( \frac{1 + \hat{u}_1}{1 - \hat{u}_1} \right)^{1/4}
\left( \frac{1 + \hat{u}_2}{1 - \hat{u}_2} \right)^{1/4}
\right]
\nonumber\\
&\qquad\qquad\quad \
+ 2 \pi i \, \delta^\prime (\hat{u}_1 + \hat{u}_2)
\, ,
\end{align}
and
\begin{align}
C^{(1)}_{\rm GG} (\hat{u}_1, \hat{u}_2) 
&
=
\frac{1 + \hat{u}_1 \hat{u}_2}{(1 - \hat{u}_1^2)(1 - \hat{u}_2^2)}
\left[
\left( \frac{1 - \hat{u}_1}{1 + \hat{u}_1} \right)^{1/4}
\left( \frac{1 + \hat{u}_2}{1 - \hat{u}_2} \right)^{1/4}
-
\left( \frac{1 + \hat{u}_1}{1 - \hat{u}_1} \right)^{1/4}
\left( \frac{1 - \hat{u}_2}{1 + \hat{u}_2} \right)^{1/4}
\right]
\nonumber\\
&
+
\frac{1 - \hat{u}_1 \hat{u}_2}{(1 - \hat{u}_1^2)(1 - \hat{u}_2^2)}
\left[
\left( \frac{1 - \hat{u}_1}{1 + \hat{u}_1} \right)^{1/4}
\left( \frac{1 - \hat{u}_2}{1 + \hat{u}_2} \right)^{1/4}
-
\left( \frac{1 + \hat{u}_1}{1 - \hat{u}_1} \right)^{1/4}
\left( \frac{1 + \hat{u}_2}{1 - \hat{u}_2} \right)^{1/4}
\right]
\, ,
\\
C^{(3)}_{\rm GG} (\hat{u}_1, \hat{u}_2) 
&
=
-
\frac{1 + \hat{u}_1 \hat{u}_2}{(1 - \hat{u}_1^2)(1 - \hat{u}_2^2)}
\left[
\left( \frac{1 - \hat{u}_1}{1 + \hat{u}_1} \right)^{1/4}
\left( \frac{1 + \hat{u}_2}{1 - \hat{u}_2} \right)^{1/4}
+
\left( \frac{1 + \hat{u}_1}{1 - \hat{u}_1} \right)^{1/4}
\left( \frac{1 - \hat{u}_2}{1 + \hat{u}_2} \right)^{1/4}
\right]
\nonumber\\
&
+
\frac{1 - \hat{u}_1 \hat{u}_2}{(1 - \hat{u}_1^2)(1 - \hat{u}_2^2)}
\left[
\left( \frac{1 - \hat{u}_1}{1 + \hat{u}_1} \right)^{1/4}
\left( \frac{1 - \hat{u}_2}{1 + \hat{u}_2} \right)^{1/4}
+
\left( \frac{1 + \hat{u}_1}{1 - \hat{u}_1} \right)^{1/4}
\left( \frac{1 + \hat{u}_2}{1 - \hat{u}_2} \right)^{1/4}
\right]
\, ,
\\
C^{(4)}_{\rm GG} (\hat{u}_1, \hat{u}_2) 
&
=
\frac{1 + \hat{u}_1 \hat{u}_2}{(1 - \hat{u}_1^2)(1 - \hat{u}_2^2)}
\left[
\left( \frac{1 - \hat{u}_1}{1 + \hat{u}_1} \right)^{1/4}
\left( \frac{1 + \hat{u}_2}{1 - \hat{u}_2} \right)^{1/4}
+
\left( \frac{1 + \hat{u}_1}{1 - \hat{u}_1} \right)^{1/4}
\left( \frac{1 - \hat{u}_2}{1 + \hat{u}_2} \right)^{1/4}
\right]
\nonumber\\
&
+
\frac{1 - \hat{u}_1 \hat{u}_2}{(1 - \hat{u}_1^2)(1 - \hat{u}_2^2)}
\left[
\left( \frac{1 - \hat{u}_1}{1 + \hat{u}_1} \right)^{1/4}
\left( \frac{1 - \hat{u}_2}{1 + \hat{u}_2} \right)^{1/4}
+
\left( \frac{1 + \hat{u}_1}{1 - \hat{u}_1} \right)^{1/4}
\left( \frac{1 + \hat{u}_2}{1 - \hat{u}_2} \right)^{1/4}
\right]
\, .
\end{align}
We used above the Hadamard regularization which is also known (to physicists) as the so-called $+$-prescription. We verified that $f_{\rm GG}^{(1)} (u_1, u_2) = f_{\rm GG}^{(2)} (u_2, u_1)$.
The above expressions coincide with the string calculation of Ref.\ \cite{Bianchi:2015vgw}.

\subsection{Fermion-gluon case}
\label{FGAppendix}

Finally, we quote the gauge-fermion phases. These are
\begin{align}
A^{(1)}_{\rm Gf} (\hat{u}_1, \hat{u}_2) 
&
=
-
\frac{1}{\hat{u}_1 - \hat{u}_2}
\left[
\left( \frac{1 - \hat{u}_1}{1 + \hat{u}_1} \right)^{1/4}
\left( \frac{\hat{u}_2 + 1}{\hat{u}_2 - 1} \right)^{1/4}
+
\left( \frac{1 + \hat{u}_1}{1 - \hat{u}_1} \right)^{1/4}
\left( \frac{\hat{u}_2 - 1}{\hat{u}_2 + 1} \right)^{1/4}
- \sqrt{2}
\right]
\nonumber\\
&
- 
\frac{1}{\hat{u}_1 + \hat{u}_2}
\left[
\left( \frac{1 - \hat{u}_1}{1 + \hat{u}_1} \right)^{1/4}
\left( \frac{\hat{u}_2 - 1}{\hat{u}_2 + 1} \right)^{1/4}
+
\left( \frac{1 + \hat{u}_1}{1 - \hat{u}_1} \right)^{1/4}
\left( \frac{\hat{u}_2 + 1}{\hat{u}_2 - 1} \right)^{1/4}
- \sqrt{2}
\right]
\, , \\
A^{(2)}_{\rm Gf} (\hat{u}_1, \hat{u}_2) 
&
=
\frac{1}{\hat{u}_1 - \hat{u}_2}
\left[
\left( \frac{1 - \hat{u}_1}{1 + \hat{u}_1} \right)^{1/4}
\left( \frac{\hat{u}_2 + 1}{\hat{u}_2 - 1} \right)^{1/4}
+
\left( \frac{1 + \hat{u}_1}{1 - \hat{u}_1} \right)^{1/4}
\left( \frac{\hat{u}_2 - 1}{\hat{u}_2 + 1} \right)^{1/4}
- \sqrt{2}
\right]
\nonumber\\
&
- 
\frac{1}{\hat{u}_1 + \hat{u}_2}
\left[
\left( \frac{1 - \hat{u}_1}{1 + \hat{u}_1} \right)^{1/4}
\left( \frac{\hat{u}_2 - 1}{\hat{u}_2 + 1} \right)^{1/4}
+
\left( \frac{1 + \hat{u}_1}{1 - \hat{u}_1} \right)^{1/4}
\left( \frac{\hat{u}_2 + 1}{\hat{u}_2 - 1} \right)^{1/4}
- \sqrt{2}
\right]
\, , \\
A^{(3)}_{\rm Gf} (\hat{u}_1, \hat{u}_2) 
&
=
\frac{1}{\hat{u}_1 - \hat{u}_2}
\left[
\left( \frac{1 - \hat{u}_1}{1 + \hat{u}_1} \right)^{1/4}
\left( \frac{\hat{u}_2 + 1}{\hat{u}_2 - 1} \right)^{1/4}
-
\left( \frac{1 + \hat{u}_1}{1 - \hat{u}_1} \right)^{1/4}
\left( \frac{\hat{u}_2 - 1}{\hat{u}_2 + 1} \right)^{1/4}
+ i \sqrt{2}
\right]
\nonumber\\
&
- 
\frac{1}{\hat{u}_1 + \hat{u}_2}
\left[
\left( \frac{1 - \hat{u}_1}{1 + \hat{u}_1} \right)^{1/4}
\left( \frac{\hat{u}_2 - 1}{\hat{u}_2 + 1} \right)^{1/4}
-
\left( \frac{1 + \hat{u}_1}{1 - \hat{u}_1} \right)^{1/4}
\left( \frac{\hat{u}_2 + 1}{\hat{u}_2 - 1} \right)^{1/4}
+
i \sqrt{2}
\right]
\, , \\
A^{(4)}_{\rm Gf} (\hat{u}_1, \hat{u}_2) 
&
=
-
\frac{1}{\hat{u}_1 - \hat{u}_2}
\left[
\left( \frac{1 - \hat{u}_1}{1 + \hat{u}_1} \right)^{1/4}
\left( \frac{\hat{u}_2 + 1}{\hat{u}_2 - 1} \right)^{1/4}
-
\left( \frac{1 + \hat{u}_1}{1 - \hat{u}_1} \right)^{1/4}
\left( \frac{\hat{u}_2 - 1}{\hat{u}_2 + 1} \right)^{1/4}
+ i \sqrt{2}
\right]
\nonumber\\
&
- 
\frac{1}{\hat{u}_1 + \hat{u}_2}
\left[
\left( \frac{1 - \hat{u}_1}{1 + \hat{u}_1} \right)^{1/4}
\left( \frac{\hat{u}_2 - 1}{\hat{u}_2 + 1} \right)^{1/4}
-
\left( \frac{1 + \hat{u}_1}{1 - \hat{u}_1} \right)^{1/4}
\left( \frac{\hat{u}_2 + 1}{\hat{u}_2 - 1} \right)^{1/4}
+
i \sqrt{2}
\right]
\, ,
\end{align}
and
\begin{align}
B^{(1)}_{\rm Gf} (\hat{u}_1, \hat{u}_2) 
&
=
\frac{\hat{u}_1 + \hat{u}_2}{4 (1 - \hat{u}_1^2)(1 - \hat{u}_2^2)}
\left[
\left( \frac{1 - \hat{u}_1}{1 + \hat{u}_1} \right)^{1/4}
\left( \frac{\hat{u}_2 + 1}{\hat{u}_2 - 1} \right)^{1/4}
+
\left( \frac{1 + \hat{u}_1}{1 - \hat{u}_1} \right)^{1/4}
\left( \frac{\hat{u}_2 - 1}{\hat{u}_2 + 1} \right)^{1/4}
\right]
\nonumber\\
&
+
\frac{\hat{u}_1 - \hat{u}_2}{4 (1 - \hat{u}_1^2)(1 - \hat{u}_2^2)}
\left[
\left( \frac{1 - \hat{u}_1}{1 + \hat{u}_1} \right)^{1/4}
\left( \frac{\hat{u}_2 - 	1}{\hat{u}_2 + 1} \right)^{1/4}
+
\left( \frac{1 + \hat{u}_1}{1 - \hat{u}_1} \right)^{1/4}
\left( \frac{\hat{u}_2 + 1}{\hat{u}_2 - 1} \right)^{1/4}
\right]
\, ,
\nonumber\\
B^{(2)}_{\rm Gf} (\hat{u}_1, \hat{u}_2) 
&
=
-
\frac{\hat{u}_1 + \hat{u}_2}{4 (1 - \hat{u}_1^2)(1 - \hat{u}_2^2)}
\left[
\left( \frac{1 - \hat{u}_1}{1 + \hat{u}_1} \right)^{1/4}
\left( \frac{\hat{u}_2 + 1}{\hat{u}_2 - 1} \right)^{1/4}
+
\left( \frac{1 + \hat{u}_1}{1 - \hat{u}_1} \right)^{1/4}
\left( \frac{\hat{u}_2 - 1}{\hat{u}_2 + 1} \right)^{1/4}
\right]
\nonumber\\
&
+
\frac{\hat{u}_1 - \hat{u}_2}{4 (1 - \hat{u}_1^2)(1 - \hat{u}_2^2)}
\left[
\left( \frac{1 - \hat{u}_1}{1 + \hat{u}_1} \right)^{1/4}
\left( \frac{\hat{u}_2 - 	1}{\hat{u}_2 + 1} \right)^{1/4}
+
\left( \frac{1 + \hat{u}_1}{1 - \hat{u}_1} \right)^{1/4}
\left( \frac{\hat{u}_2 + 1}{\hat{u}_2 - 1} \right)^{1/4}
\right]
\, ,
\nonumber\\
B^{(3)}_{\rm Gf} (\hat{u}_1, \hat{u}_2) 
&
=
-
\frac{\hat{u}_1 + \hat{u}_2}{4 (1 - \hat{u}_1^2)(1 - \hat{u}_2^2)}
\left[
\left( \frac{1 - \hat{u}_1}{1 + \hat{u}_1} \right)^{1/4}
\left( \frac{\hat{u}_2 + 1}{\hat{u}_2 - 1} \right)^{1/4}
-
\left( \frac{1 + \hat{u}_1}{1 - \hat{u}_1} \right)^{1/4}
\left( \frac{\hat{u}_2 - 1}{\hat{u}_2 + 1} \right)^{1/4}
\right]
\nonumber\\
&
+
\frac{\hat{u}_1 - \hat{u}_2}{4 (1 - \hat{u}_1^2)(1 - \hat{u}_2^2)}
\left[
\left( \frac{1 - \hat{u}_1}{1 + \hat{u}_1} \right)^{1/4}
\left( \frac{\hat{u}_2 - 	1}{\hat{u}_2 + 1} \right)^{1/4}
-
\left( \frac{1 + \hat{u}_1}{1 - \hat{u}_1} \right)^{1/4}
\left( \frac{\hat{u}_2 + 1}{\hat{u}_2 - 1} \right)^{1/4}
\right]
\, ,
\nonumber\\
B^{(4)}_{\rm Gf} (\hat{u}_1, \hat{u}_2) 
&
=
\frac{\hat{u}_1 + \hat{u}_2}{4 (1 - \hat{u}_1^2)(1 - \hat{u}_2^2)}
\left[
\left( \frac{1 - \hat{u}_1}{1 + \hat{u}_1} \right)^{1/4}
\left( \frac{\hat{u}_2 + 1}{\hat{u}_2 - 1} \right)^{1/4}
-
\left( \frac{1 + \hat{u}_1}{1 - \hat{u}_1} \right)^{1/4}
\left( \frac{\hat{u}_2 - 1}{\hat{u}_2 + 1} \right)^{1/4}
\right]
\nonumber\\
&
+
\frac{\hat{u}_1 - \hat{u}_2}{4 (1 - \hat{u}_1^2)(1 - \hat{u}_2^2)}
\left[
\left( \frac{1 - \hat{u}_1}{1 + \hat{u}_1} \right)^{1/4}
\left( \frac{\hat{u}_2 - 	1}{\hat{u}_2 + 1} \right)^{1/4}
-
\left( \frac{1 + \hat{u}_1}{1 - \hat{u}_1} \right)^{1/4}
\left( \frac{\hat{u}_2 + 1}{\hat{u}_2 - 1} \right)^{1/4}
\right]
\, ,
\nonumber\\
\end{align}
with
\begin{align}
C^{(1)}_{\rm Gf} (\hat{u}_1, \hat{u}_2) 
&
=
\frac{1 + \hat{u}_1 \hat{u}_2}{(1 - \hat{u}_1^2)(1 - \hat{u}_2^2)}
\left[
\left( \frac{1 - \hat{u}_1}{1 + \hat{u}_1} \right)^{1/4}
\left( \frac{\hat{u}_2 + 1}{\hat{u}_2 - 1} \right)^{1/4}
-
\left( \frac{1 + \hat{u}_1}{1 - \hat{u}_1} \right)^{1/4}
\left( \frac{\hat{u}_2 - 1}{\hat{u}_2 + 1} \right)^{1/4}
\right]
\nonumber\\
&
+
\frac{1 - \hat{u}_1 \hat{u}_2}{(1 - \hat{u}_1^2)(1 - \hat{u}_2^2)}
\left[
\left( \frac{1 - \hat{u}_1}{1 + \hat{u}_1} \right)^{1/4}
\left( \frac{\hat{u}_2 - 	1}{\hat{u}_2 + 1} \right)^{1/4}
-
\left( \frac{1 + \hat{u}_1}{1 - \hat{u}_1} \right)^{1/4}
\left( \frac{\hat{u}_2 + 1}{\hat{u}_2 - 1} \right)^{1/4}
\right]
\, ,
\nonumber\\
C^{(2)}_{\rm Gf} (\hat{u}_1, \hat{u}_2) 
&
=
-
\frac{1 + \hat{u}_1 \hat{u}_2}{(1 - \hat{u}_1^2)(1 - \hat{u}_2^2)}
\left[
\left( \frac{1 - \hat{u}_1}{1 + \hat{u}_1} \right)^{1/4}
\left( \frac{\hat{u}_2 + 1}{\hat{u}_2 - 1} \right)^{1/4}
-
\left( \frac{1 + \hat{u}_1}{1 - \hat{u}_1} \right)^{1/4}
\left( \frac{\hat{u}_2 - 1}{\hat{u}_2 + 1} \right)^{1/4}
\right]
\nonumber\\
&
+
\frac{1 - \hat{u}_1 \hat{u}_2}{(1 - \hat{u}_1^2)(1 - \hat{u}_2^2)}
\left[
\left( \frac{1 - \hat{u}_1}{1 + \hat{u}_1} \right)^{1/4}
\left( \frac{\hat{u}_2 - 	1}{\hat{u}_2 + 1} \right)^{1/4}
-
\left( \frac{1 + \hat{u}_1}{1 - \hat{u}_1} \right)^{1/4}
\left( \frac{\hat{u}_2 + 1}{\hat{u}_2 - 1} \right)^{1/4}
\right]
\, ,
\nonumber\\
C^{(3)}_{\rm Gf} (\hat{u}_1, \hat{u}_2) 
&
=
-
\frac{1 + \hat{u}_1 \hat{u}_2}{(1 - \hat{u}_1^2)(1 - \hat{u}_2^2)}
\left[
\left( \frac{1 - \hat{u}_1}{1 + \hat{u}_1} \right)^{1/4}
\left( \frac{\hat{u}_2 + 1}{\hat{u}_2 - 1} \right)^{1/4}
+
\left( \frac{1 + \hat{u}_1}{1 - \hat{u}_1} \right)^{1/4}
\left( \frac{\hat{u}_2 - 1}{\hat{u}_2 + 1} \right)^{1/4}
\right]
\nonumber\\
&
+
\frac{1 - \hat{u}_1 \hat{u}_2}{(1 - \hat{u}_1^2)(1 - \hat{u}_2^2)}
\left[
\left( \frac{1 - \hat{u}_1}{1 + \hat{u}_1} \right)^{1/4}
\left( \frac{\hat{u}_2 - 	1}{\hat{u}_2 + 1} \right)^{1/4}
+
\left( \frac{1 + \hat{u}_1}{1 - \hat{u}_1} \right)^{1/4}
\left( \frac{\hat{u}_2 + 1}{\hat{u}_2 - 1} \right)^{1/4}
\right]
\, ,
\nonumber\\
C^{(4)}_{\rm Gf} (\hat{u}_1, \hat{u}_2) 
&
=
\frac{1 + \hat{u}_1 \hat{u}_2}{(1 - \hat{u}_1^2)(1 - \hat{u}_2^2)}
\left[
\left( \frac{1 - \hat{u}_1}{1 + \hat{u}_1} \right)^{1/4}
\left( \frac{\hat{u}_2 + 1}{\hat{u}_2 - 1} \right)^{1/4}
+
\left( \frac{1 + \hat{u}_1}{1 - \hat{u}_1} \right)^{1/4}
\left( \frac{\hat{u}_2 - 1}{\hat{u}_2 + 1} \right)^{1/4}
\right]
\nonumber\\
&
+
\frac{1 - \hat{u}_1 \hat{u}_2}{(1 - \hat{u}_1^2)(1 - \hat{u}_2^2)}
\left[
\left( \frac{1 - \hat{u}_1}{1 + \hat{u}_1} \right)^{1/4}
\left( \frac{\hat{u}_2 - 	1}{\hat{u}_2 + 1} \right)^{1/4}
+
\left( \frac{1 + \hat{u}_1}{1 - \hat{u}_1} \right)^{1/4}
\left( \frac{\hat{u}_2 + 1}{\hat{u}_2 - 1} \right)^{1/4}
\right]
\, .
\end{align}


\end{document}